\documentclass[fleqn,usenatbib]{mnras}

\usepackage{newtxtext,newtxmath}
 
\usepackage[T1]{fontenc}
\usepackage{ae,aecompl}

\usepackage{graphicx}	
\usepackage{amsmath}	
\usepackage{amssymb}	
\usepackage{mwe}
\usepackage{todonotes}
\usepackage{xcolor}
\usepackage{hyperref}

\usepackage{siunitx}
\usepackage{enumitem}
\usepackage{lscape}
          \usepackage{pdflscape}
          \usepackage{morefloats}
			\usepackage{booktabs} 
              \usepackage{multirow}
              \usepackage{array}
              \usepackage{longtable}
               \usepackage{tabu}
                \usepackage{threeparttable} 
                 \usepackage{threeparttablex} 
                 \usepackage{rotating}
                  \usepackage{xtab}
                   \usepackage{multicol}
                   \usepackage{adjustbox}
                    
\usepackage{upgreek}
\usepackage{units}
\usepackage{caption}
\usepackage{subcaption} 

\newcommand{\kms}{${\rm km~s}^{-1}$}

\newcommand{\mstar}{$M_{\ast}$}

\DeclareRobustCommand{\ion}[2]{%
\relax\ifmmode
\ifx\testbx\f@series
{\mathbf{#1\,\mathsc{#2}}}\else
{\mathrm{#1\,\mathsc{#2}}}\fi
\else\textup{#1\,{\mdseries\textsc{#2}}}%
\fi}

\title {Kinematics of molecular gas in star-forming galaxies with large-scale ionised outflows}

\author[L. M. Hogarth et al.]{\Large \parbox{\textwidth}{
L. M. Hogarth$^{1}$\thanks{E-mail: l.hogarth.18@ucl.ac.uk},
A. Saintonge$^{1}$\thanks{E-mail: a.saintonge@ucl.ac.uk}
and T. A. Davis$^{2}$}
\\
\\
$^{1}$University College London, Department of Physics and Astronomy, Gower Street, London, WC1E 6BT, UK \\
$^{2}$Cardiff Hub for Astrophysics Research \&\ Technology, School of Physics \&\ Astronomy, Cardiff University, Queens Buildings, Cardiff, CF24 3AA, UK}
\date{Accepted XXX. Received YYY; in original form ZZZ}

\pubyear{2021}

\begin{document}
\label{firstpage}
\pagerange{\pageref{firstpage}--\pageref{lastpage}}
\maketitle

\begin{abstract}
We investigate the kinematics of the molecular gas in a sample of seven edge-on (i>60$^\circ$) galaxies identified as hosting large-scale outflows of ionised gas, using ALMA CO(1-0) observations at $\sim$~1~kpc resolution. We build on \citet{hogarth21} (H21), where we find that molecular gas is more centrally concentrated in galaxies which host winds than in control objects. We perform full 3-dimensional kinematic modelling with multiple combinations of kinematic components, allowing us to infer whether these objects share any similarities in their molecular gas structure. We use modelling to pinpoint the kinematic centre of each galaxy, in order to interpret their minor- and major-axis position velocity diagrams (PVDs). From the PVDs, we find that the bulk of the molecular gas in our galaxies is dynamically cold, tracing the rotation curves predicted by our symmetric, rotation-dominated models, but with minor flux asymmetries. Most notably, we find evidence of radial gas motion in a subset of our objects, which demonstrate a characteristic ``twisting'' in their minor-axis PVDs generally associated with gas flow along the plane of a galaxy. In our highest S/N object, we include bi-symmetric radial flow in our kinematic model, and find (via the Bayesian Information Criterion) that the presence of radial gas motion is strongly favoured. This may provide one mechanism by which molecular gas and star formation are centrally concentrated, enabling the launch of massive ionised gas winds. However, in the remainder of our sample, we do not observe evidence that gas is being driven radially, once again emphasising the variety of physical processes that may be powering the outflows in these objects, as originally noted in H21. 
\end{abstract}




\section{Introduction}
\label{sec: intro}

The processes that trigger angular momentum to collapse inwards towards the inner regions of a galaxy are relatively well understood on kiloparsec scales. Generally, we understand that the formation of an asymmetric component in a galaxy generates a forcing frequency which exerts a torque and destabilises material in the disc. In the case of weak perturbations (like spiral arms) this can cause streaming motions, with gas typically flowing inwards along spiral arms \citep{sellwood21}. In stronger disturbances, material is forced on to a series of resonant orbits, where they do not experience a torque from the rotation of the asymmetric component. Primarily, these orbits fall at Lindblad resonances, which describes a series of closed stable orbits where an object's epicyclic frequency is an integer-multiple of the forcing frequency \citep{combes91, bertin14}. The strongest resonance is at the co-rotation radius; where the orbital velocity of the stars/gas is equal to the velocity of the rotating asymmetric component, often coincident with the turnover radius of the disc's velocity profile.   

Strongly asymmetric components in a galaxy can take the form of bars or distortions generated by interactions or mergers \citep[which can subsequently cause the formation of a bar, ][]{eliche-moral11}. Bars in particular are known to be vital in bulge formation by driving gas into the inner regions of galaxies, which enhances the central mass concentration and subsequently leads to starburst activity \citep{spinoso17}. In edge-on galaxies, the presence of a bar can be inferred from the existence of multiple components in the kinematics of the gaseous tracers, describing rings of gas at the resonant orbits of a disc subjected to a pattern frequency \citep{bureau99}. The surface profile of this gas would also appear more centralised as angular momentum is driven inwards. The consequences of bar-induced inflow are an increase in star formation activity, possibly leading to the launching of outflows \citep{spinoso17}, the build-up of a pseudo bulge, and the subsequent suppression of star-formation activity \citep{gavazzi15,zolotov15, tacchella16}.

\begin{figure}
    \centering
    \includegraphics[scale=0.45]{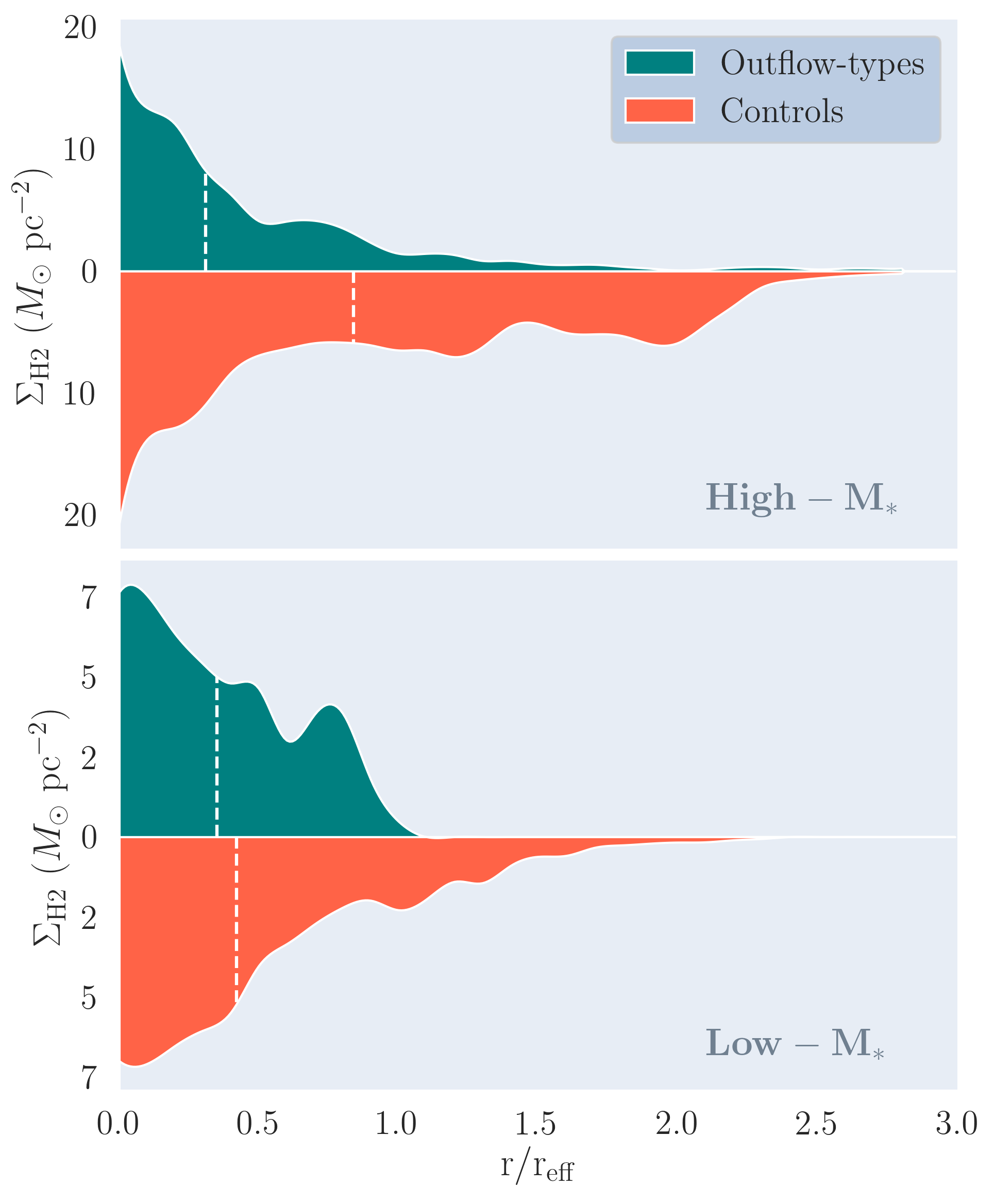}
     \caption{Radial $\rm \Sigma_{H2}$ profiles of molecular gas in the high and low-$M_*$ outflow-type samples from ALMA and derived from SAMI samples (converted to molecular gas density using the $\Sigma_{SFR}$-$\Sigma_{H2}$ relation from \citet{leroy13}) defined in H21. \textbf{Top:} The high-$\rm M_*$ sample of outflow-type objects given by the teal vertical density profile and the high-$M_*$ SAMI sample in orange. \textbf{Bottom:} The low-$\rm M_*$ sample of outflow-type objects given by the teal vertical density profile and the low-$M_*$ SAMI sample in orange. In both panels the vertical white dashed lines indicate the radius within which half of the integral under the curves lies to give an indication of the different extents of the gas.}
     \label{fig: viogas}
\end{figure}

We explored these ideas in \citet[][hereafter H21]{hogarth21} by analysing ALMA CO(1-0) observations of ``normal" star-forming galaxies revealed by optical integral field spectroscopy to be hosting large-scale ionised gas outflows \citep{ho16}. The key finding from H21 is that the molecular gas (and star-formation) in galaxies hosting large-scale ionised outflows is more centralised than in control objects, as also independently observed by \citet{bao21}. This result is illustrated in Figure \ref{fig: viogas}. The figure shows the molecular gas mass surface density ($\Sigma_{H_2}$) as a function of normalised radius along the major axis, after the galaxies hosting strong ionised-gas outflows are divided into low- ($\rm \log ( M_*/M_{\odot}) <10$) and high-mass ($\rm \log ( M_*/M_{\odot} ) >10$) sub-samples. The mean $\Sigma_{H_2}$ profile of each sub-sample is compared to that of a control sample matched in \mstar\ and SFR, and selected from the SAMI Galaxy Survey\footnote{As CO maps are not available for all of these SAMI control objects, the $\Sigma_{H_2}$ profiles are derived from  $\rm \Sigma _{SFR}$ using the results of \citet{leroy13} (see details and validation of this procedure in H21)} \citep{croom12, bryant15, green18, scott18, croom21}. While the central gas surface densities (and SFR surface densities) are similar in both outflow and control galaxies, the gas is significantly more centrally concentrated in the outflow sample, especially in the high-mass sub-sample. 

Despite this trend, the molecular gas profiles and kinematics of individual galaxies show significant diversity. In H21, we speculated that this diversity points towards a range of physical mechanisms that can affect the kinematics and distribution of the molecular gas, while at the same time enabling the galaxies to launch large-scale ionised gas outflows. In this paper, we conduct kinematic modelling of the outflowing galaxies from H21 to (1) explore the diversity of the molecular gas kinematic structures across the sample, and (2) test whether a single, dominant mechanism such as bar instabilities can explain the central gas and SFR concentrations. Ultimately, we wish to pinpoint the process (or processes) by which large-scale ionised outflows can be powered and whether the mechanism that drives the apparent centralisation of gas is responsible.

Throughout this paper we adopt a standard $\Lambda$CDM cosmology with $H_0=70\ $\kms~Mpc$^{-1}$, $\Omega _{m_0}=0.3$, $\Omega _{\Lambda}=0.7$ and a \citet{chabrier03} IMF.

\section{Sample selection and data}
\label{sec: selectionanddata}

A full description of our sample selection and data products is outlined in H21; we provide a cursory overview here, and refer to that publication for more details. Our objects are selected from the SAMI Galaxy Survey, an optical integral field  spectroscopic survey of 3068 spatially resolved galaxies at $z \lesssim 0.1$ \citep{croom12, bryant15, green18, scott18, croom21}. We use the two diagnostics developed by \cite{ho16} to identify edge-on galaxies (i.e. $i \gtrapprox 70^{\circ}$) harbouring large-scale ionised winds, which we refer to as ``outflow-types''. We targeted 9 of these galaxies, as well as 7 control objects, for follow-up CO(1-0) observations with ALMA in Cycle 5. These observations were done in Band 3 with a synthesised beam of 1\arcsec\ ($\approx 0.5$~kpc - $1$~kpc) and a spectral resolution of $\approx$ 10~\kms, with an on-source integration time between 45-53 minutes for each object. 

We use standard CASA (Common Astronomy Software Applications) pipeline subroutines \citep{mcmullin07} to reduce the ALMA data. We then extract total CO(1-0) emission (and total molecular gas masses), moment maps and position velocity diagrams (PVDs) as described in H21. 

\begin{figure*} 
   
    \centering
        \includegraphics[scale=0.75]{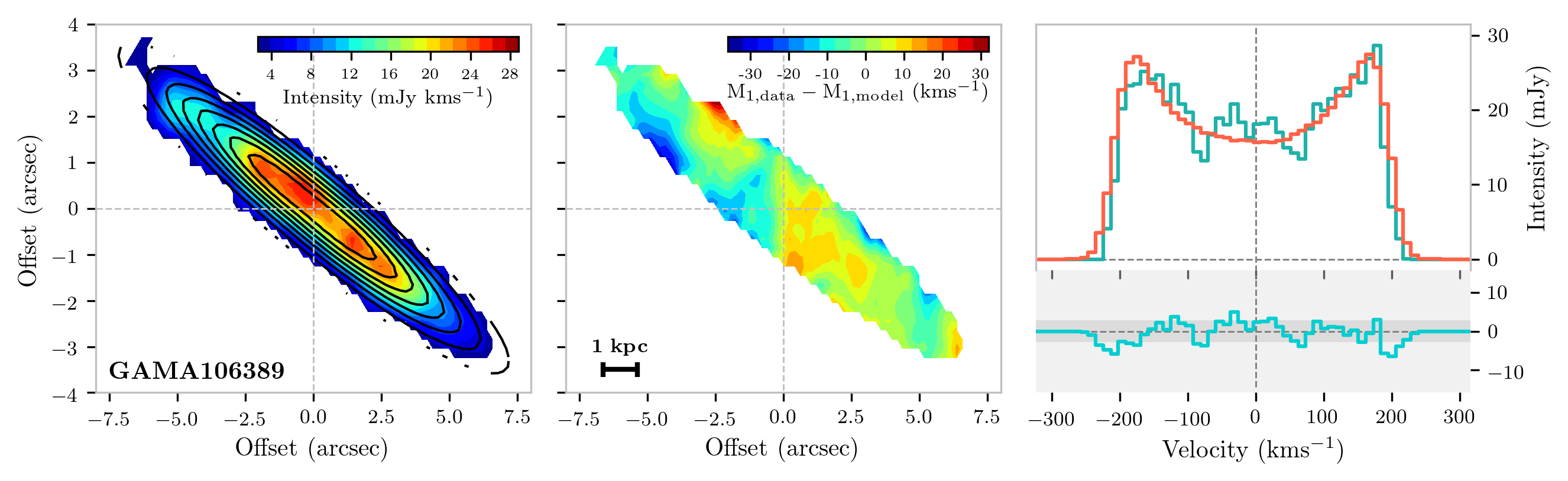}
    \centering
        \includegraphics[scale=0.75]{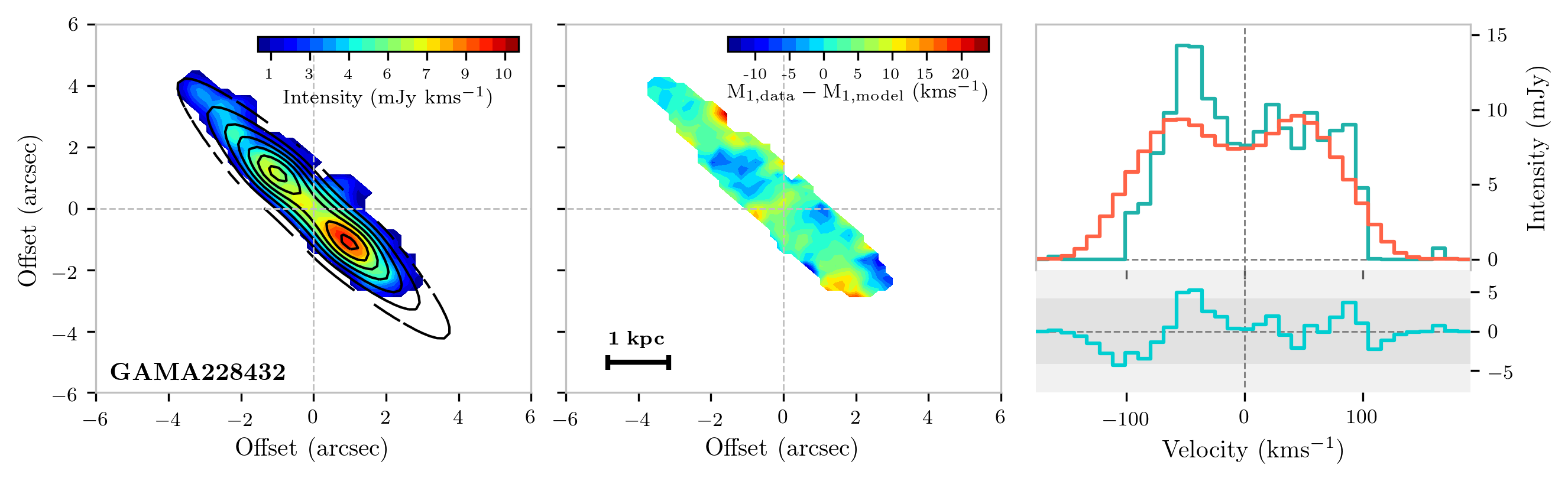}
    \centering
        \includegraphics[scale=0.75]{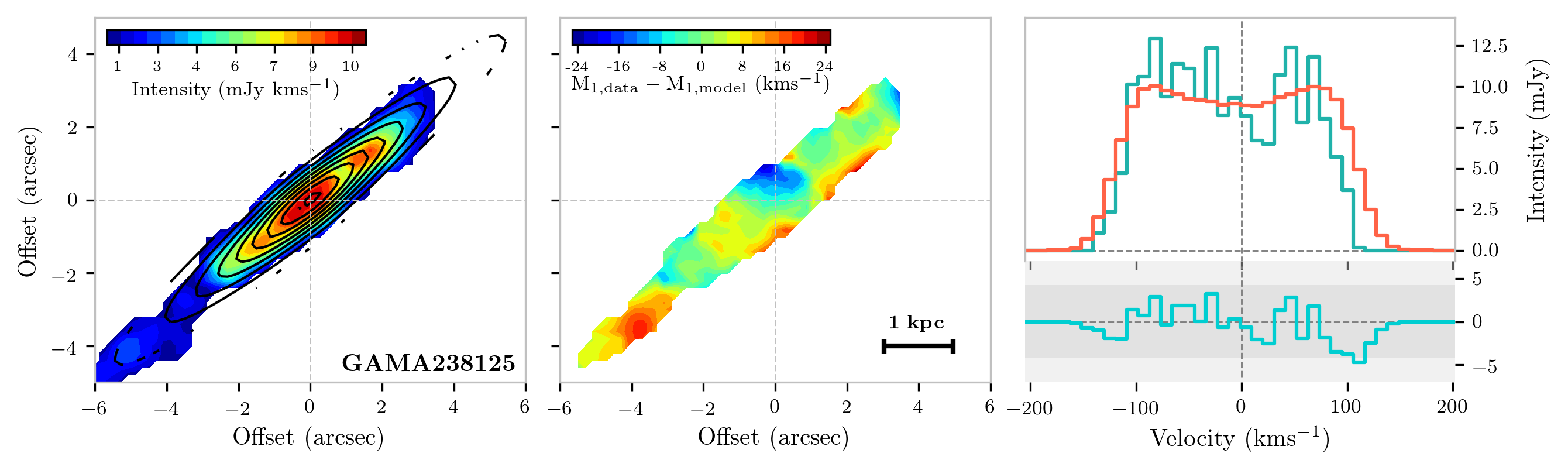}
    \centering
        \includegraphics[scale=0.75]{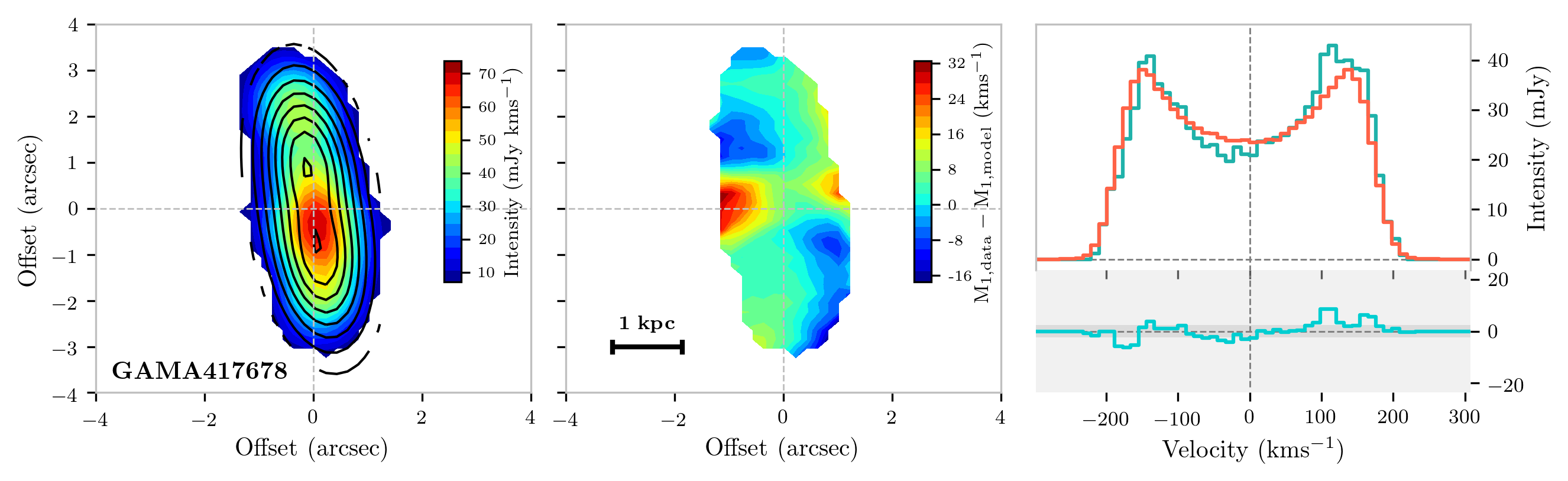}

    \caption{Best-fit models created using KinMS for seven of our outflow-type objects. Each row of panels in the figure gives details of the kinematic fit for each object respectively. In each row of the three panels: \textbf{left}, the zeroth moment of the CO(1-0) data cube (filled coloured contours) with the zeroth moment of the model overlaid in the black contours; \textbf{middle}, the residual velocity map obtained by subtracting the observed and modelled intensity weighted mean velocity maps (i.e. $\rm M_{1,data}-M_{1,model}$); \textbf{right}, the spectrum extracted from the data when collapsed over its spatial axes (teal line) with the spectrum from the model (red line) and the residual of the two beneath (with the RMS noise level of the data given by the dark shaded region).}
    \label{fig: modplots}
\end{figure*}

\begin{figure*} 
   
    \centering
        \includegraphics[scale=0.75]{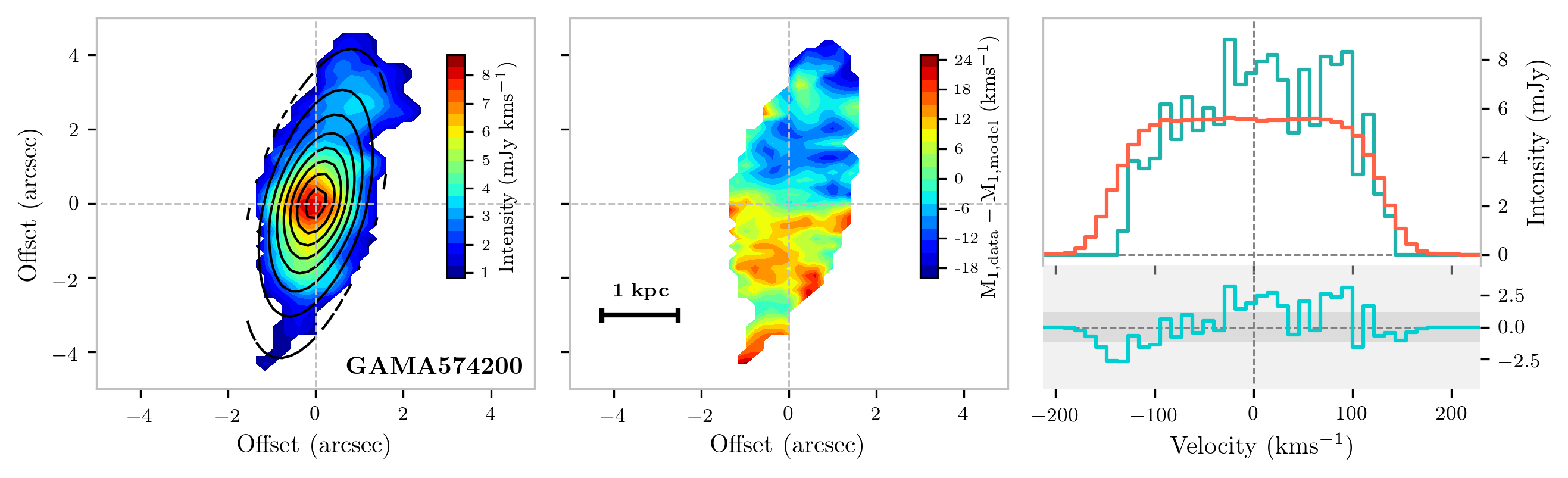}
    \centering
        \includegraphics[scale=0.75]{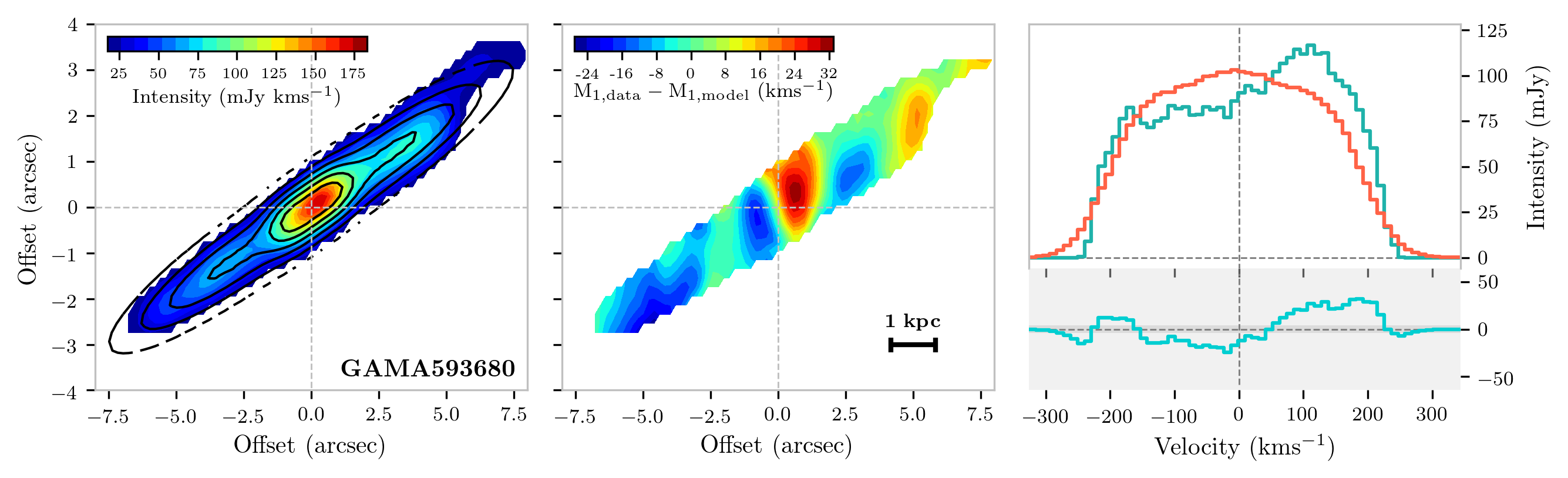}
    \centering
        \includegraphics[scale=0.75]{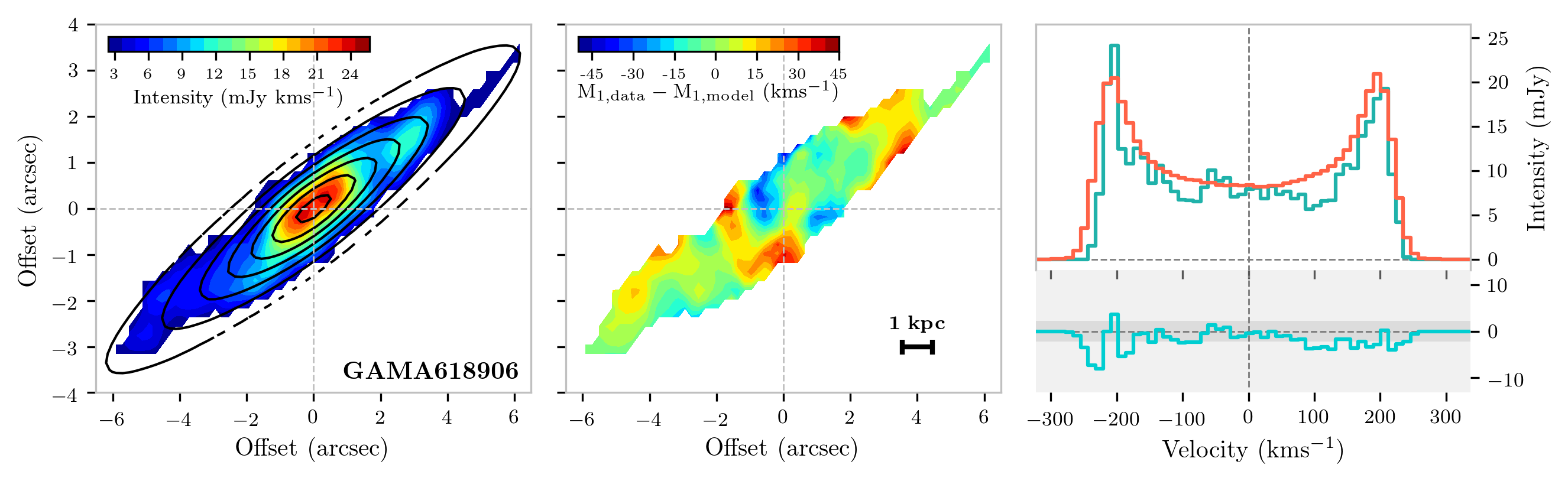}
  
    \contcaption{}
\end{figure*}

\section{Results}
\label{sec: results}

\subsection{Kinematic Modelling}
\label{subsec: Kin}

\begin{table} 
    \centering
    \caption{Surface brightness models used to construct KinMS models of the CO(1-0) gas in our ALMA outflow-type objects.} 
    \begin{tabular}{c c l c }
    	\hline
	    GAMA ID & $\rm {{S/N}_{CO}}^a$ & SB Profile$^b$ & Free Parameters \\[2pt]
	    \hline
        \hline
    	106389 & 33  &  gauss         &  11       \\[1pt]
        228432 & 15  &  exp + hole    &  10        \\[1pt]
        238125 & 14  &  exp           &  9        \\[1pt]
        31452  & 19  &  $\ldots$      &  $\ldots$ \\[1pt]
        417678 & 53  &  gauss         &  11       \\[1pt]
        567624 & <3  &  $\ldots$      &  $\ldots$ \\[1pt]
        574200 & 18  &  exp           &  9        \\[1pt]
        593680 & 100 &  gauss + gauss &  14       \\[1pt]
        618906 & 10  &  exp           &  10        \\[1pt]
	    \hline
	\end{tabular}
	\label{tab: fits}
	\begin{tablenotes}
	    \item[a] $a$ S/N ratio calculated over the integrated CO(1-0) spectra.
	    \item[b] $b$ The type of surface brightness (SB) profile used in the KinMS model, where ``exp'' indicates an exponential disc, and ``gauss'' a Gaussian ring. Multiple profile names indicate a superposition of profiles.
	    \item[Notes]Additional notes: Where rows are left blank, we either had no detection of CO(1-0) for the objects or the gas was too diffuse to model.
	    
	\end{tablenotes}
\end{table}

We perform kinematic modeling of the ALMA data to determine the kinematic centres of the CO(1-0) emission (see Figure~\ref{fig: modplots}). By finding the kinematic centre of our objects through modelling, we can more accurately centre the PVDs and assess kinematic features. Due to the high inclination of our objects, we have to use full 3D kinematic modelling to overcome projection effects. We employ the KINematic Molecular Simulation \citep[KinMS; tool of ][]{davis13, davis20}\footnote{ \url{https://github.com/TimothyADavis/KinMSpy}}\footnote{\url{https://kinms.space}}. The KinMS tool allows us to create simple simulated interferometric data cubes by defining  arbitrary surface-brightness and velocity profiles. The tool accounts for the dispersion of the gas as well as both projection and disk-thickness, while also including observational effects such as beam-smearing in the output simulation. We employ the \texttt{KinMS\_fitter}\footnote{\url{https://github.com/TimothyADavis/KinMS_fitter}} wrapper for KinMS, which acts as a front end for the most common fitting tasks within KinMS and provides a simple interface for defining surface brightness and velocity profiles. It also provides an interface to the GAStimator package\footnote{\url{https://github.com/TimothyADavis/GAStimator}}, which implements a \texttt{Python} MCMC Gibbs-sampler with adaptive stepping employed to fit the mock interferometric data cubes generated by KinMS, with predefined surface-brightness and velocity profiles, to the original data cubes.

\par Intrinsically, KinMS assumes that the gas in our observations is dynamically cold and relaxed, so the models are subsequently symmetric about their spatial and spectral kinematic centres. In order to capture the rotation-dominated gas, we use an arctangent velocity curve in each case. While arctangent velocity profiles have been widely used in the literature, they are not physically motivated, and real galaxies rotation curves are often significantly more complicated. In our case, however, excellent fits are obtained using this simple rotation curve form. By using a velocity curve model, we can effectively separate gas in our objects which has non-circular dynamics from that captured by the pure circular motion model. For each object, we fit multiple different surface brightness profiles and combinations of surface brightness profiles with the arctan velocity profile (i.e. exponential disc, exponential disc + central hole, Gaussian ring, two exponential discs, exponential disc + Gaussian ring, two Gaussian rings). The total number of parameters used in a given model is, therefore, dictated entirely by the surface brightness model used. Prior to using the MCMC fitting process, we perform a least squares fit to the data in order to obtain first guesses at the model parameters. Once we have implemented MCMC fits using all prospective surface brightness models, we select a best-fit model that minimises the Bayesian Information Criterion (BIC) (i.e. the model with the greatest likelihood with respect to a penalty incurred for larger numbers of parameters).

\par In Figure~\ref{fig: modplots}, we present the KinMS models of our objects identified as outflow-types. In Table~\ref{tab: fits}, we list the best-fit surface brightness profiles and total number of free parameters used in the fit. In addition to the parameters associated with the surface brightness profiles and arctan velocity profile, we also fit the position angle, RA centre, Dec centre, velocity centre, inclination, total flux and velocity dispersion of the gas (see Appendices~\ref{appendix: corner} \& \ref{appendix: params}). However, in the cases of GAMA228432, GAMA238125 and GAMA574200, we fix the total flux parameter in the models to our initial guess (calculated by integrating the flux of the cube) due to overestimation of this parameter when the gas profiles are significantly asymmetric.

\par There is no ubiquitous or defining feature that exists in all of the outflow-type objects. Most are best modelled by a single-component surface brightness profile, but we find no preference towards a disc-type profile as opposed to a ring-type profile. However, using these kinematically symmetric models that simulate pure rotation, we can begin to interpret any potentially dynamically disturbed features in the PVDs extracted from our outflow-type objects in Section~\ref{subsec: PVDs}.

\subsection{Circular Velocity}
\label{subsec: mge}

\begin{figure} 
    \centering
    \includegraphics[scale=0.5]{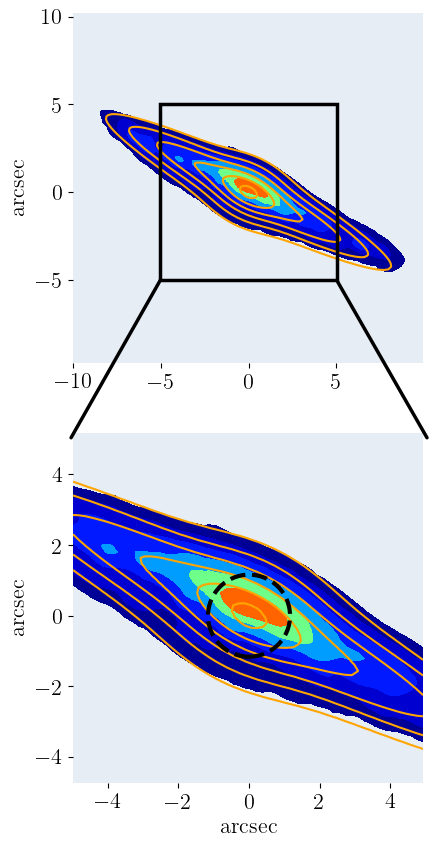}
     \caption{Example of MGE model derived from r-band HSC photometry of GAMA106389. The r-band data is given by the coloured filled isophotes and the model by the orange isophotes, both spaced by $0.5~mag~{arcsec}^{-2}$. The bottom panel is zoomed image of the central regions of the top panel. The black dashed circle represents the HSC resolution ($\rm \upsigma_{PSF} \approx 1 \arcsec$) to indicate which Gaussian components would be unresolved by our data. The MGE model centre is offset from the flux centre in this instance due to the dust lane obscuring flux on the near-side of the edge-on galaxy.}
     \label{fig: mge}
\end{figure}

Another method of determining whether the CO(1-0) gas observed in our galaxy sample is dynamically disturbed is by comparing its kinematics to the circular velocity of the objects predicated by the gravitational potential of the stellar content. We attempt this by extracting Multi-Gaussian Expansion (MGE) models \citep{emsellem94} using r-band photometry from HSC \citep[Hyper Suprime-Cam Subaru Strategic Program; ][]{aihara19}. Our MGE models are derived using the \texttt{MgeFit} package of \citet{cappellari02}. An example of the MGE models produced by implementing the \texttt{MgeFit} package on the r-band photometry on our objects is given in Figure~\ref{fig: mge}, where we contrast the original photometry for GAMA106389 with the output surface brightness model.  

\par Having obtained MGE surface brightness models for the r-band photometry of our galaxies, we extract circular velocity curves using the \texttt{mge\_vcirc} function in the \texttt{JamPy} package \citep{cappellari08}. The \texttt{mge\_vcirc} function assumes a constant mass-to-light ratio (M/L) within a galaxy, which we derive for each galaxy using the colour calibration outlined in \citet{bell01}:
\begin{equation}
    B-V = 0.98 (g-r) + 0.22\ , 
\label{eq: bv}
\end{equation}
\begin{equation}
   \log M/L = -0.66 + 1.222 (B-V) \pm 0.1~dex\ ,
 \label{eq: ml}
\end{equation}
\hspace{10pt}where $g$ and $r$ are the g-band and r-band magnitudes for our objects from SDSS DR16 \citep{ahumada20}, which using SDSS conversions in Equation~\ref{eq: bv}, we transform into $B-V$ colours.   

\par As all of the objects in our sample are highly inclined ($i>60^\circ$), the disc inclination parameter in most cases falls below the lower limit of the inclination prior built into \texttt{JamPy} when analytically deprojecting our 2D MGE models into 3D space \citep[in all cases we use the inclination from GAMA DR3,][]{baldry18}. Below this lower limit, \texttt{JamPy} cannot deproject the MGE model. Using the method outlined in \citet{cappellari09} and \citet{smith19}, we circularise the MGE model components by transforming the input to \texttt{JamPy} from $[S_i,\upsigma _i,q_{obs,i}]$ \textrightarrow $[S_i,\upsigma _i \sqrt{q_{obs,i}},1]$, where $S_i$, $\upsigma _i$ and $q_{obs,i}$ are the surface brightness, width and axis ratio of the $i^{th}$ Gaussian component respectively. By performing this conversion, which makes only a small modification to the MGE models, the input no longer falls below the lower bound of the inclination required for deprojection. We also perform a first-order dust-correction of the r-band images by masking pixels containing dust regions by-eye; assuming the dust acts as a screen in front of the stellar emission \citep{carollo97,  cappellari02}. As the galaxies are edge-on, the dust is easily identified and masked by-eye within dust lanes. This procedure for dust-correcting is widely used in the literature \citep[e.g. see][]{scott13,davis17,thater17,davis18}. The MGE model generated for GAMA106389 is illustrated in Figure~\ref{fig: mge}.

\begin{figure*} 
    \centering
        \includegraphics[scale=0.63]{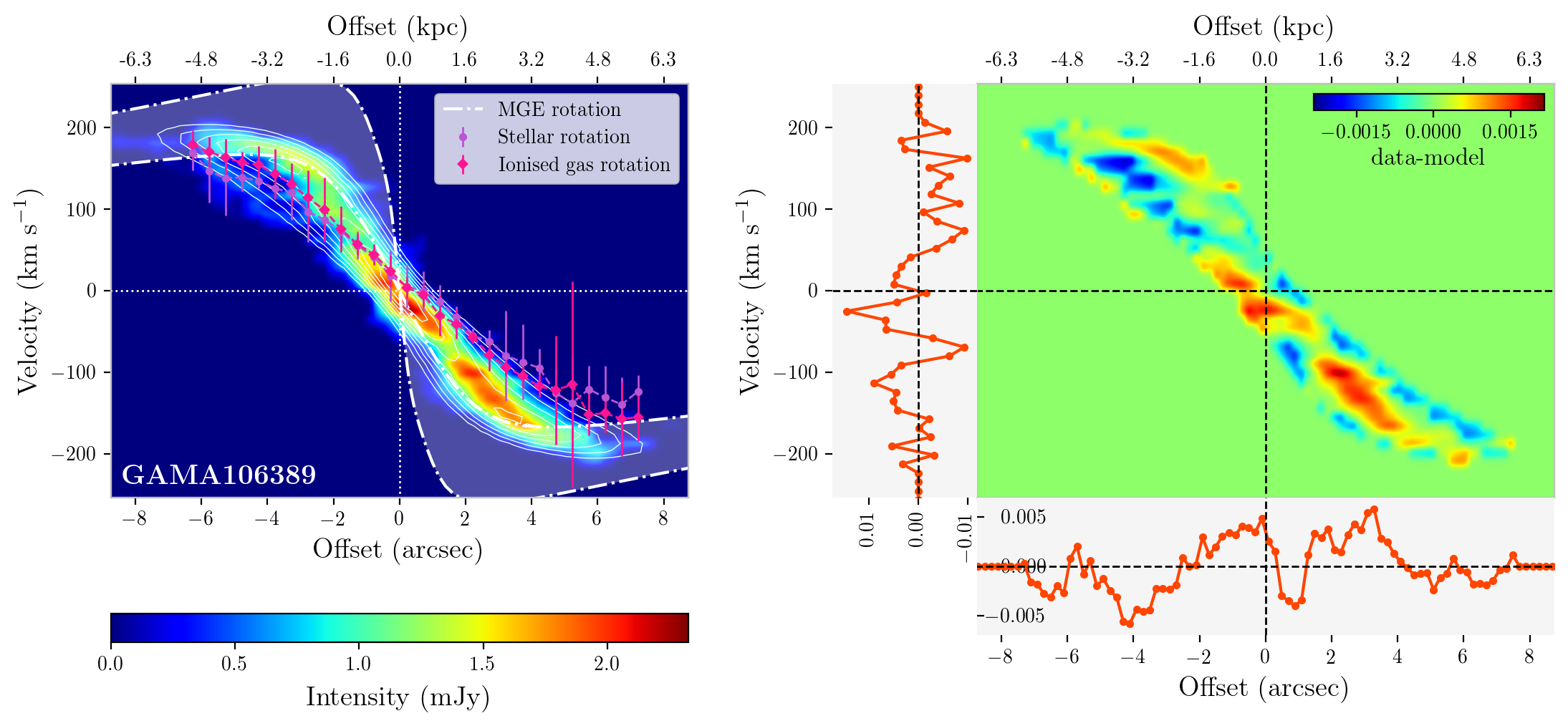}\\
    \centering
        \includegraphics[scale=0.63]{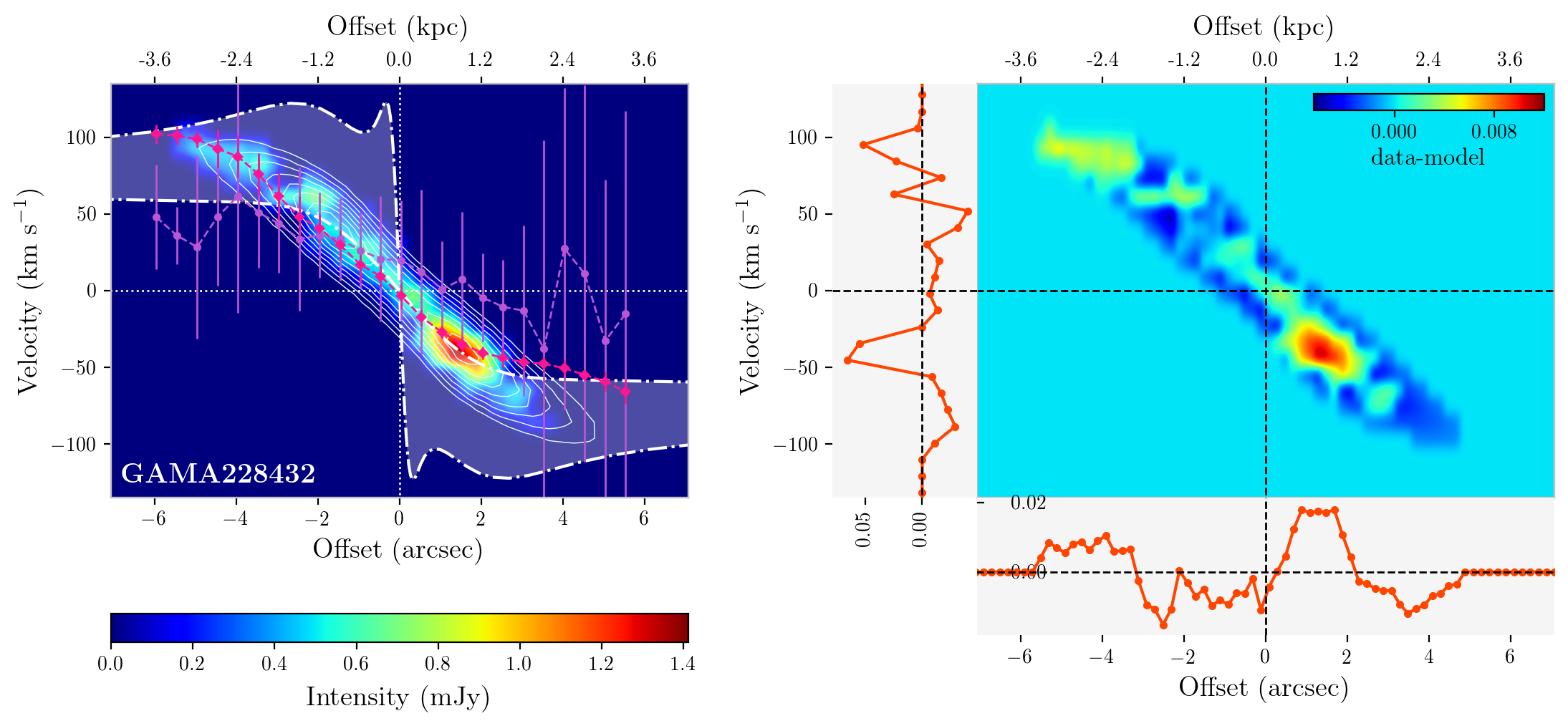}\\
  
    \caption{PVDs extracted along the major axes of our outflow-type objects with their residuals when compared to their respective KinMS model. In each of the panels on the left, we give the PVD extracted from a 2~kpc-wide slit centred on the kinematic centre and orientated along the position angle, both determined by the KinMS modelling (given by the filled coloured contours). The PVD extracted from the KinMS model is given by the pale blue contours over the data, representing 10\% of the maximum value in the model PVD to 90\% in steps of 10\% of the maximum value. For comparison, we also plot the velocity curves extracted from the SAMI stellar and ionised gas velocity maps, extracted and averaged along the minor axes in the same 2~kpc-wide slit used to create the CO(1-0) PVDs. The stellar and ionised gas rotation curves are given by the purple circles and magenta diamonds respectively. The uncertainties represent one standard error in the mean when averaging over the minor axes (this is the dominant error). In each of these panels, we also plot the rotation curves generated by our MGE fits to HSC r-band photometry. We give an upper and lower limit of the these rotation curves by the white dashed lines (see text for more details). In the panels on the right side, we give the residual of the data and model both normalised by their total fluxes (i.e. $\rm data_{ij}/\Sigma_{i,j} data_{ij}-model_{ij}/\Sigma_{i,j} model_{ij}$, where $\rm i$ and $\rm j$ are the PVD axes) with the collapsed residuals for the velocity and offset axes.}
    \label{fig: majpvds}
\end{figure*}

\begin{figure*}
    \centering
        \includegraphics[scale=0.63]{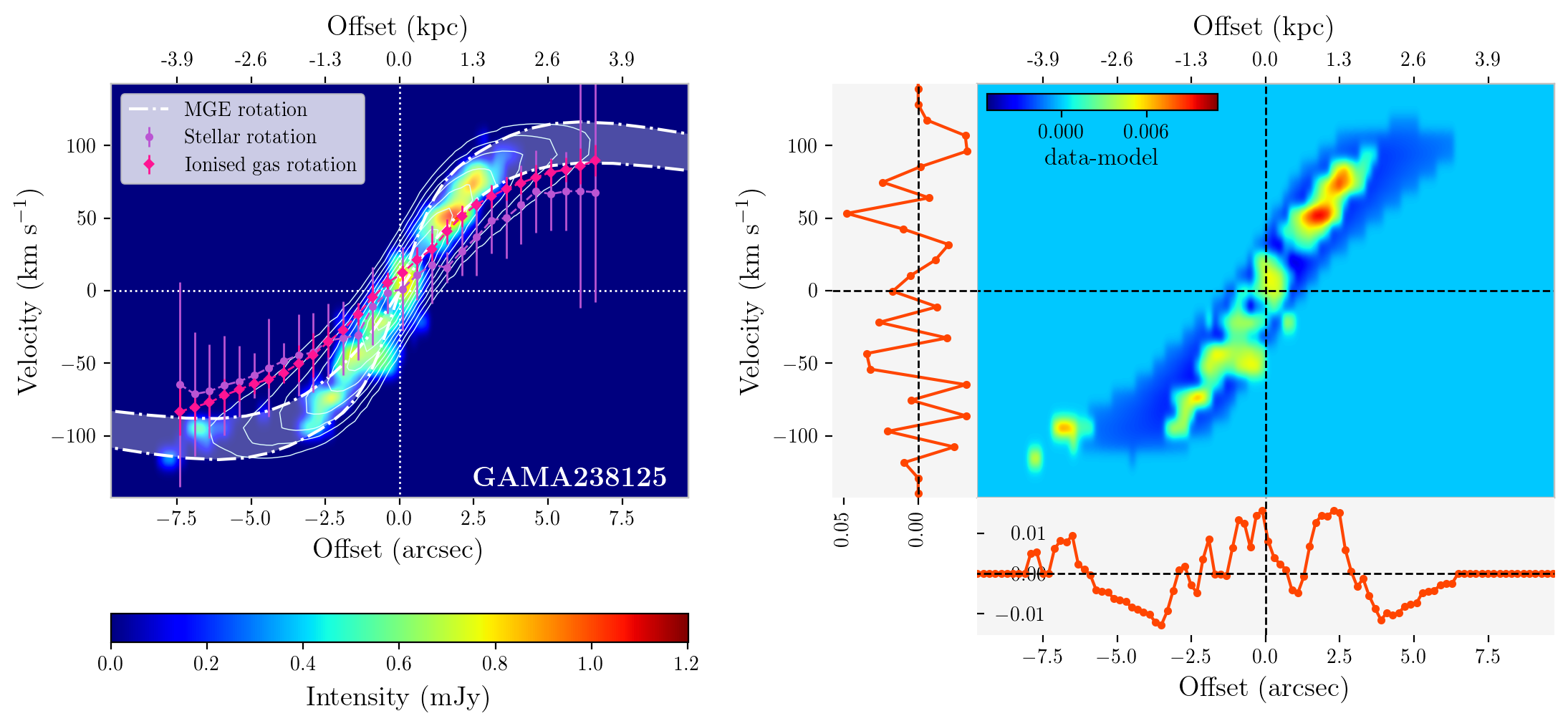}\\
    \centering
        \includegraphics[scale=0.63]{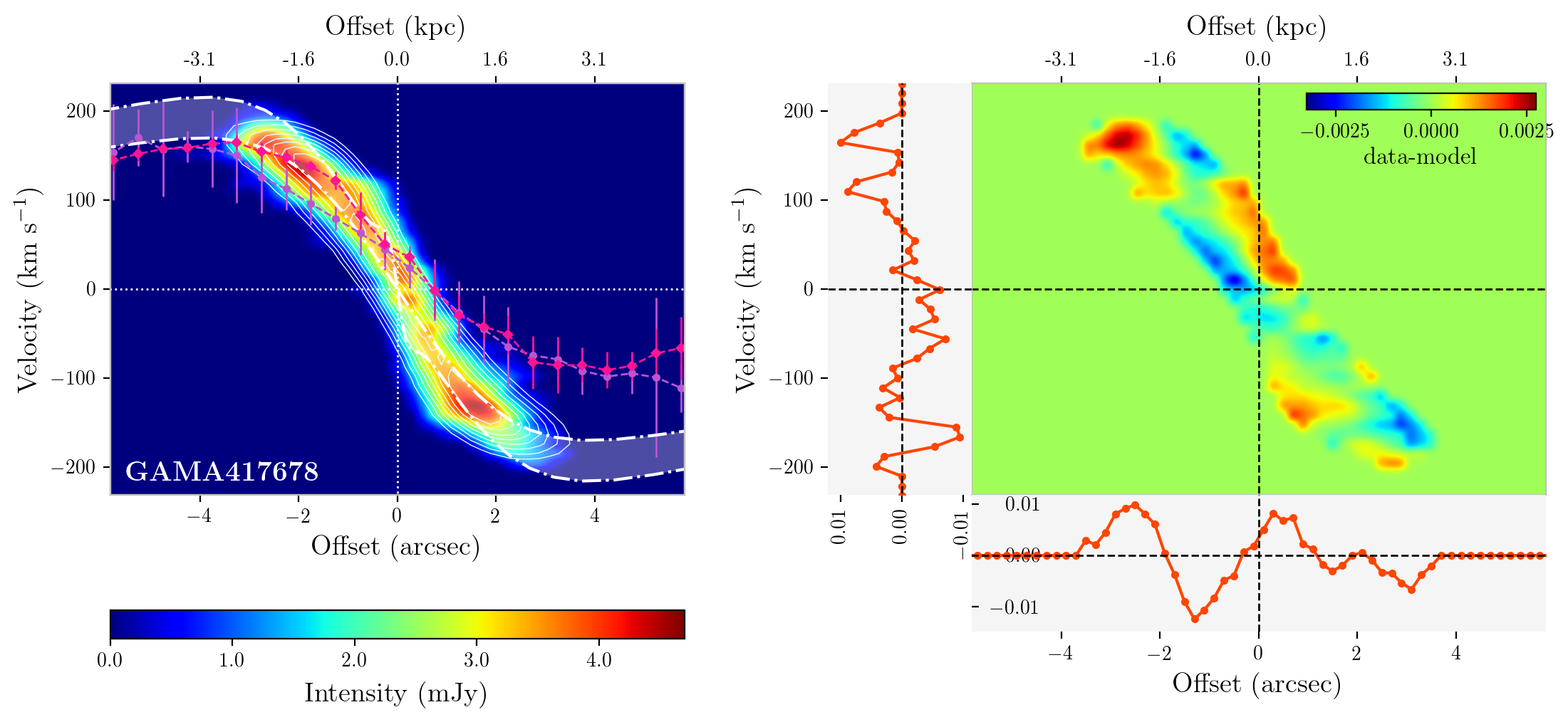}\\
    \centering
        \includegraphics[scale=0.63]{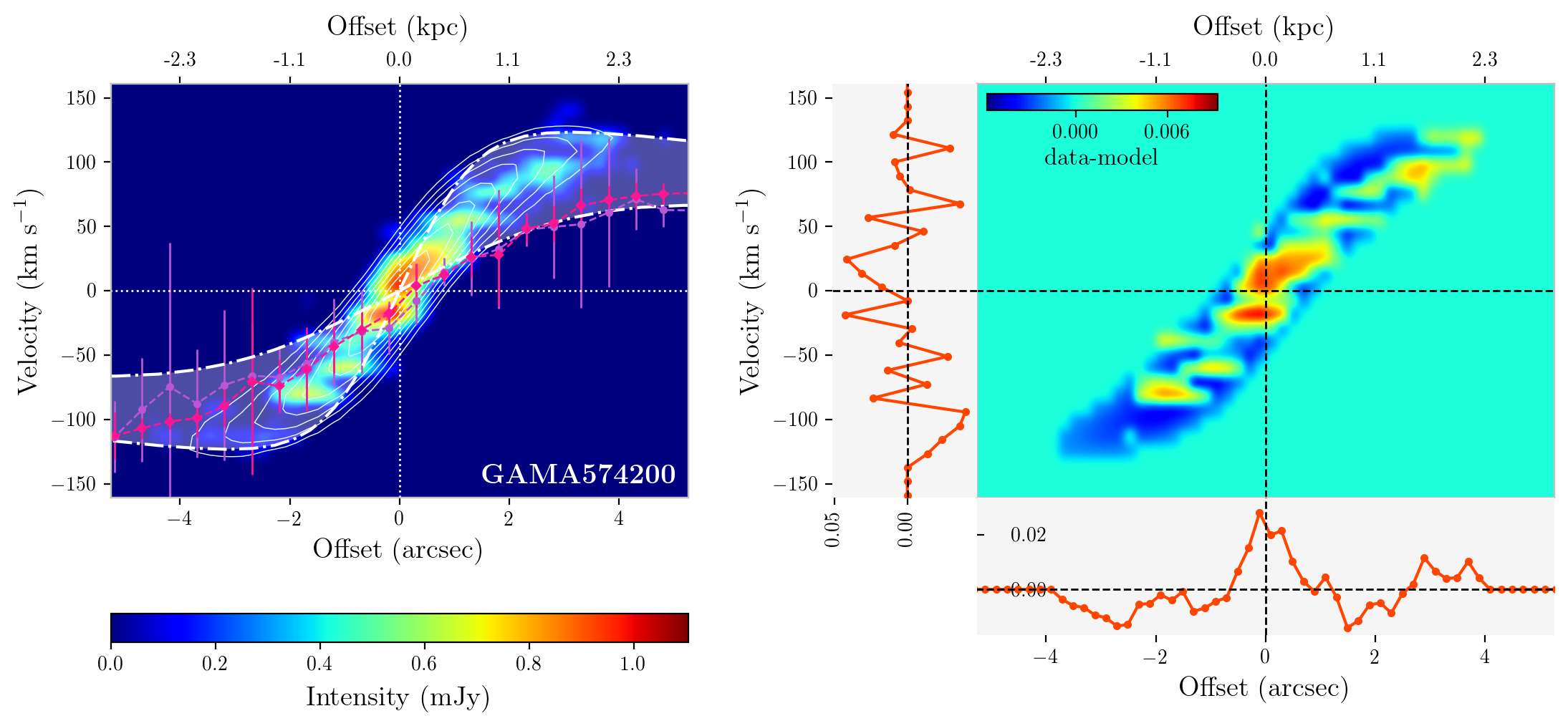}\\
    \contcaption{}
\end{figure*}

\begin{figure*}
    \centering
        \includegraphics[scale=0.63]{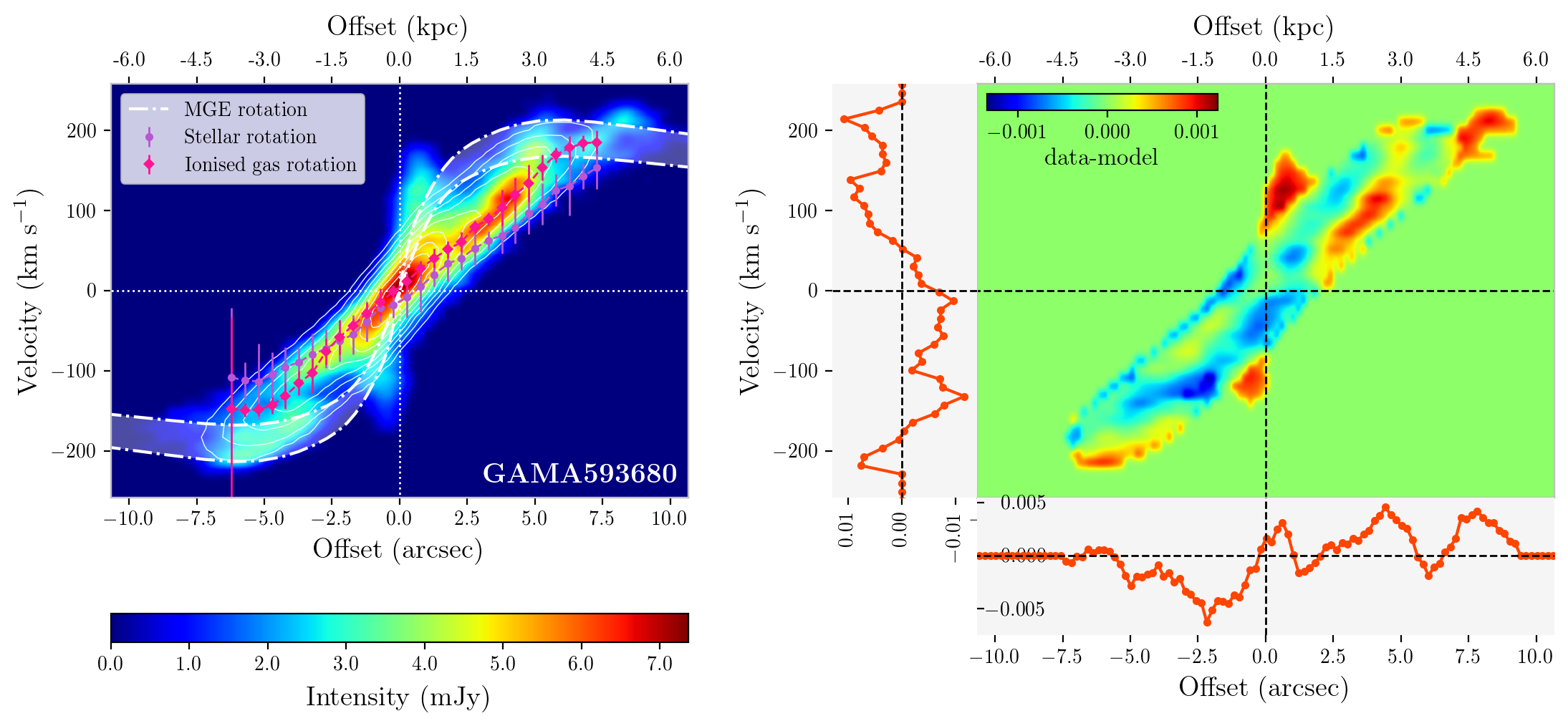}\\
    \centering
        \includegraphics[scale=0.63]{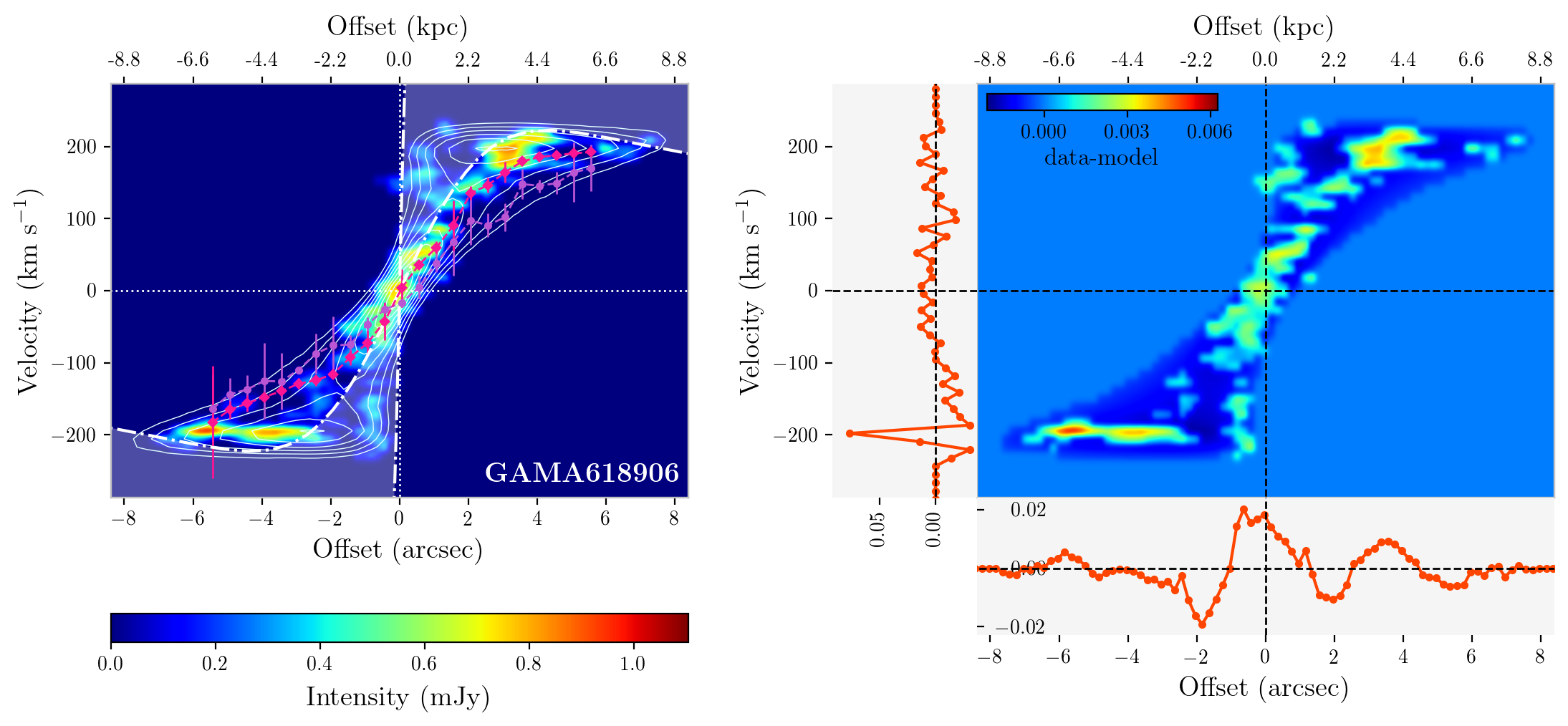}\\
    \contcaption{}
\end{figure*}

\par We find in many of our MGE models that the central Gaussian components are similar in size or smaller than the spatial resolution of the HSC r-band data. Again using the methodology outlined in \citet{smith19}, we speculate that the central components that are unresolved by the HSC data may be associated with optical emission from AGN. Consequently, those components do not contribute to the stellar mass and could artificially steepen our velocity curves in the central regions. We discuss this further in Section~\ref{subsec: PVDs}, where we use the velocity curves extracted from our MGE models including and excluding these unresolved central components as the upper and lower limits of the component of the rotational velocity due to the stellar mass distribution, respectively. For the lower limits, we exclude MGE components that have $\rm \upsigma _i \lesssim \upsigma_{PSF}$, where $\rm \upsigma _{i}$ are the dispersions of the MGE Gaussian components and $\rm \upsigma _{PSF}$ is the dispersion of the HSC PSF (see Figure~\ref{fig: mge}). We follow a similar reasoning of that employed in \citet{davis13}, where we anticipate that a substantial deviation of our ALMA data PVDs from the circular velocity curve would imply that the gas in the objects is dynamically disrupted with respect to the stellar disc.

\section{Results}
\label{subsec: PVDs}
\subsection{Major Axis Position Velocity Diagrams}
\label{subsubsec: majorPVD}

We use the method outlined in \citet{davis13} to construct and interpret the major axis PVDs from our ALMA CO(1-0) datacubes, which are presented in Figure~\ref{fig: majpvds}. We extract our PVDs from the original ALMA cubes by rotating the spatial axes by the position angle found by our KinMS fitting procedure (see Section~\ref{subsec: Kin}) and centred on the kinematic centre also found using the results from our fit. Each PVD is extracted using a 2~kpc-wide slit, which includes all the emission for these edge-on galaxies. We produce two panels for each object; one with the major axis PVD on the left and one with the residual between the PVDs of the data and the KinMS model on the right. In each left-hand panel, we overlay the PVDs extracted from the respective KinMS models using the same position angle and slit width. Additionally, we show rotation curves for both the ionised gas and stars by rotating the 2D SAMI maps by the KinMS position angle and averaging the velocity along the major axis in the same slit used to extract the CO(1-0) PVDs. As detailed in Section~\ref{subsec: mge}, we also show the velocity curve of the stellar disc calculated using our MGE models in each panel in Figure~\ref{fig: majpvds}. We create an upper and lower limit for these velocity curves by including and excluding MGE components unresolved by the HSC data respectively (these limits also include the uncertainty created by our M/L ratio estimates given by Equation~\ref{eq: ml}). These limits serve as the range of possible velocities predicted by the MGE models. While we attempted to account for dust attenuation in our MGE models, errors related to this dust correction can be included within the error budget associated with the M/L ratio.    

\par Using the KinMS models, we can centre our PVDs appropriately and, consequently, can more accurately assess kinematic features and observe features that depart from the dynamically relaxed models. In Figure~\ref{fig: majpvds}, we find that the bulk of the CO(1-0) gas in most instances falls within the velocity range predicted by our MGE models. We also find that the velocity of the CO(1-0)-emitting gas lies above that observed for the ionised gas and stellar rotation in all cases. This is consistent with the principle of asymmetric drift, where stellar and gaseous discs with large dispersions rotate slower than dynamically cold components. In addition, we find well-converged solutions using our pure rotation models for each object (see Appendix~\ref{appendix: corner}). In roughly half of our objects, our models are in good agreement with the rotation curves extracted from our MGE models, suggesting that the bulk motion of the molecular gas is driven by the stellar potential. In all cases, the gas at large offsets is in agreement with the MGE-derived rotation curves (radii $\gtrapprox$ effective radius). However, we note that in the three galaxies where we see some disagreement between the molecular gas rotation and that predicted by the MGE (GAMA106389, GAMA228432 and GAMA593680), the molecular gas is distributed in a ring/s, or has a central hole, which alters the rotation curve to follow solid body rotation along our line of sight. This may artificially flatten the rotation of the molecular gas in the inner regions (radii $\lesssim$ ring/hole radius) of the galaxies compared to the rotation curve generated from the stellar potential. Furthermore, these models have generally low gas velocity dispersions ($\upsigma _{vel}\lesssim 30$~\kms), with the exception of GAMA593680 (see Appendix~\ref{appendix: params}). However, this exception is likely driven by the central velocity flare in this object, which is unaccounted for in our simple model. The combination of these factors suggests that the CO(1-0) gas in our observations is largely dynamically cold and relaxed in a disc/ring, so that its kinematics is primarily driven by the stellar potential of the host galaxy (and at these radii, the stellar contribution to the total potential is dominant). 

\par Despite finding evidence that the molecular gas (as observed with the CO(1-0) line) is dynamically cold, in some cases we observe features that are not captured by the pure rotation model. More generally, we also see indications of flux asymmetry in many of our PVD diagrams. To exemplify this more clearly, in Figure~\ref{fig: majpvds} we show the residuals of the PVDs extracted from our CO(1-0) observations with those from the KinMS models, which are both normalised by their respective total flux (normalised to ease the interpretation of flux asymmetry with respect to the model). As the KinMS models are by nature spatially and spectrally symmetric about their kinematic centres, asymmetry or unmodelled features in our data should be highlighted when compared against them. 

As each object has a unique kinematic structure and a model that captures that structure with different degrees of success, we will consider each object in Figure~\ref{fig: majpvds} in isolation:

\begin{figure*} 
    \centering
        \includegraphics[scale=0.63]{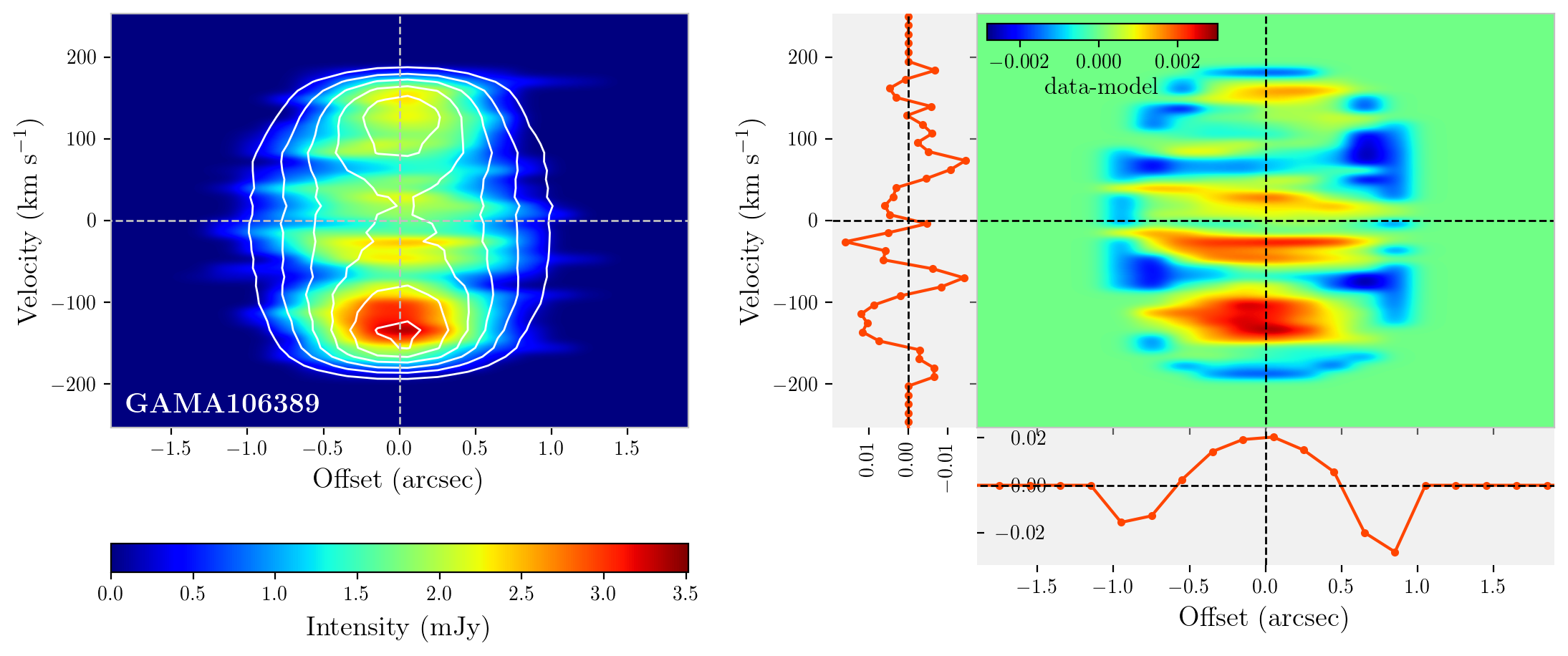}
    \centering
        \includegraphics[scale=0.63]{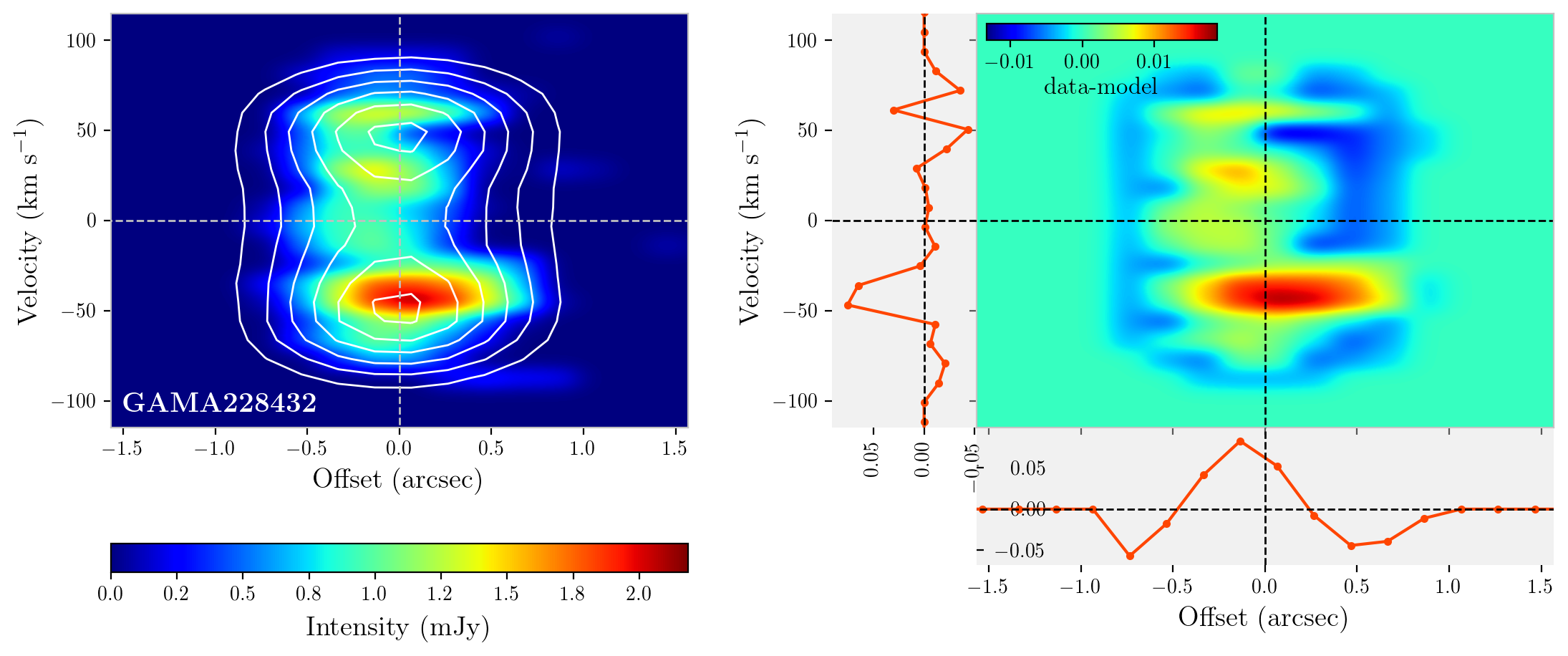}
   
     \caption{PVDs extracted along the minor axes of our outflow-type objects and their residuals with the KinMS models. For each object, we give the minor-axis PVD extracted from a slit with a total width equal to the effective radius of the objects centred on the kinematic centre determined by the KinMS kinematic modelling and orientated orthogonal to the position angle we also determine from the models in the left panel. We overlay the model with blue contours over the data, where the contour levels represent 10\% of the maximum value in the model minor-axis PVD to 90\% in steps of 10\% of the maximum value. In the right panel, we give the residual of the data and model both normalised by their total fluxes (i.e. $\rm data_{ij}/\Sigma_{i,j} data_{ij}-model_{ij}/\Sigma_{i,j} model_{ij}$, where $\rm i$ and $\rm j$ are the PVD axes) with the collapsed residuals for the velocity and offset axes.}
     \label{fig: minpvds}
\end{figure*}

\begin{figure*}
    \centering
        \includegraphics[scale=0.63]{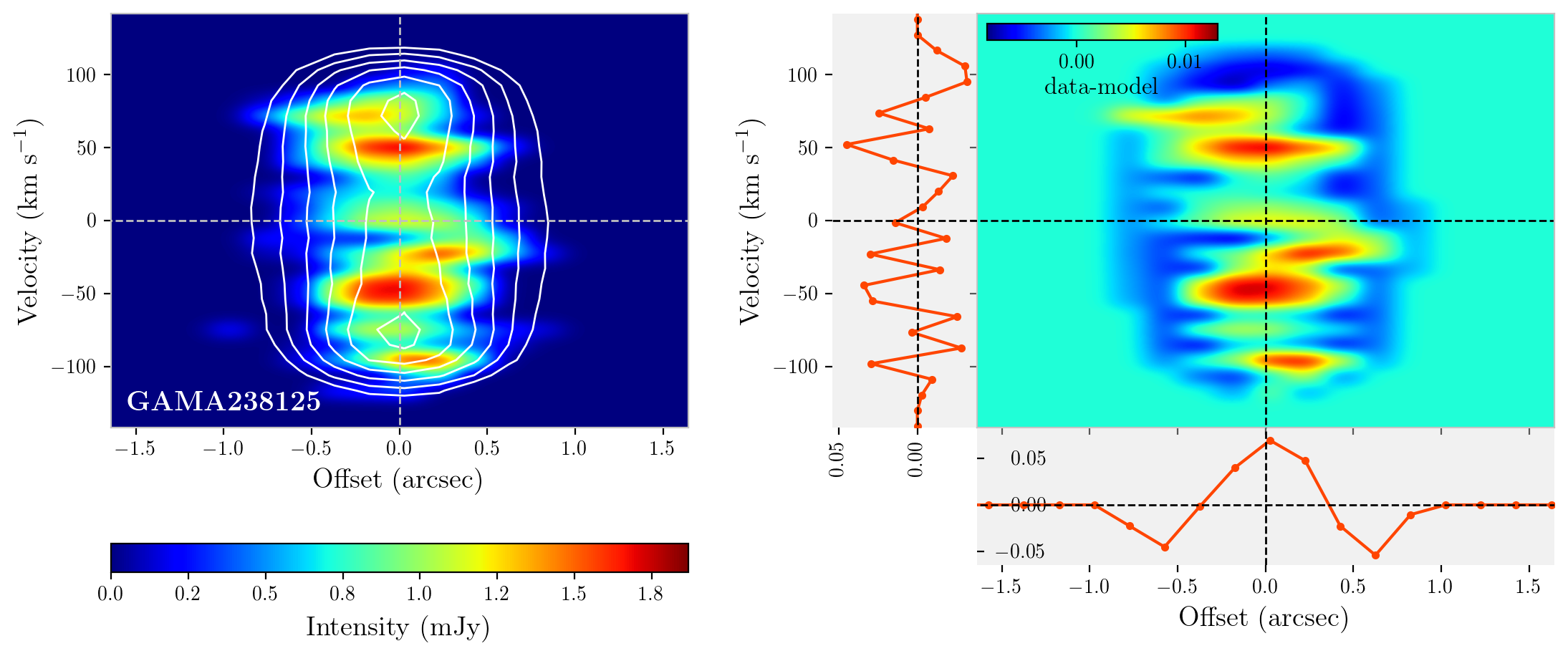}\\
    \centering
        \includegraphics[scale=0.63]{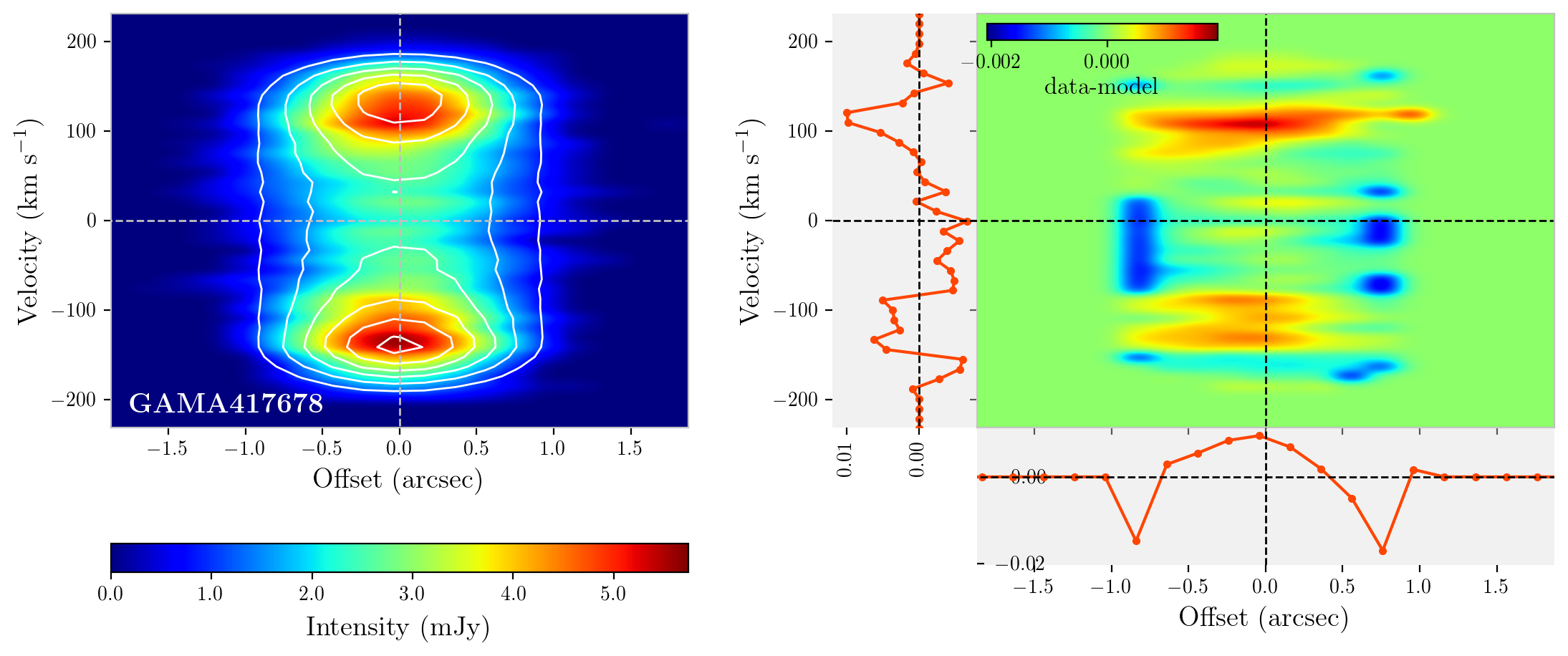}\\
     \centering
        \includegraphics[scale=0.63]{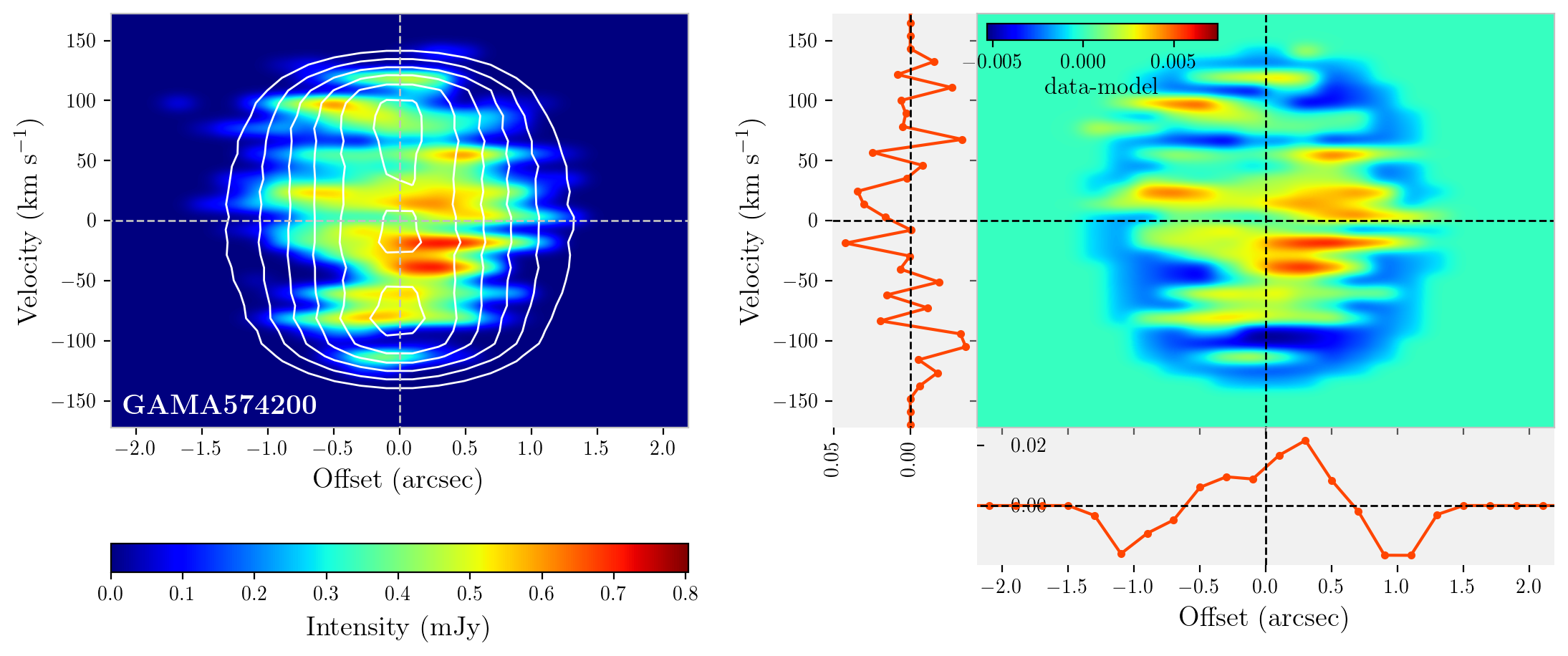}\\
    \contcaption{}
\end{figure*}

\begin{figure*}
    \centering
        \includegraphics[scale=0.63]{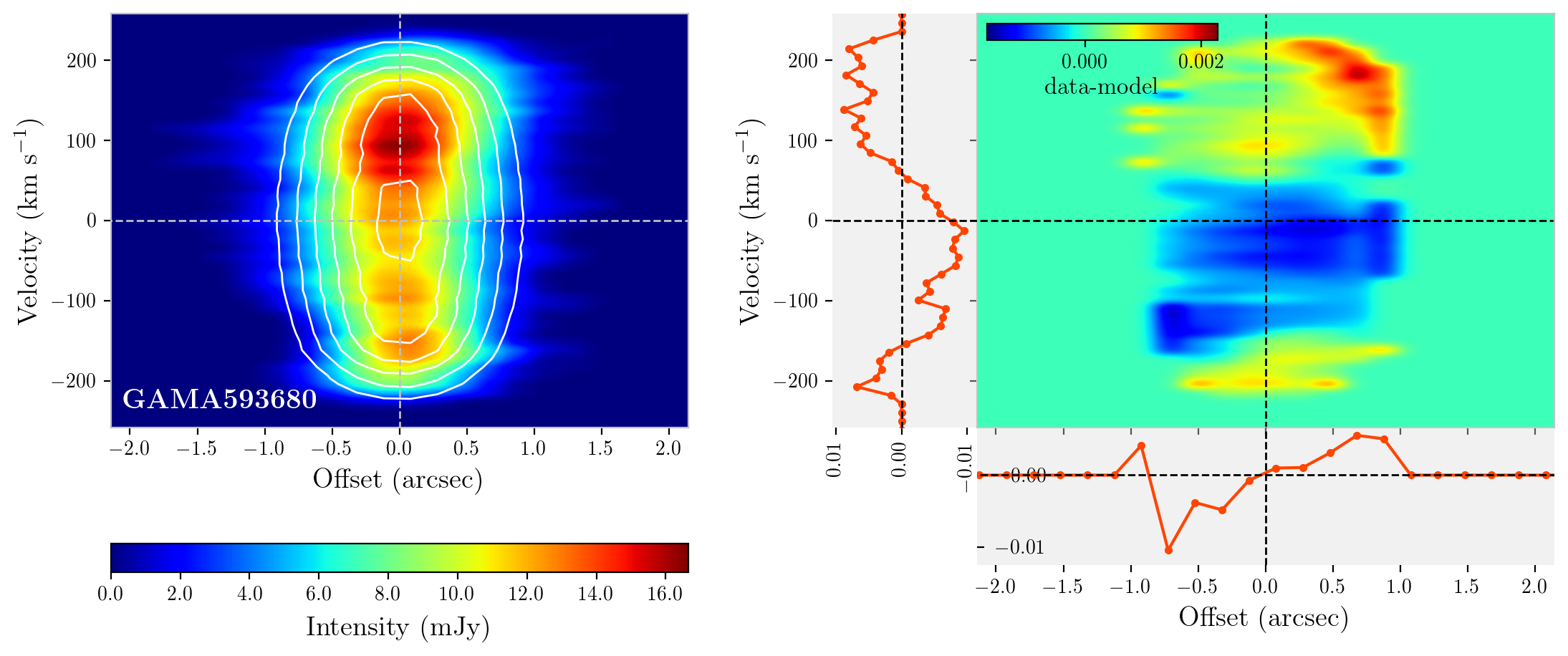}\\
    \centering
        \includegraphics[scale=0.63]{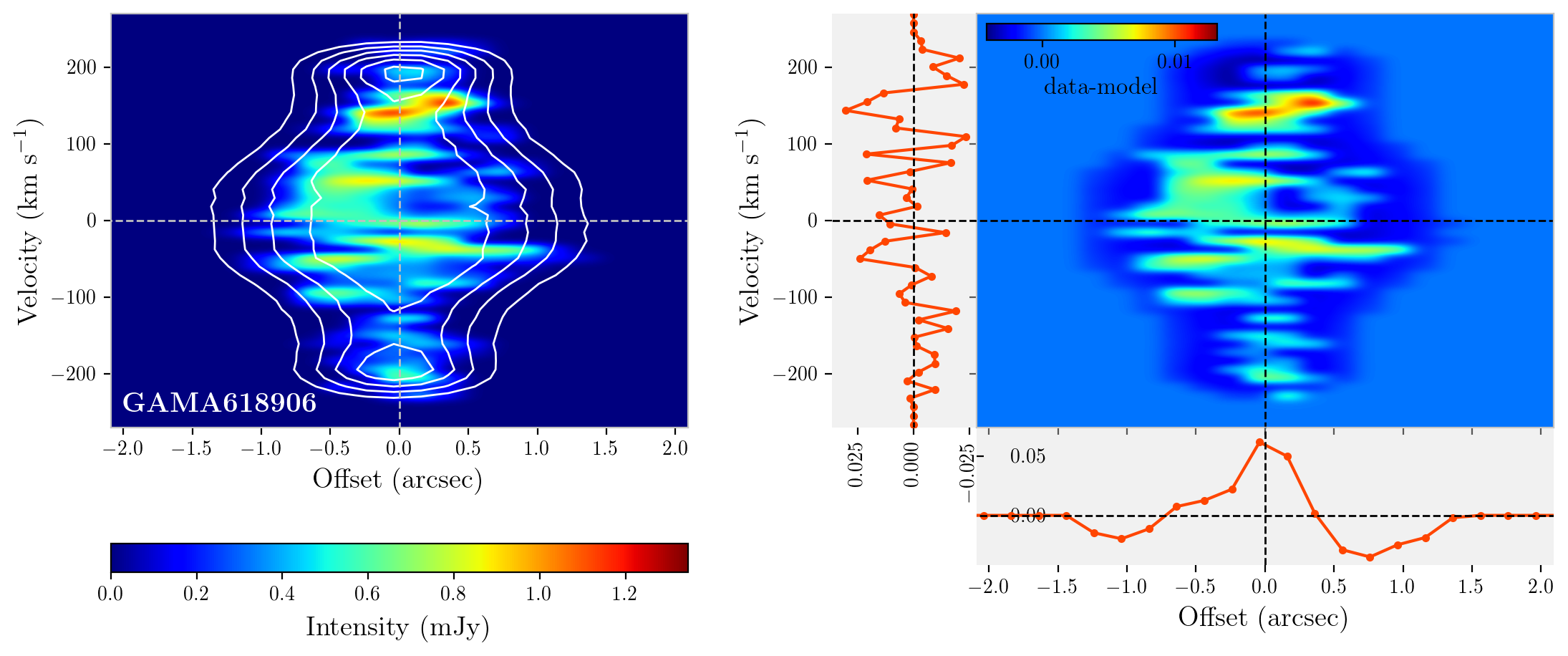}\\
    \contcaption{}
\end{figure*}

\begin{itemize}
    \item \textbf{GAMA106389:} the molecular gas distribution is best modelled by a Gaussian ring, as opposed to an exponential disc (see Section~\ref{subsec: Kin}). The central emission in the PVD also appears to follow the solid-body rotation pattern (i.e. where velocity increases linearly with radius), rather than the steeper central stellar rotation predicted by the MGE model. This is what we would expect to observe for a gas ring viewed in projection. The most likely interpretation of this object's kinematic profile is the existence of a resonance system, which may be driven by the forcing frequency associated with a bar. The pattern frequency of a bar causes gas to accumulate on stable, closed orbital paths where the gas rotates at a rate that is commensurate with the rate at which it encounters the forcing. In this case, it is probable that either the inner Lindblad resonance or X2 orbits are populated by gas (those at and within co-rotation, which usually coincides with the turnover radius). The object appears to have a flux asymmetry in the approaching velocity side of its PVD, suggesting the gas is not evenly distributed around this ring. This asymmetry is also present in the residual plot for the object. 

    \item \textbf{GAMA228432} is modelled by an exponential disc with a central hole. This profile could again be indicative of a resonance system where the inner orbits have been evacuated (the linear shape of the inner region of the PVD is also what would be expected for a ring viewed in projection). Furthermore, there is a flux enhancement in the approaching side of its PVD (which we also observe in the ionised gas distribution) and the receding side extends to higher velocities and offsets than the approaching end with respect to the kinematic centre.
    
    \item \textbf{GAMA238125} is modelled by an exponential disc with a velocity curve in very close agreement with the MGE-derived curve. Like GAMA228432, the spatial distribution of the gas is asymmetric with respect to the kinematic centre (see Figure~\ref{fig: modplots}) which can be observed in the approaching side of the PVD.

    \item \textbf{GAMA417678} is modelled by a Gaussian ring surface brightness profile with a rotation curve that is in close agreement with the MGE-derived rotation curve. This ring-type profile may again be indicative of a resonance system, with gas occupying the inner Lindblad resonance, within the co-rotation radius. Again, with reference to Figure~\ref{fig: majpvds}, we observe a gas enhancement at receding velocities. We also note a slight flaring of the velocities towards the kinematic centre of the profile, which can be more clearly observed in the residual plot, where there are residual flux enhancements corresponding to this central velocity flare.
    
    \item \textbf{GAMA574200} is modelled by an exponential disc and appears relatively symmetric with regard to the kinematic centre. The object does extend further at receding velocities, which can be observed in Figure~\ref{fig: modplots}, where the object is asymmetric with respect to the kinematic centre.  
    
    \item \textbf{GAMA593680} appears to be the most complex in terms of kinematic structure and is modelled by a superposition of two Gaussian profiles; an inner Gaussian disc-like profile (we say disc-type here as the mean position of the ring is located at the kinematic centre) and outer Gaussian ring-type profile. The inner region of the PVD also follows a linear rotation pattern, emulating solid-body rotation, instead of the more sharply rising central rotation predicted by our MGE model. As previously noted, this is what we would expect to observe when viewing gas rings in projection. Again, we interpret the multiple ring structure of this object as indicative of an axi-asymmetric forcing, most likely by a bar. In this instance, we may have inner Lindblad resonances/X2 orbits occupied in addition to the co-rotation radius or outer resonances, as the gas profile appears to extend beyond the turnover radius. The bulk of the profile is in reasonable agreement with the MGE-derived rotation (there is some indication that the turnover radius of the model is larger than that of the MGE-derived rotation curve). However, the gas clearly has a central flare in the velocity direction. This central flare is not captured by either the MGE-rotation or by that of the model, which is evident in the residual. We can not, therefore, attribute this flare to the inner Gaussian ring in this profile assuming the dynamics are purely driven by circular rotation. Furthermore, in the residual panel the receding side of the PVD is clearly more gas rich than the approaching side, implying a flux asymmetry in this object. 

    \item \textbf{618906} is modelled by an exponential disc and appears relatively symmetric with respect to the kinematic centre, but with some indications of flux asymmetry at velocities close to the MGE-predicted turnover radius.
\end{itemize}

\par Our galaxies, which all harbour large-scale ionised gas winds, demonstrate a degree of flux asymmetry in their molecular gas when compared to their symmetric KinMS models (we will discuss the significance of the asymmetry in Section~\ref{subsec: cold}). We also find some instances of unmodelled kinematic features (e.g. central velocity flaring). There may be evidence, therefore, that the molecular gas is not perfectly kinematically settled in our outflow-type galaxies.

\subsection{Minor Axis PVDs} 
\label{subsubsec: minorPVD}

We also extract PVDs along the minor axes of our objects. These minor axis PVDs may reveal radial motion of the CO(1-0) gas, otherwise difficult to identify in the major axis PVDs. Radial motion can be detected by a ``twisting/bending'' of the minor axis PVD about the kinematic centre, indicating gas flow in the plane of the disc. For these PVDs we use slit width equal to the effective radius ($r_e$) of the objects (where we use $r_e$ values from the GAMA DR3 catalogue), so that the slit extends half an $r_e$ either side of the kinematic centre along the major axis. By using a wide slit for these PVDs, we hope to capture the bulk of the CO(1-0) gas, in order to detect widespread radial motion (it should be noted, referring back to Figure~\ref{fig: viogas}, that the molecular gas is centrally concentrated, meaning this slit covers most of the CO emission). In Figure~\ref{fig: minpvds}, as in Figure~\ref{fig: majpvds}, we include two panels for each object; one with the minor axis PVD drawn from our data overlaid with the model PVD and the other the residual between the PVDs of the data and the model (calculated in the same manner as the residual plots in Figure~\ref{fig: majpvds}). Again, we consider each galaxy separately:

\begin{itemize}
    \item \textbf{GAMA106389}: there is evidence of gas enhancement at negative offsets and at negative velocities, but no ``bending/twistiing'' indicative of radial motion.
    
    \item \textbf{GAMA228432}: we observe a clear ``bending'' in this PVD, represented as spatial and velocity asymmetries, possibly indicative of radial motion and inflow or outflow of gas along the rotation plane. As detailed in Section~\ref{subsubsec: majorPVD}, the central hole in the exponential profile used to model this object may suggest the presence of a bar, which would exert gravitational torques on the surrounding gas and, subsequently, internal radial motion.
    
    \item \textbf{GAMA238125}: we do not observe asymmetry in this object, but we note there are deviations from the minor-axis PVD extracted from the model. The low S/N of this object may restrict our ability to discern kinematic features clearly.
    
    \item \textbf{GAMA417678}: we can infer from the residual a gas enhancement at negative offsets and positive velocities, but no ``bending/twistiing'' indicative of radial motion.
    
    \item \textbf{GAMA574200}: the lower S/N is this object is again limiting, but we do not observe notable asymmetry in this object.
    
    \item \textbf{GAMA593680}: the presence of a ``bend'' in the PVD towards the positive offsets is very striking in this case. The residual panel validates this observations by showing a clear gas enhancement at positive velocities. In this instance, given the ring-type component used to model this object and the observations of the kinematic features of the major-axis PVD made in Section~\ref{subsubsec: majorPVD}, we again can infer a resonance system with radial motion induced by a bar.
    
    \item \textbf{GAMA618906}: We do not observe significant asymmetry in this object apart from a slight enhancement at positive velocities, but the low-S/N of this object limits further analysis.

\end{itemize}

\par We will reflect on these observations and their implications in Section~\ref{sec: discussion}, and discuss how this interpretation relates to previous results.

\section{Discussion}
\label{sec: discussion}

\subsection{Dynamically Cold Discs}
\label{subsec: cold}

A ubiquitous feature of our sample of SAMI-selected galaxies with galactic-scale ionised gas winds, is that the molecular gas appears to be dynamically cold. We find good agreement between the PVDs we extract from our CO(1-0) datacubes and our MGE models in about half of objects (with some MGE models being steeper than our CO(1-0) data at low offsets, see Section~\ref{subsubsec: majorPVD}). However, the objects where we see disagreement (GAMA106389, GAMA228432 and GAMA593680) all have ring-type molecular gas profiles, which produce flattened rotation curves when viewed in projection. We, therefore, do not necessarily interpret disagreement here between the rotation of the molecular gas and that from the stellar potential. In all cases, however, the data agrees reasonably with the MGE-derived rotation curves at large offsets ($\rm \gtrapprox r_e$). We also find excellent agreement between our KinMS models and the observed CO(1-0) emission (see Section~\ref{subsec: Kin}). As detailed in \citet{davis13}, KinMS models make the assumption that the gas is dynamically cold and symmetrically distributed about its kinematic centre. We also find low velocity dispersions for each object, further supporting the inference that the molecular gas is dynamically cold and rotation-dominated. 

\par
\begin{figure}
    \centering
        \includegraphics[scale=0.42]{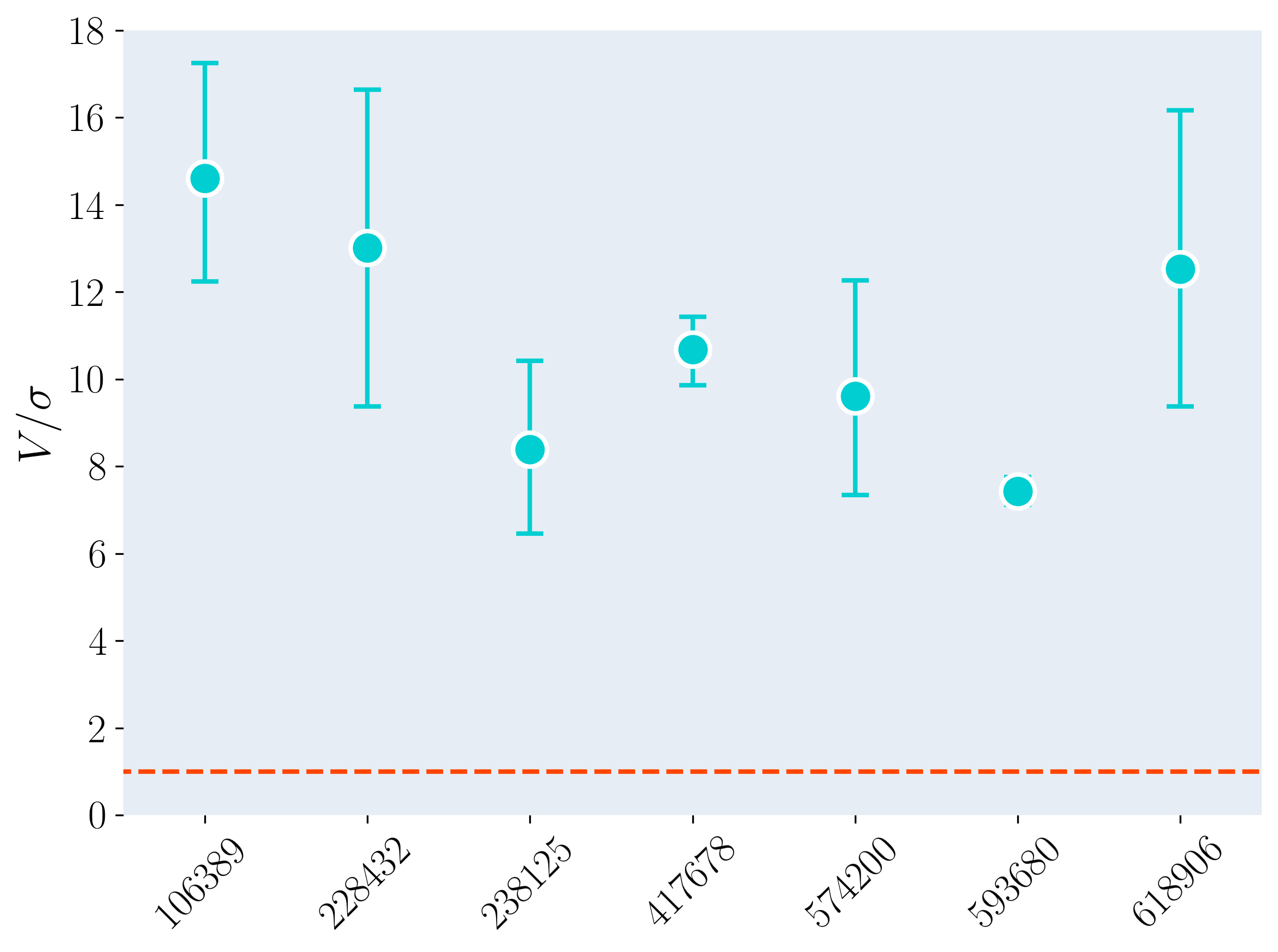}
     \caption{Plot illustrating the $\rm V/\upsigma$ ratios for the outflow-type galaxies in our sample. In each case, we use the maximum velocity and velocity dispersion values from the best fit models generated from KinMS (see Section~\ref{subsec: Kin} and Appendix~\ref{appendix: params}). We mark $\rm V/\upsigma = 1$ with an orange dashed line to mark the rotation-dominated ($\rm V/\upsigma$ > 1) and dispersion-dominated ($\rm V/\upsigma < 1$) regimes.}
     \label{fig: vsig}
\end{figure}

 We can test the hypothesis that the gas in our galaxies is largely rotation-dominated by calculating the velocity to velocity dispersion ($\rm V/\upsigma$) ratio using the parameters from our models. In Figure~\ref{fig: vsig}, we present the $\rm V/\upsigma$ values for our outflow-type objects. The whole sample falls consistently in the very highly rotation-dominated regime (i.e. $\rm V/\upsigma>>1$), implying again that the molecular gas in these objects is largely dynamically cold and rotation-supported. This is further substantiated by Figure~\ref{fig: majpvds}, where the bulk of the CO(1-0) emission follows the stellar kinematics. 

\par While it is evident that the objects in our sample are largely rotation-dominated and dynamically relaxed, we do observe some  asymmetries when examining our PVDs in Section~\ref{subsubsec: majorPVD}. These could be due to either {\it velocity asymmetries} (i.e. deviations of the kinematics from circular motion) or {\it flux asymmetries} (i.e. asymmetries in the gas distribution in an otherwise axi-symmetric potential). 

In an extended analysis, in order to distinguish between flux and velocity asymmetries, we look for the presence of velocity asymmetry by transforming our major-axis PVDs so that where gas is present (according to our masking process) the intensity is $=1$ and otherwise $=0$ (see Appendix~\ref{appendix: velocity}). The PVDs are then rotated by 180$^\circ$ and subtracted from the un-rotated, binary PVDs. This process aims to highlight asymmetry in the shape of a PVD with respect to the kinematic centre, which is an indication of instability over dynamical timescales ($\uptau_{dyn}$). In all cases, and especially for the high mass galaxies, this analysis reveals little evidence for velocity asymmetries. This suggests that the kinematics of the gas in these objects is stable over a timescale $\rm \uptau_{dyn}$ ($\equiv 2\uppi \rm R_{max}/ \rm V_{R_{max}} \approx 300~Myr$).

\begin{figure}
    \centering
        \includegraphics[scale=0.5]{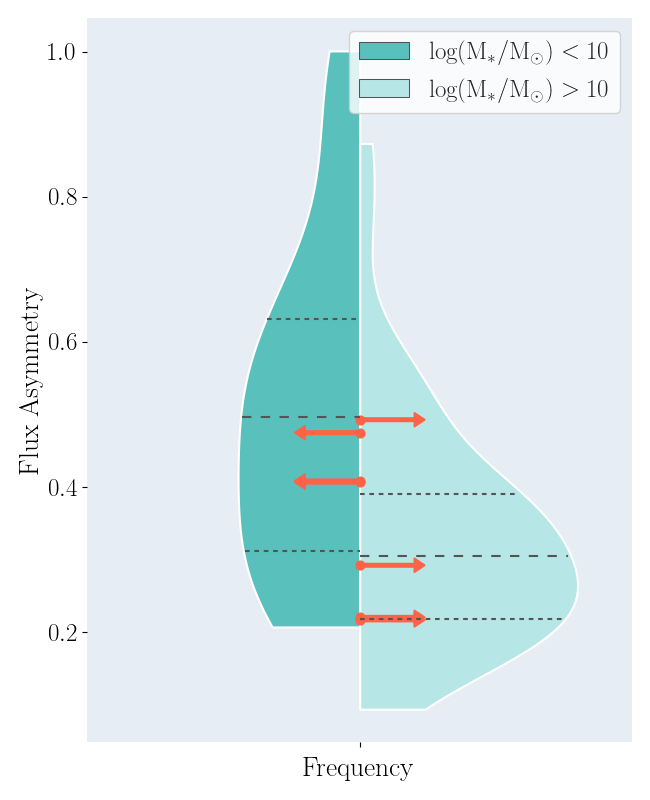}
     \caption{Violin plot depicting the distributions of flux asymmetry (see text for definition) in the PHANGS-ALMA sample \citep{leroy21}. The right side of the violin plot gives the KDE for the distribution of flux asymmetry for objects with high stellar mass $\rm \log (M_*/M_\odot) > 10$), while the left side illustrates the distribution for objects with low stellar mass ($\rm \log (M_*/M_\odot) < 10$). The dashed lines illustrate the inner quartiles for each respective distribution. The flux asymmetry for each object in our ALMA outflow-type sample is marked by orange circles in the figure. Each circle has an arrow pointing to the right or left indicating whether the object is in the high or low stellar mass sample respectively.}
     \label{fig: flux_asym}
\end{figure}

Next, we employ a similar technique to quantify the significance of the flux asymmetries seen in the major-axis PVDs. Whereas velocity asymmetry by itself can indicate on-going dynamical instability, flux asymmetry without velocity asymmetry is more likely associated with a settled disc that has undergone past instability (or potentially chance alignments of molecular clouds along our line-of-sight).

We define flux asymmetry using a similar method as the Asymmetry Index \citep[A; ][]{schade95, abraham96}, so that:

\begin{equation}
    \rm Flux\ Asymmetry = \frac{\Sigma_{ij}|{PVD}_{ij}-{PVD}_{rot,ij}|}{2 \Sigma_{ij}{PVD}_{ij}}\ ,
    \label{eq: flux}
\end{equation}

\hspace{10pt}where $\rm {PVD}_{ij}$ and $\rm {PVD}_{rot,ij}$ are the PVDs and 180$^\circ$-rotated PVDs respectively. To compare the flux asymmetry we observe in our objects to a control sample, we use the PHANGS-ALMA \citep{leroy21} sample to create a distribution of flux asymmetry. We degrade the PHANGS-ALMA CO(2\textrightarrow1) resolution to be comparable to our physical resolution of $\rm\approx 1~kpc$ and perform the same procedure summarised by Equation~\ref{eq: flux}. The results of this analysis are illustrated in Figure~\ref{fig: flux_asym}, where we split the PHANGS-ALMA flux asymmetries into high-$\rm M_*$ and low-$\rm M_*$ subsamples and overlay our outflow-type flux asymmetries. We find that our objects fall within the inner (i.e. 25-75\%) quartiles of the high-$\rm M_*$ and low-$\rm M_*$ KDE profiles extracted from PHANGS-ALMA, with the exception of GAMA618906 in the high-$\rm M_*$ distribution. 

Ultimately, we do not find evidence to suggest that the flux asymmetry of the molecular gas content in our outflow-type objects is extraordinary compared to a sample of galaxies not undergoing large-scale outflows of ionised gas. However, the generally lower flux asymmetries we observe in our higher $\rm M_*$ objects is consistent with the findings of \citet{davis22}, who find a similar result between the Asymmetry Parameter values from CO zeroth moment maps and total galaxy $\rm M_*$. The explanation they postulate is that higher-$\rm M_*$ galaxies are more likely to host dense stellar bulges which form deep potentials at the centre of these galaxies and, as a result, keep the molecular gas more stable (i.e. the stellar potential dominates over the gas self-gravity). \citet{davis22} find the stellar mass surface density is the most influential global galaxy property that dictates the morphology of the molecular gas content and has a stabilising effect on its kinematics. This is highly relevant to our results, where we have a sample of objects hosting molecular gas with dynamics largely dominated by the stellar potential, which is very dynamically stable.    

\begin{figure*}
    \centering
        \includegraphics[scale=0.78]{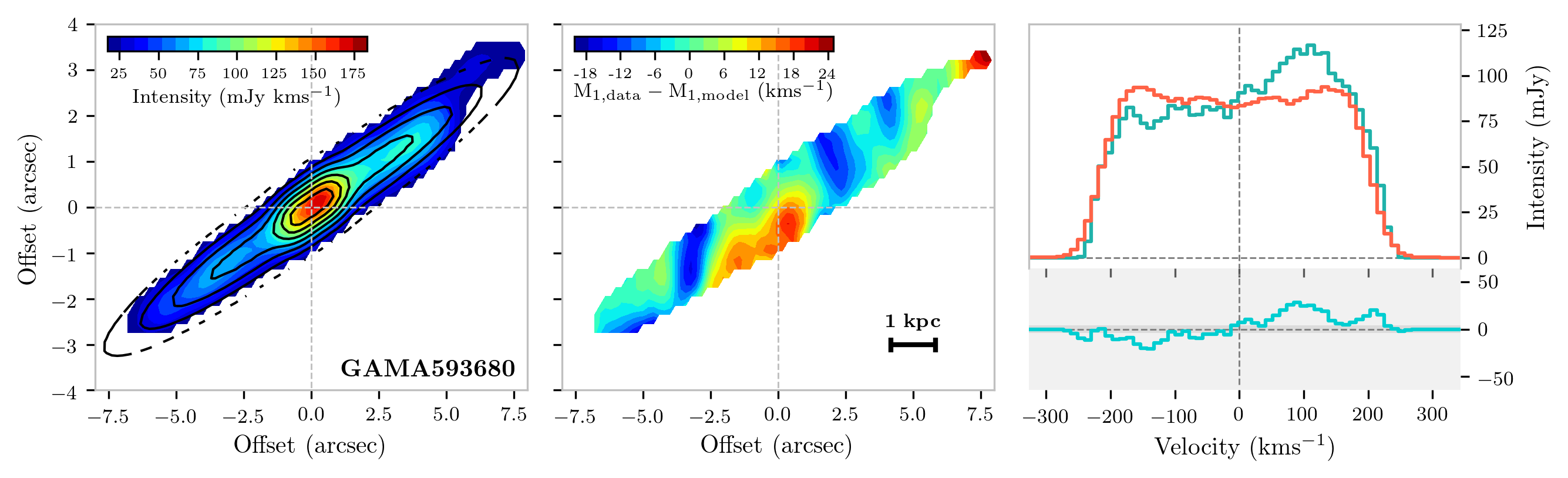}\\
    \centering
        \includegraphics[scale=0.63]{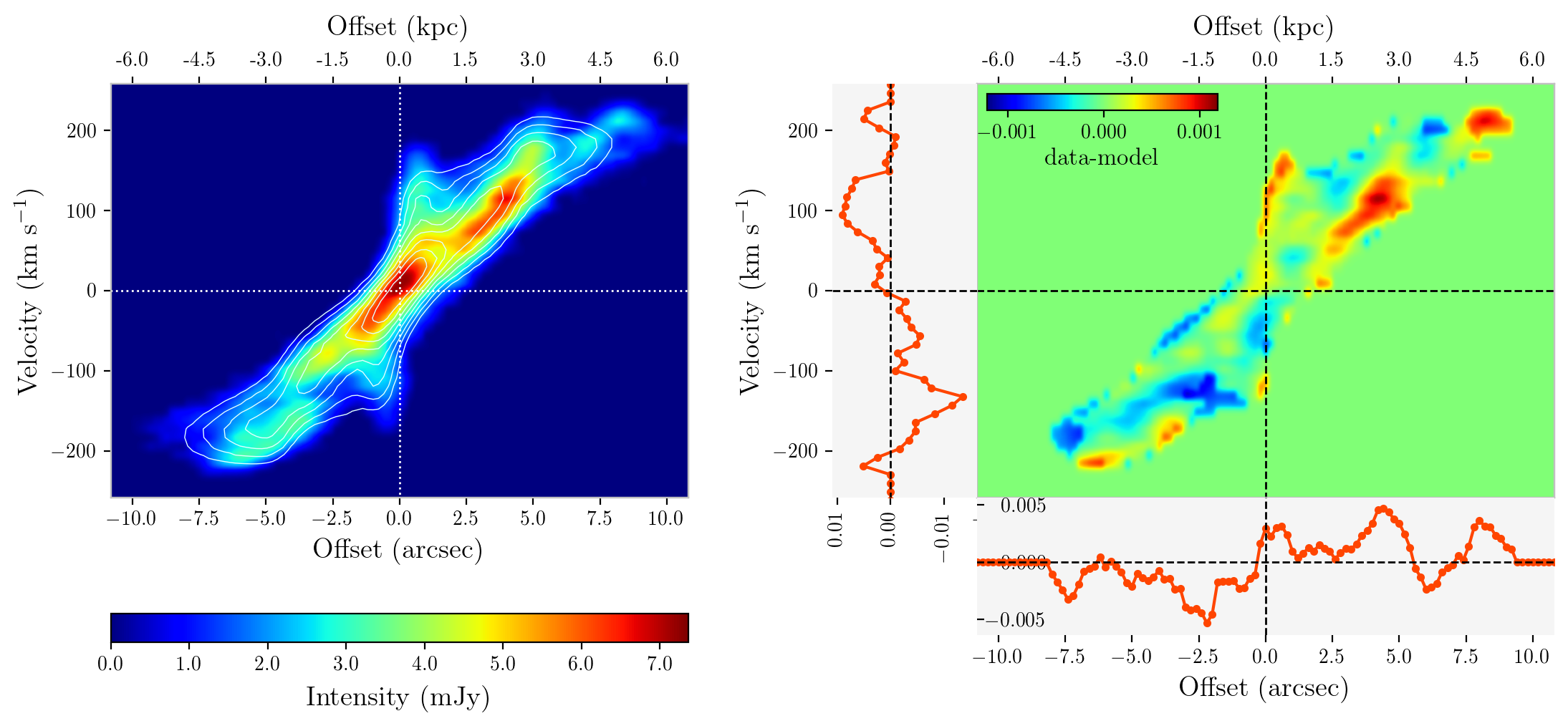}\\
    \centering
        \includegraphics[scale=0.63]{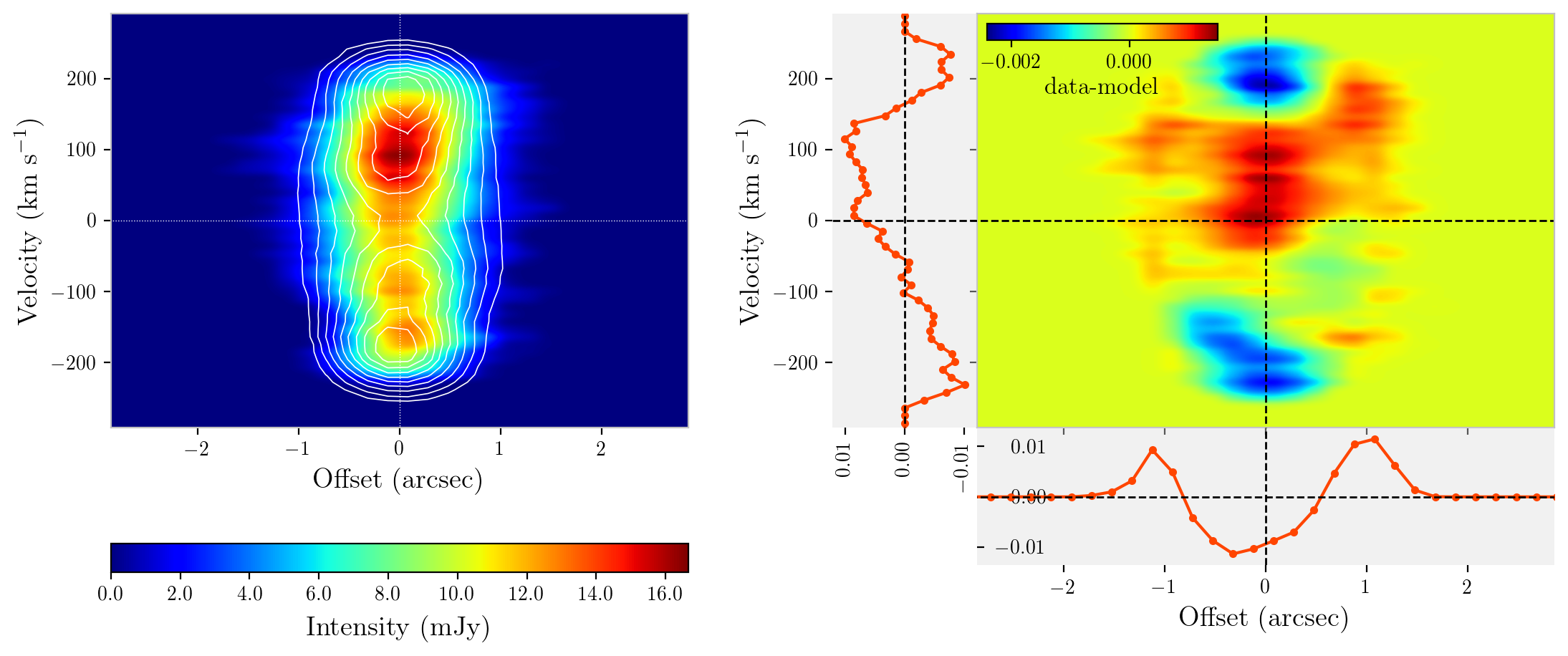}\\
    \caption{\textbf{Top}: Major-axis PVD extracted from GAMA593680 (colour contours) overlaid with the PVD from the model simulating both rotational and radial motion (the format of the left and right panels is the same that in Figure~\ref{fig: majpvds}). \textbf{Bottom}: Minor-axis PVD extracted from GAMA593680 overlaid with the PVD from the model.}
    \label{fig: 593680_rad}
\end{figure*}

\subsection{Radial Gas Motion}
\label{subsec: inflows}

In Section~\ref{subsubsec: minorPVD}, we presented minor axes PVDs to look for signatures of radial motion. We note a clear ``twisting'' feature in only two of our objects; GAMA228432 and GAMA593680. This ``twisting'' manifests itself as an excess of gas either above or below the disc and either at approaching or receding velocity (so as to create a bend in the PVD profile). This feature is often associated with radial inflows or outflows along the disc.

\par By modelling the kinematically settled bulk of the gas using an arctan rotation curve, we find multiple instances of ring-type gas distributions, which are often associated with the presence of a resonance system, such as a bar \citep[e.g. see][]{combes91, comeron14, fraser-mckelvie20, chiba21}. As previously discussed, these ring, or multiple ring, structures can emerge from the presence of Lindblad resonances, where an orbit's epicyclic frequency is commensurate with the pattern frequency driven by the rotation of a central bar. In order to test the hypothesis that the molecular gas in our two objects where we identify ``twisting'' features in their PVDs is undergoing radial motion, we re-model the highest S/N object of the two, GAMA593680, with a radial bi-symmetric bar-flow component (we are unable to produce a compelling solution with the lower S/N object GAMA228432 due to the number of parameters required). GAMA593680 has the most compelling evidence for radial bar-flow (i.e. ring-type profiles, un-modelled kinematic components and ``twisting'' minor-axis PVD). To model radial motion induced by a central bar, we use the \texttt{radial$\_$barflow} function within the \texttt{KinMS$\_$fitter} wrapper, which is based on the non-axisymmetric models outlined in \citet{spekkens07}. This radial motion model assumes that the bar extends from the galaxy's centre to a radius $R_{bar}$, with a phase $\phi_{bar}$ and that the gas has a fixed radial and transverse velocity ($V_r$, $V_t$). In both models, we use the same surface brightness profile used in the preliminary modelling (see Table~\ref{tab: fits}) and an arctan rotation curve. The results of this process are given in Figure~\ref{fig: 593680_rad}, where we present PVDs extracted from both the major and minor axes. The BIC chooses the model with radial motion as the best-fit model for GAMA593680, despite the penalisation for the four additional parameters introduced by the radial bar-flow (see Appendix~\ref{appendix: corner}). Visually, the radial bar-flow provides the flaring in the velocity direction that we observe in the major-axis PVDs. There is clear improvement in the model when comparing against the major and minor-axis PVDs in Figure~\ref{fig: 593680_rad}. The remaining flux asymmetries we observe in the residual panels may simply trace flux asymmetries in the object, as opposed to unmodelled kinematic features. However, the improvement in the model at least suggests that radial motion plays a part in this object's molecular gas kinematics.

\par We can infer from our models of GAMA593680 (a superposition of a Gaussian disc-like profile and Gaussian ring-type profile), that a central bar may be responsible for driving radial gas flow towards resonant orbits commensurate with the bar pattern frequency. We do observe a ring-type gas structures in GAMA106389, GAMA228432 and GAMA417678, however, only see visual evidence of radial motion in GAMA228432. We are unable to identify whether these objects have optical bars due to their high inclination. Although ring structures such as these are typically caused by a bar, it is possible in some cases for minor mergers to cause the formation of inner discs and rings in a galaxy \citep{eliche-moral11}. Unlike in the cases of GAMA228432 and GAMA593680, the other two galaxies identified as having ring-type molecular gas profiles, GAMA106389 and GAMA417678, do not show any indication of ongoing radial motion of gas. This does not preclude past gas motion which has led to a centralised molecular gas distribution in these objects or gas flows that small enough as to be beyond the sensitivity of our observations. However, objects GAMA238125, GAMA574200 and GAMA618906 show no evidence of a ring-type gas distribution, which further hinders the argument for a consistent kinematic process occurring in all of our outflow-type objects, as we find no evidence in the gas kinematics for a resonance system. These findings reinforce much of the discussion in H21, where we emphasise the variety in the properties of this sample, despite finding the consistent result of centrally concentrated molecular gas. Our results here imply that there is no one consistent kinematic process by which gas is driven into the centre of these galaxies undergoing large-scale ionised outflow. 

\par As noted in Section~\ref{subsubsec: majorPVD}, we do see some indication of flux asymmetry in the major axis PVDs drawn from our sample, implying that although the bulk of the gas in our objects appears dynamically cold, there may be some component of the gas that has undergone a disturbance. However, with reference to Section~\ref{subsec: cold} and Figure~\ref{fig: flux_asym}, the flux asymmetry we see in our outflow-types generally falls well within the bulk of the distribution derived from PHANGS-ALMA galaxies of their approximate stellar mass. This does not preclude the possibility that the flux asymmetry we observe results from a minor merger in the objects' histories, but it does suggest that either the objects have largely dynamically re-stabilised following the merger in their histories or the merger occurred in such a way as not to significantly disturb the objects (e.g. gas accretion along the plane of the disc).

\par Minor mergers can create morphological distortions with respect to the disc, creating a pattern frequency and subsequently resonating, stable orbits in the centre of the disc. Again with reference to GAMA593680, we see a significant warp in the dust lane of this object in the HSC imagery (see H21), suggestive of a morphological disturbance in this object's history. If our outflow-types have all experienced a minor-merger event, it may be that this has led to a resonant ring structure in some and and less obvious morphological instabilities in others. This may also aid in explaining the kinematic complexity we note in GAMA593680, as merging events are likely to cause a degree of dynamic turbulence.

\par In H21, we find that our outflow-type objects (i.e. galaxies hosting large-scale outflows of ionised gas) have more centrally-concentrated molecular gas compared to a control sample. Star-formation in our outflow-types is also restricted to the same central region ($<r_e$). This observation is also confirmed by \citet{bao21}, who find older stellar populations in the centre of their sample of normal star-forming galaxies, suggesting the stars in the centre of these objects formed prior to the outer stellar population. This is further reinforced by the enhanced central metallicites found by \citet{rodriguez-baras19} and \citet{bao21}. From these combined results, we can infer an ``inside-out'' process, whereby stars primarily form in the galactic centre. However, we note that \citet{bao21} observe an elevation is SFR density compared to their control sample, whereas we do not see see evidence of an enhancement in our outflow-types. The outflow-types studied by \citet{bao21} appear more powerful, in terms of the strength of their outflows, than those considered in our sample \citep[where we detect outflows using the technique outlined in ][]{ho16}. We conjecture that our samples are, therefore, not entirely analogous or are perhaps at different stages of their outflow activity. In terms of finding a process whereby molecular gas and star-formation is centralised, we only find evidence that two of the objects in our sample are undergoing a process whereby gas could be being driven radially inwards. In the remainder of our sample, the galaxies appear to possess a range of surface brightness profiles (see Section~\ref{subsec: Kin}); some hosting ring-type gas distributions indicative of a driving pattern frequency and others hosting disc-like profiles. The variety of these results implies that there is not a consistent process responsible for centralising molecular gas and star-formation is galaxies hosting large-scale ionised outflows. However, we note the small size of our sample limits our ability to draw firm conclusions. Larger scale surveys of main-sequence star-forming galaxies are requiring to expand on our initial findings.

\subsection{Molecular Gas Outflows}
\label{subsec: molecular outflows}

The original motivation of this work (prior to the findings of H21), was to assess whether galaxies on the main-sequence of star-formation, which harbour galactic-scale stellar winds, entrain molecular gas in their ionised gas outflows. Molecular gas outflows for galaxies with extreme star-formation activity or with active galactic nuclei (AGN) appear relatively common at all redshifts \citep[e.g. see][]{veilleux05, cicone14, fluetsch19, lutz20}, but many simulations require molecular outflows in more ``normal'' star-forming galaxies in order to regulate star-formation according to the ``Equilibrium Model'' \citep{lilly13}. These simulations have been successful in reproducing a large number of critical observations \citep[e.g.][]{bouche10, dave11}, but this is without direct observational evidence. In both this work and in H21, we attempt to reconcile this assumption of molecular gas outflows in main-sequence galaxies with observations of such galaxies with large-scale ionised outflows. In our investigation, while we do not find direct evidence of extra-planar molecular gas entrained in the ionised winds of these objects, we do find evidence that that gas is more centralised and provide a potential mechanism for gas to be driven inwards in a subset of these galaxies. These findings are, however, subtle and if there is molecular gas being expelled by the ionised wind, it is at levels below the sensitivity of our observations. This is similar to the conclusion of \citet{roberts-borsani20b}, who did not detect evidence of CO or HI outflows at the detection limit of the deepest data available in a sample of nearby galaxies showing both ionised (H$\upalpha$) and neutral (NaD) outflows. These observations may be consistent with the findings of \citet{tacchella16}, who report that galaxies move above and below the star-formation main-sequence through cycles of compaction and depletion on far longer timescales at $\rm z=0$ ($\rm\sim 5\ Gyr$) compared to their counterparts at higher redshifts. This may result from greater disc stability at low redshifts and lower gas accretion rates \citep{zolotov15, tacchella16}. It is possible, therefore, that our outflow-types are undergoing a compaction stage, but the subsequent depletion process, where molecular gas may be expelled, is occurring over far longer timescales than galaxies that harbour extreme star-formation or AGN.

\section{Summary and Conclusions}
\label{sec: conclusions}

In this work, we build on the main finding of H21 that the molecular gas in galaxies hosting large-scale outflows of ionised gas is more centrally concentrated compared to objects without these intense ionised winds. In Section~\ref{subsec: Kin}, we perform a kinematic analysis of our outflow-type galaxies using the \texttt{KinMS} package, which allows us to model the dynamic structure of their molecular gas content. It also enables us to interpret the kinematic features we see in the major and minor-axis PVDs extracted from our objects, by comparing the data with dynamically stable models. Our key findings are:

\begin{itemize}

    \item Four of the seven outflow-type objects we model with \texttt{KinMS} appear to possess ring-type gas distributions, which are indicative of a resonance system. We infer that these objects may host a bar or an axi-asymmetric feature caused by a minor merger, which could drive a resonant ring structure.
    
    \item The bulk of the gas in our objects is consistent with our dynamically stable models. Furthermore, the gas rotation generally traces the velocity curves generated from MGE models derived from r-band photometry. {These} findings imply that the molecular gas is largely dynamically cold and rotation-dominated (where the rotation is driven primarily by the gravity well of the stellar disc).
    
    \item We do not find a consistent model that best fits the molecular gas in our sample galaxies; a variety of different ring-type and disc-type (and superpositions) are selected for each object. This suggests that a range of kinematic processes may exist in these objects; with no one prevailing mechanism facilitating the centralisation of the molecular gas in these galaxies hosting large-scale ionised outflows.
    
    \item Despite finding that the bulk of the molecular gas in our outflow-type objects is dynamically cold, we do observe some kinematic features not captured by our models when examining our major-axis PVDs. In some cases, we observe a clear central flaring of the gas in the velocity direction. More ubiquitously, we observe minor flux asymmetries embedded in the dynamically cold bulk of the gas. This could indicate historical dynamical disruption. However, the flux asymmetry in these objects is not higher than that observed in a control sample of galaxies.
    
    \item We observe a ``twisting'' feature in the two of the minor-axis PVDs we draw from our data, whereby there is an asymmetric gas enhancement both spatially and on the velocity axis about the kinematic centre. This feature is associated with radial gas motion either from radial inflows or outflows. If this radial motion describes gas moving towards the centre of our objects, it could supply the mechanism through which molecular gas becomes centralised in a subset of our outflow-types, which subsequently powers the ionised gas winds. However, we note that we do not see evidence of radial gas motion in the remainder of our sample.
    
    \item We test the hypothesis of radial gas flow by simulating a bar in our KinMS model for our highest S/N object with signatures of radial gas flow (GAMA593680). The addition of radial motion driven by a bar improves the model according to the BIC. We infer from these observations and the literature, that the centralisation of the molecular gas in this object may be driven by axi-asymmetric kinematic structures, such as a bar or minor merger (where gas flow is largely confined to the plane of the disc).

\end{itemize}

\par Our foremost conclusion from this work is that we do see some signatures of radial motion of molecular gas in our outflow-type objects and are able to directly model radial bar-flow in our highest S/N object. This radial flow could cause the increased central concentration of molecular gas and star formation we reported in H21. However, we only observe evidence for radial gas motion in a small subset of our sample in addition to finding a large variety of surface brightness profiles in the sample as a whole. The lack of a universal best-fit model for the spatial gas distribution in our galaxies, in addition to finding signatures of radial gas motion in only two objects, suggests that our sample is composed of galaxies that are governed by a range of kinematic processes. We also find that the bulk of the molecular gas appears dynamically cold, with some of minor flux asymmetries embedded in a relaxed disc. We hypothesise that the existence of radial flow in a subset of our objects may be indicative of bar instabilities or minor mergers, which can drive gas inwards. Furthermore, we can infer from the small flux asymmetries that there was some past dynamic disruption, such as a minor merger, but that the gas has subsequently relaxed. We do find a potential mechanism by which gas is driven radially inwards in a subset of our objects, by the possible presence of a bar or disc instability caused by a minor merger. Based on the results of H21, a centralised molecular gas distribution appears to be a necessary condition for galaxies to power large-scale ionised gas winds. However, this mechanism is not present in the majority of our objects, implying they are at different stages of their evolution, have radial gas motion at levels below our sensitivity or are undergoing an entirely separate kinematic process by which molecular gas is centralised. The variety of molecular gas kinematic profiles we model in this work suggests that galaxies hosting galactic-scale ionised winds do not share one ubiquitous dynamical process that centralises their gas and subsequently launches their stellar winds.

\par The conclusions of this study are limited by sample size and depth of the observations, but mostly by the high inclination of the galaxies, as required by the diagnostics of \citet{ho16} to identify ionised gas winds. Extending this type of analysis to larger samples with a broader range of inclinations should enable us to better understand the conditions required for normal star-forming galaxies to power large-scale outflows, a crucial component of the baryon cycle.

\section*{Acknowledgements}
\label{sec: acknowledgements}
LMH and AS acknowledge support from the Royal Society. TAD acknowledges support from the UK Science and Technology Facilities Council through grants ST/S00033X/1 and ST/W000830/1. This paper makes use ALMA data from project {S/JAO.ALMA\#2011.0.01234.S}.  ALMA is a partnership of ESO (representing its member states), NSF (USA) and NINS (Japan), together with NRC (Canada), MOST and ASIAA (Taiwan), and KASI (Republic of Korea), in cooperation with the Republic of Chile. The Joint ALMA Observatory is operated by ESO, AUI/NRAO and NAOJ.

\section*{Data Availability}

All data products used in this work are available in the SAMI public database (\url{https://sami-survey.org/data}) and the ALMA science archive (\url{https://almascience.eso.org/asax/}).



\bibliography{refs.bib}
\bibliographystyle{mnras}


\appendix

\section{KinMS Corner Plots}
\label{appendix: corner}

Presenting all corner plots produced by our KinMS modelling.

\begin{figure*}
    \begin{subfigure}{\textwidth}
        \centering
        \includegraphics[width=.6\linewidth]{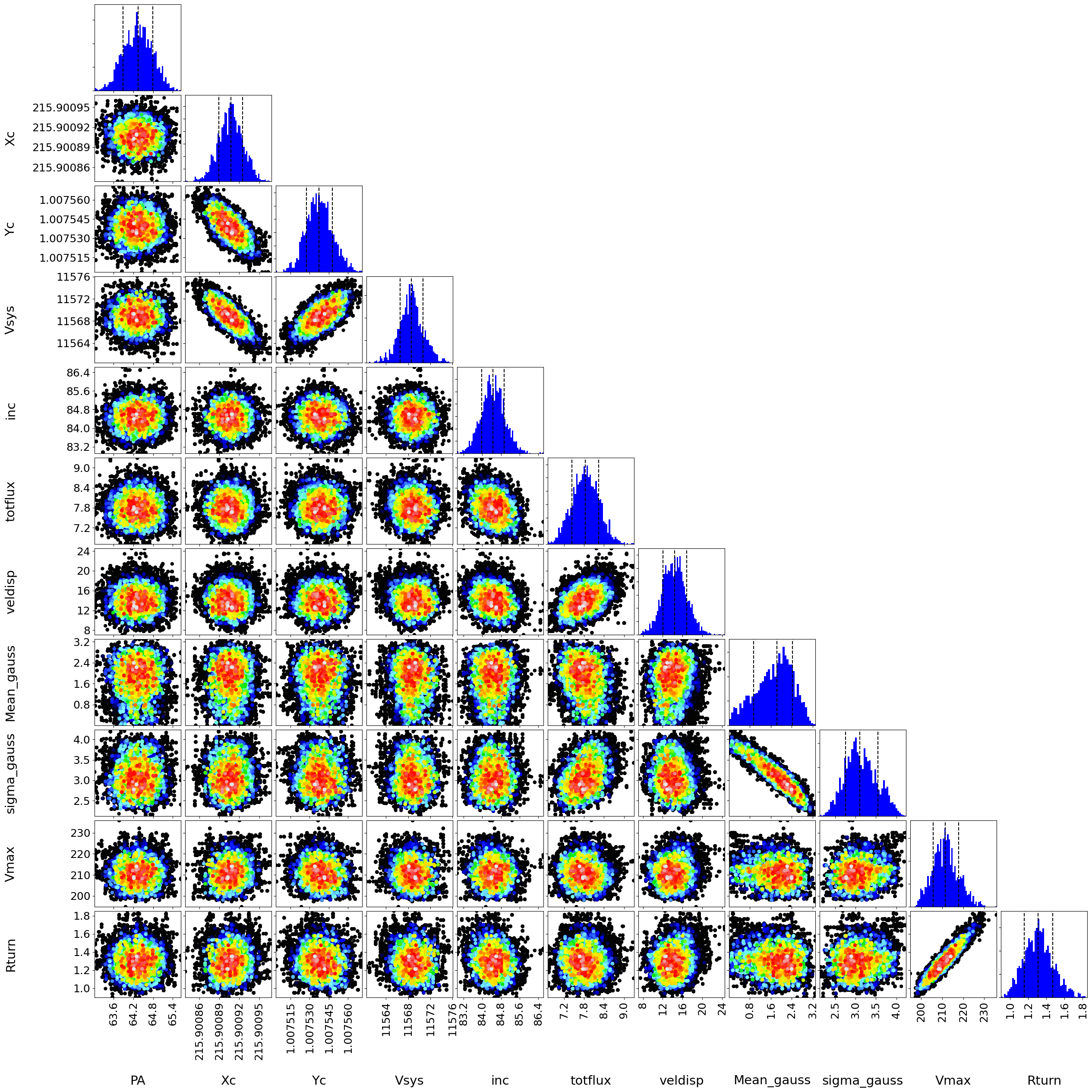}
        \caption{GAMA106389}
    \end{subfigure}
    
    \begin{subfigure}{\textwidth}
        \centering
        \includegraphics[width=.6\linewidth]{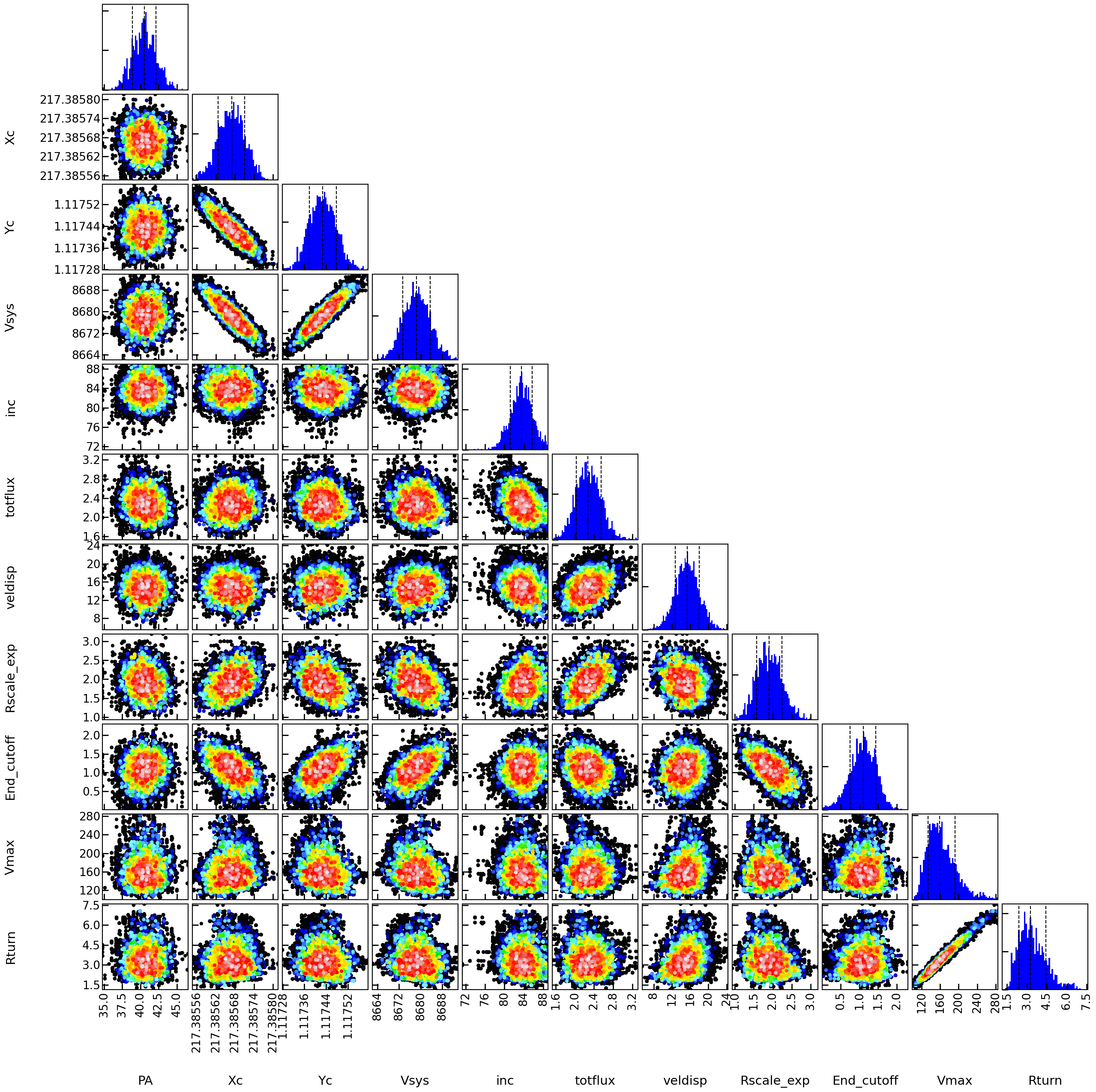}
        \caption{GAMA228432}
    \end{subfigure}
    
    \caption{Corner plots from the implementation of of the MCMC tool GAStimator on the models created using KinMS. Each panel in the figure gives the unique parameter spaces used in the fit for each of our outflow-type objects.}
    \label{fig: corners}
    
\end{figure*}

\begin{figure*}
    
    \begin{subfigure}{\textwidth}
        \centering
        \includegraphics[width=.6\linewidth]{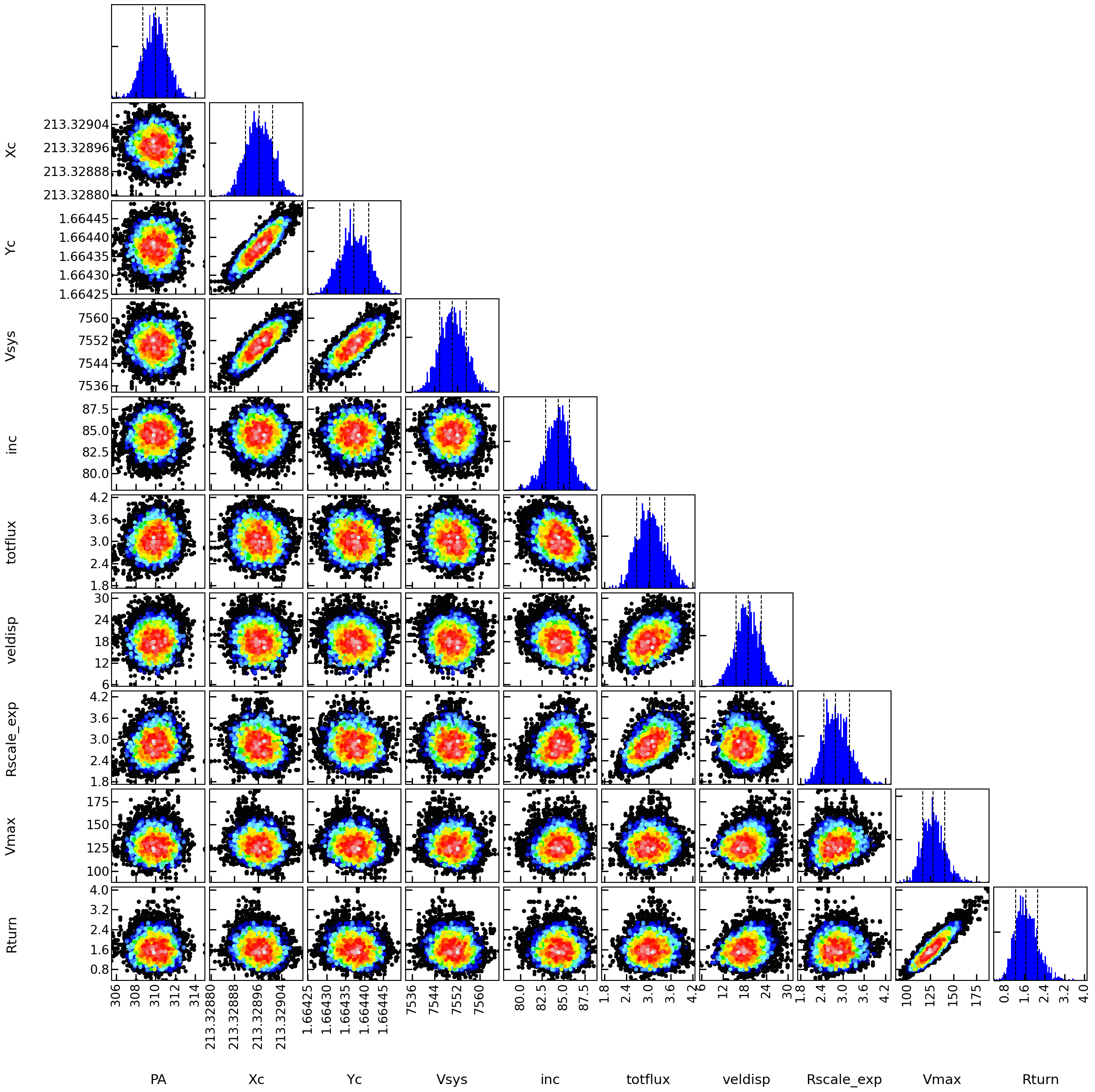}
        \caption{GAMA238125}
    \end{subfigure}

    \begin{subfigure}{\textwidth}
        \centering
        \includegraphics[width=.6\linewidth]{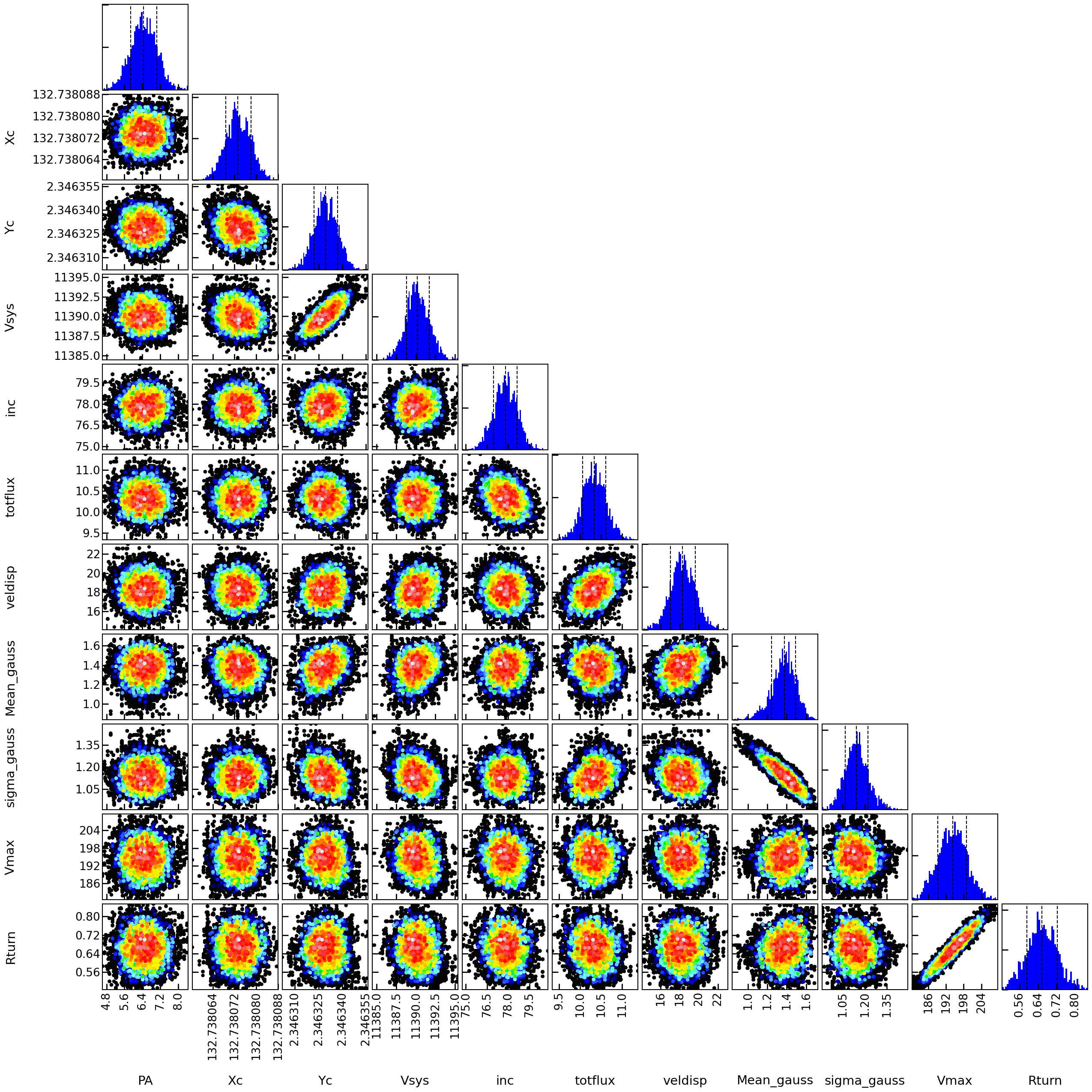}
        \caption{GAMA417678}
    \end{subfigure}
    
    \contcaption{}
\end{figure*}

\begin{figure*}
    \begin{subfigure}{\textwidth}
        \centering
        \includegraphics[width=.6\linewidth]{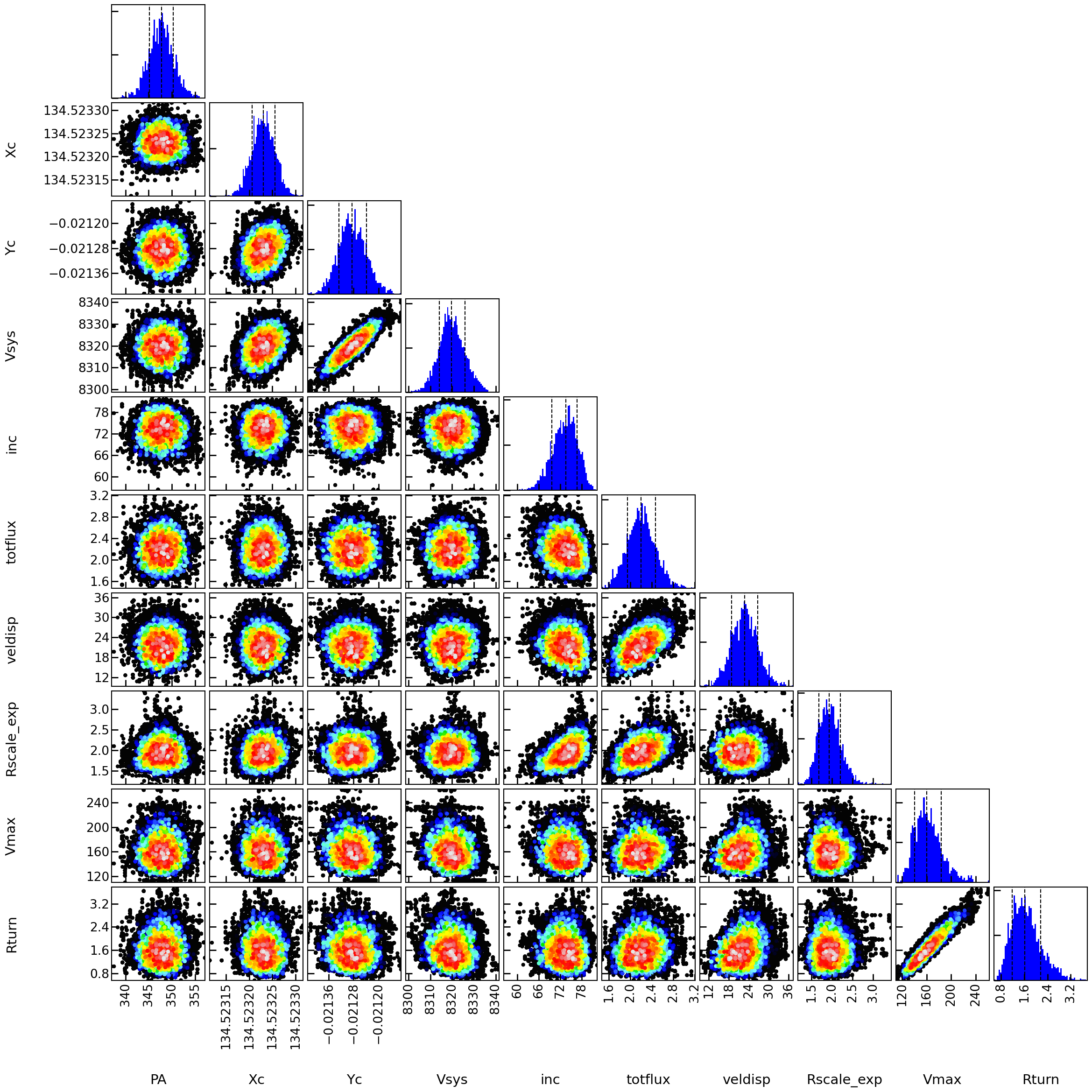}
        \caption{GAMA574200}
    \end{subfigure}
    
    \begin{subfigure}{\textwidth}
        \centering
        \includegraphics[width=.6\linewidth]{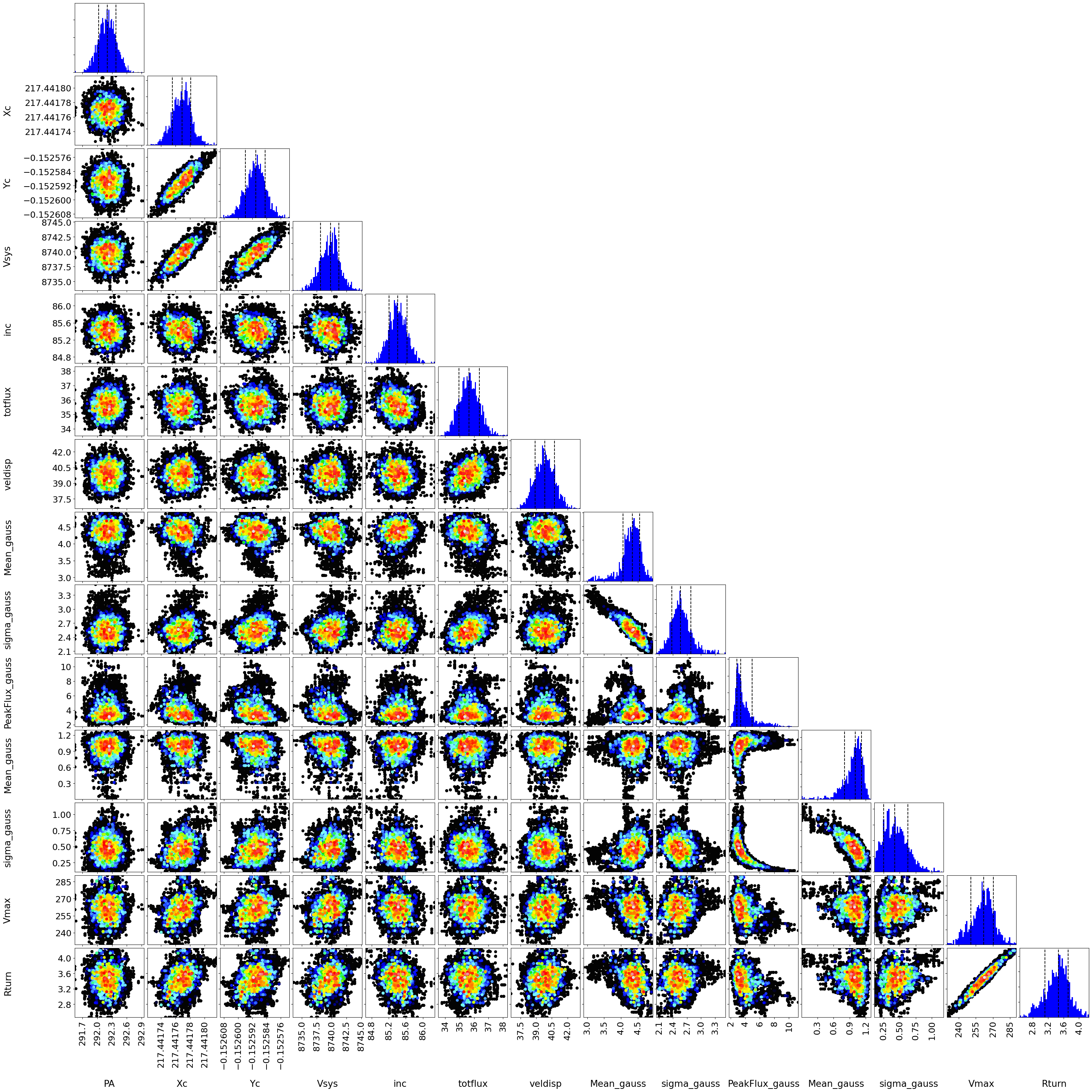}
        \caption{GAMA593680}
    \end{subfigure}
    
    \contcaption{}
\end{figure*}

\begin{figure*}

    \begin{subfigure}{\textwidth}
        \centering
        \includegraphics[width=.6\linewidth]{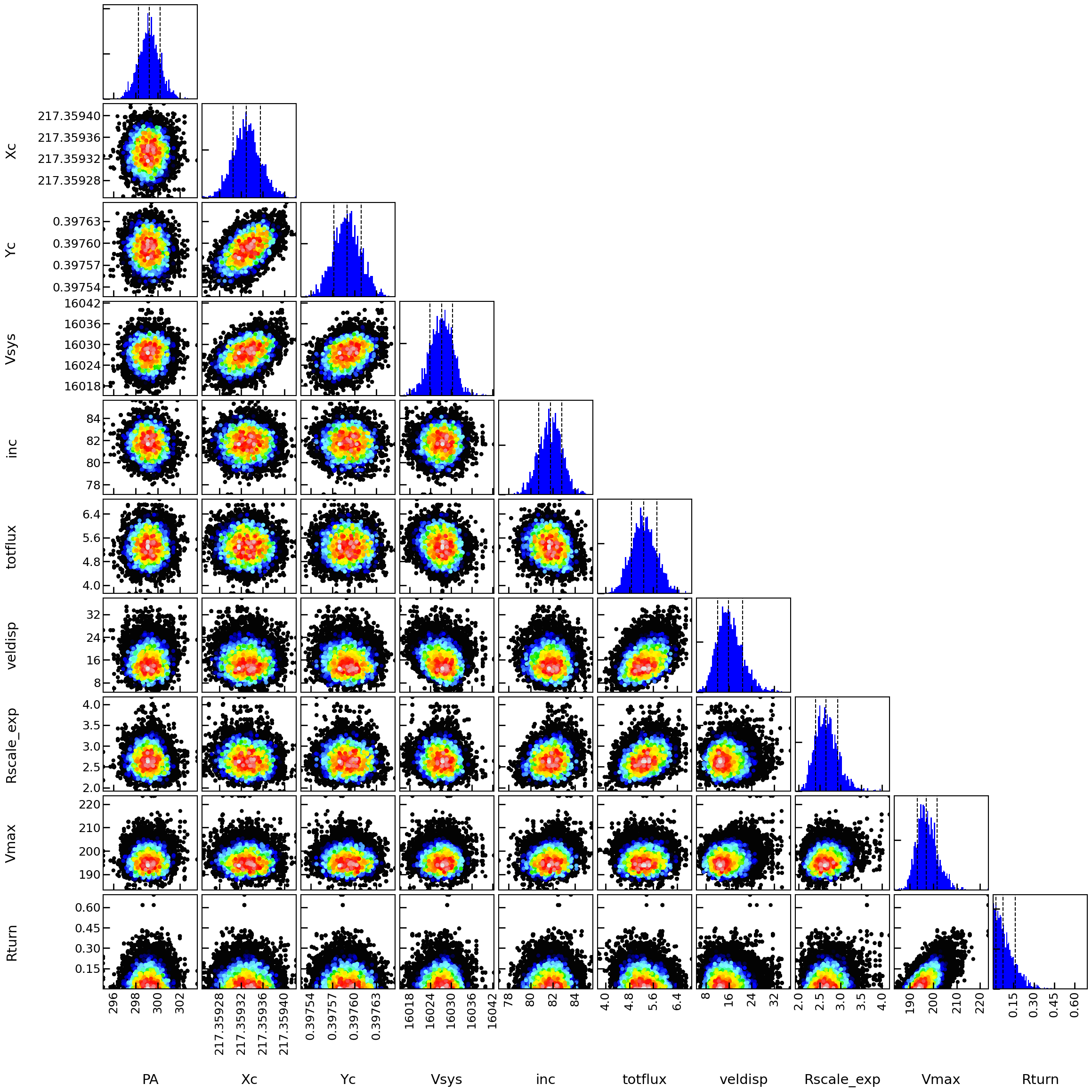}
        \caption{GAMA618906}
    \end{subfigure}

    \contcaption{}
\end{figure*}

\clearpage
\newpage

\begin{figure*}

\centering
    \includegraphics[width=.6\linewidth]{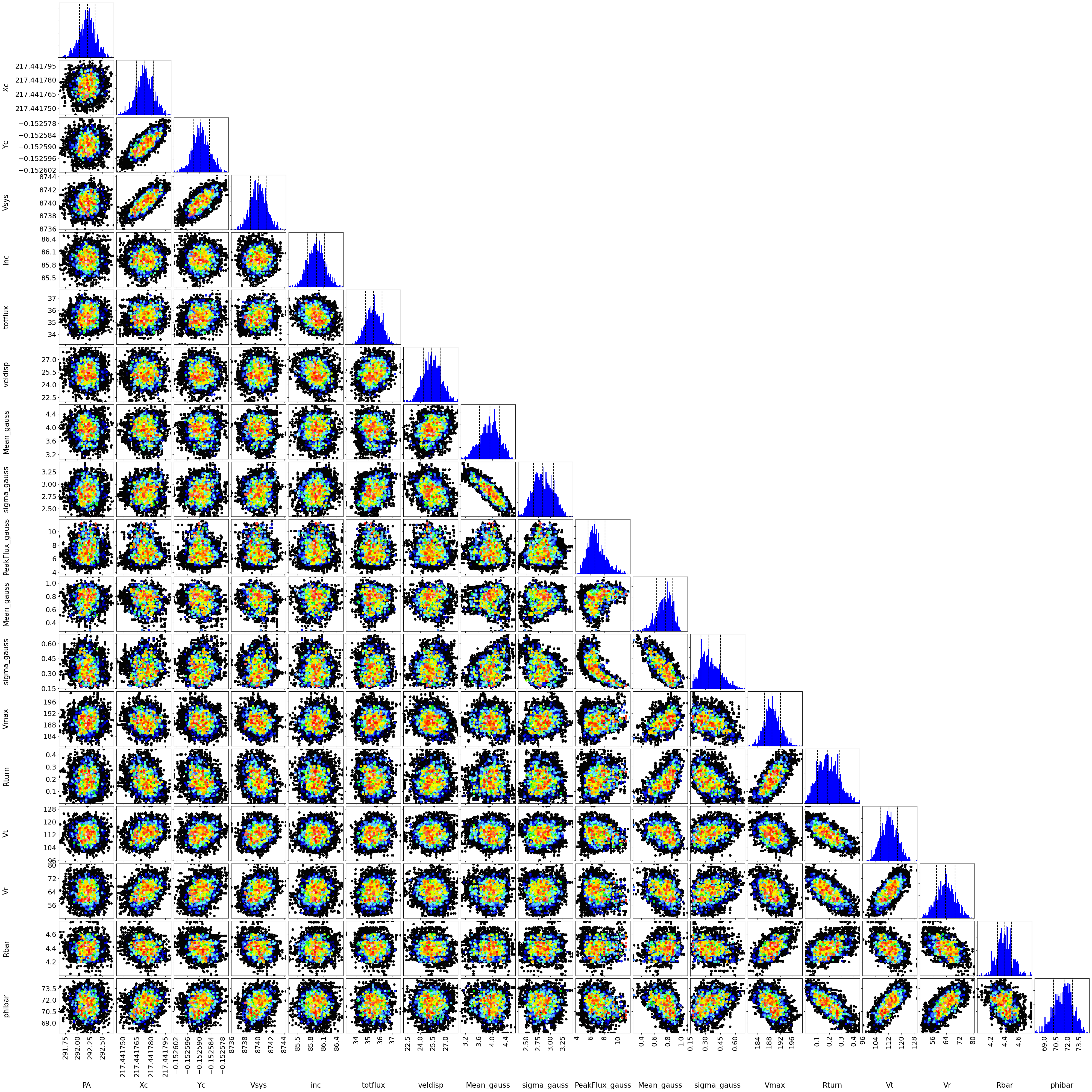}
    \caption{Corner plot produced by KinMS model modified to include radial bar-flow for GAMA593680.}
\label{fig: corners}
    
\end{figure*}

\clearpage
\newpage

\section{KinMS Parameters}
\label{appendix: params}

Presenting all initial and final parameters from our fitting procedure and the boxcar priors used in each case.

\begin{table*} 
    \centering
    \caption{Details of fitting parameters used for GAMA106389.} 
    \begin{tabular}{l l l l}
    	\hline
	    Parameter & Initial Guess & [lower bound, upper bound] & Best fit \\[2pt]
	    \hline
        \hline \\
        PA ($^\circ$) & 59.9 & [0.0, 360.0] & $64.3\substack{+0.5\\-0.4}$\\[2pt]
        $\rm X_c$ ($^\circ$) & 215.90102 & [215.89755, 215.90449] & $215.90090 \pm 0.00002$\\[2pt]
        $\rm Y_c$ ($^\circ$) & 1.00763 & [1.00416, 1.01110] & $1.00754 \pm 0.00001$\\[2pt]
        $\rm V_{sys}$ (\kms) & 11572 & [11277, 11866] & $11569 \pm 2$ \\[2pt]
        $\rm inc$ ($^\circ$) & 77.7 & [1.0, 89.0] & $84.4 \pm 0.5$ \\[2pt]
        totflux (Jy) & 7.5 & [0.0, $3\times$guess] & $7.8 \pm 0.4$ \\[2pt]
        $\rm vel_{disp}$ (\kms) & 10 & [0, 50] & $14\substack{+3\\-2}$ \\[2pt]
        $\rm Peak_{Flux, exp}$ & 1 & Fixed & 1 \\[2pt]
        $\rm Mean_{gauss}$ (\arcsec) & 2.6 & [0.0, 12.3] & $2.0\substack{+0.6\\-0.7}$ \\[2pt]
        $\rm Sigma_{gauss}$ (\arcsec) & 2.4 & [0.0, 12.3] & $3.0 \pm 0.3$ \\[2pt]
        $\rm V_{max}$ (\kms) & 189 & [0, 1000] & $211\substack{+6\\-5}$ \\[2pt]
        $\rm R_{turn}$ (\arcsec) & 0.8 & [0, 12.3] & $1.2\substack{+0.2\\-0.1}$ \\[2pt]
	    \hline
	\end{tabular}
	\label{tab: 106389 params}
\end{table*}

\begin{table*} 
    \centering
    \caption{Details of fitting parameters used for GAMA228432.} 
    \begin{tabular}{l l l l}
    	\hline
	    Parameter & Initial Guess & [lower bound, upper bound] & Best fit \\[2pt]
	    \hline
        \hline \\
        PA ($^\circ$) & 40 & [0, 360] & $41 \pm 2$ \\[2pt]
        $\rm X_c$ ($^\circ$) & 217.38572 & [217.38080, 217.39064] & $217.38566\substack{+0.00004\\-0.00005}$\\[2pt]
        $\rm Y_c$ ($^\circ$) & 1.11741 & [1.11249, 1.12233] & $1.11744 \pm 0.00005$\\[2pt]
        $\rm V_{sys}$ (\kms) & 8678 & [8519, 8865] & $8680\substack{+6\\-5}$ \\[2pt]
        $\rm inc$ ($^\circ$) & 87 & [1, 89] & $86 \pm 2$ \\[2pt]
        totflux (Jy) & 1.64892 & Fixed & 1.64892 \\[2pt]
        $\rm vel_{disp}$ (\kms) & 10 & [0,50] & $13 \pm 2$ \\[2pt]
        $\rm Peak_{Flux, exp}$ & 1 & Fixed & 1 \\[2pt]
        $\rm R_{scale, exp}$ (\arcsec) & 1.4 & [0,15.7] & $1.6 \pm 0.3$ \\[2pt]
        $\rm Start_{cutoff}$ (\arcsec) & 0 & Fixed & 0 \\ [2pt]
        $\rm End_{cutoff}$ (\arcsec) & 1.1 & [0.0, 7.9] & $1.3\substack{+0.4\\-0.3}$ \\[2pt]
        $\rm V_{max}$ (\kms) & 100 & [0, 1000] & $147\substack{+30\\-20}$ \\[2pt]
        $\rm R_{turn}$ (\arcsec) & 1.4 & [0.0, 15.7] & $2.9\substack{+1.0\\-0.8}$ \\[2pt]
       
	    \hline
	\end{tabular}
	\label{tab: 228432 params}
\end{table*}

\begin{table*} 
    \centering
    \caption{Details of fitting parameters used for GAMA238125.} 
    \begin{tabular}{l l l l}
    	\hline
	    Parameter & Initial Guess & [lower bound, upper bound] & Best fit \\[2pt]
	    \hline
        \hline \\
        PA ($^\circ$) & 311 & [0, 360] & $310 \pm 1$ \\[2pt]
        $\rm X_c$ ($^\circ$) & 213.32889 & [[213.32386, 213.33392] & $213.32898 \pm 0.00004$\\[2pt]
        $\rm Y_c$ ($^\circ$) & 1.66443 & [1.65940, 1.66946] & $1.66438 \pm 0.00003$\\[2pt]
        $\rm V_{sys}$ (\kms) & 7551 & [7362, 7748] & $7551 \pm 4$ \\[2pt]
        $\rm inc$ ($^\circ$) & 86 & [1, 89] & $86 \pm 2$ \\[2pt]
        totflux (Jy) & 2.11234 & Fixed & 2.11234 \\[2pt]
        $\rm vel_{disp}$ (\kms) & 10 & [0,50] & $15 \pm 3$ \\[2pt]
        $\rm Peak_{Flux, exp}$ & 1 & Fixed & 1 \\[2pt]
        $\rm R_{scale, exp}$ (\arcsec) & 2.4 & [0.0, 16.1] & $2.3\substack{+0.3\\-0.2}$ \\[2pt]
        $\rm V_{max}$ (\kms) & 122 & [0, 1000] & $124 \pm 10$ \\[2pt]
        $\rm R_{turn}$ (\arcsec) & 1.3 & [0, 16.1] & $1.4 \pm 0.4$ \\[2pt]
       
	    \hline
	\end{tabular}
	\label{tab: 238125 params}
\end{table*}

\begin{table*} 
    \centering
    \caption{Details of fitting parameters used for GAMA417678.} 
    \begin{tabular}{l l l l}
    	\hline
	    Parameter & Initial Guess & [lower bound, upper bound] & Best fit \\[2pt]
	    \hline
        \hline \\
        PA ($^\circ$) & 6.6 & [0.0, 360.0] & $6.5 \pm 0.6$ \\[2pt]
        $\rm X_c$ ($^\circ$) & 132.73820 & [132.73561, 132.74078] & $132.73807 \pm 0.00001$\\[2pt]
        $\rm Y_c$ ($^\circ$) & 2.34620 & [2.34362, 2.34879] & $2.34633 \pm 0.00001$\\[1pt]
        $\rm V_{sys}$ (\kms) & 11393 & [11120, 11679] & $11390 \pm 2$ \\[2pt]
        $\rm inc$ ($^\circ$) & 77.5 & [1.0, 89.0] & $77.8 \pm 0.8$ \\[2pt]
        totflux (Jy) & 10.3 & [0.0, $3\times$guess] & $10.3 \pm 0.3$ \\[2pt]
        $\rm vel_{disp}$ (\kms) & 10 & [0, 50] & $18 \pm 1$ \\[2pt]
        $\rm Peak_{Flux, gauss}$ & 1 & Fixed & 1 \\[2pt]
        $\rm Mean_{gauss}$ (\arcsec) & 1.1 & [0.0, 9.1] & $1.4 \pm 0.1$ \\[2pt]
        $\rm Sigma_{gauss}$ (\arcsec) & 1.30 & [0.00, 9.11] & $1.10 \pm 0.08$ \\[2pt]
        $\rm V_{max}$ (\kms) & 179 & [0, 1000] & $195 \pm 5$ \\[2pt]
        $\rm R_{turn}$ (\arcsec) & 0.45 & [0.00, 9.11] & $0.70 \pm 0.07$ \\[2pt]

	    \hline
	\end{tabular}
	\label{tab: 417678 params}
\end{table*}

\begin{table*} 
    \centering
    \caption{Details of fitting parameters used for GAMA574200.} 
    \begin{tabular}{l l l l}
    	\hline
	    Parameter & Initial Guess & [lower bound, upper bound] & Best fit \\[2pt]
	    \hline
        \hline \\
        PA ($^\circ$) & 349 & [0, 360] & $348 \pm 3$ \\[2pt]
        $\rm X_c$ ($^\circ$) & 134.52334 & [134.52070, 134.52598] & $134.52322 \pm 0.00002$\\[2pt]
        $\rm Y_c$ ($^\circ$) & -0.02112 & [-0.02376, -0.01848] & $-0.02129 \pm 0.00004$\\[2pt]
        $\rm V_{sys}$ (\kms) & 8326 & [8123, 8540] & $8318 \pm 6$ \\[2pt]
        $\rm inc$ ($^\circ$) & 70 & [1, 89] & $75\substack{+3\\-4}$ \\[2pt]
        totflux (Jy) & 1.41666 & Fixed & 1.41666 \\[1pt]
        $\rm vel_{disp}$ (\kms) & 10 & [0, 50] & $17\substack{+4\\-3}$ \\[2pt]
        $\rm Peak_{Flux, exp}$ & 1 & Fixed & 1 \\[2pt]
        $\rm R_{scale, exp}$ (\arcsec) & 1.4& [0.0, 7.5] & $1.5 \pm 0.2$ \\[2pt]
        $\rm V_{max}$ (\kms) & 107 & [0, 1000] & $162\substack{+30\\-20}$ \\[2pt]
        $\rm R_{turn}$ (\arcsec) & 0.5 & [0.0, 7.5] & $1.3\substack{+0.6\\-0.4}$ \\[2pt]
       
	    \hline
	\end{tabular}
	\label{tab: 574200 params}
\end{table*}

\begin{table*} 
    \centering
    \caption{Details of fitting parameters used for GAMA593680.} 
    \begin{tabular}{l l l l}
    	\hline
	    Parameter & Initial Guess & [lower bound, upper bound] & Best fit \\[2pt]
	    \hline
        \hline \\
        PA ($^\circ$) & 294.8 & [0.0, 360.0] & $292.2 \pm 0.2$ \\[2pt]
        $\rm X_c$ ($^\circ$) & 217.44187 & [217.43801, 217.44573] & $217.44177 \pm 0.00001$\\[2pt]
        $\rm Y_c$ ($^\circ$) & -0.15236 & [-0.15622, -0.14850] & $-0.15259 \pm 0.00001$\\[2pt]
        $\rm V_{sys}$ (\kms) & 8768 & [8438, 9068] & $8740\substack{+1\\-2}$ \\[2pt]
        $\rm inc$ ($^\circ$) & 80.9 & [1.0, 89.0] & $85.4 \pm 0.2$ \\[2pt]
        totflux (Jy) & 36.2 & [0.0, $3\times$guess] & $35.6 \pm 0.7$ \\[2pt]
        $\rm vel_{disp}$ (\kms) & 10.0 & [0.0, 50.0] & $39.7\substack{+1.0\\-0.9}$ \\[2pt]
        $\rm Peak_{Flux, gauss}$ & 1 & Fixed & 1 \\[2pt]
        $\rm Mean_{gauss}$ (\arcsec) & 4.8 & [0.0, 13.7] & $4.4\substack{+0.2\\-0.3}$ \\[2pt]
        $\rm Sigma_{gauss}$ (\arcsec) & 3.9 & [0.0, 13.7] & $2.4 \pm 0.2$ \\[2pt]
        $\rm Peak_{Flux, gauss}$ & 3.2 & [0.0, 500.0] & $3.3\substack{+0.9\\-0.5}$ \\[2pt]
        $\rm Mean_{gauss}$ (\arcsec) & 0.1 & [0.0, 13.7] & $1.0\substack{+0.1\\-0.2}$ \\[2pt]
        $\rm Sigma_{gauss}$ (\arcsec) & 2.4 & [0.0, 13.7] & $0.5 \pm 0.2$ \\[2pt]
        $\rm V_{max}$ (\kms) & 187 & [0, 1000] & $256\substack{+12\\-9}$ \\[2pt]
        $\rm R_{turn}$ (\arcsec) & 1.4 & [0.0, 13.7] & $3.6\substack{+0.4\\-0.2}$ \\[2pt]
       
	    \hline
	\end{tabular}
	\label{tab: 593680 params}
\end{table*}

\begin{table*} 
    \centering
    \caption{Details of fitting parameters used for GAMA618906.} 
    \begin{tabular}{l l l l}
    	\hline
	    Parameter & Initial Guess & [lower bound, upper bound] & Best fit \\[2pt]
	    \hline
        \hline \\
        PA ($^\circ$) & 300.1 & [0.0, 360.0] & $299.1\substack{+1.0\\-0.9}$ \\[2pt]
        $\rm X_c$ ($^\circ$) & 217.35939 & [217.35642, 217.36236] & $217.35933 \pm 0.00003$\\[2pt]
        $\rm Y_c$ ($^\circ$) & 0.39759 & [[0.39462,  0.40056] & $0.39759 \pm 0.00002$\\[2pt]
        $\rm V_{sys}$ (\kms) & 16028 & [15745, 16334] & $16028 \pm 3$ \\[2pt]
        $\rm inc$ ($^\circ$) & 80 & [1, 89] & $82 \pm 1$ \\[2pt]
        totflux (Jy) & 3.7 & [0.0, $3\times$guess] & $5.0 \pm 0.4$ \\[2pt]
        $\rm vel_{disp}$ (\kms) & 10 & [0, 50] & $16\substack{+5\\-4}$ \\[2pt]
        $\rm Peak_{Flux, gauss, 1}$ & 1 & Fixed & 1 \\[2pt]
        $\rm R_{scale, exp}$ (\arcsec) & 2.7 & [0.0, 8.5] & $2.9\substack{+0.4\\-0.3}$ \\[2pt]
        $\rm V_{max}$ (\kms) & 209 & [0, 1000] & $196 \pm 4$ \\[2pt]
        $\rm R_{turn}$ (\arcsec) & 0.41 & [0.00, 8.50] & $0.03\substack{+0.09\\-0.03}$ \\[2pt]
    
	    \hline
	\end{tabular}
	\label{tab: 618906 params}
\end{table*}

\begin{table*} 
    \centering
    \caption{Details of fitting parameters used for GAMA593680 with radial motion.} 
    \begin{tabular}{l l}
    	\hline
	    Parameter & Best fit \\[2pt]
	    \hline
        \hline \\
        PA ($^\circ$) & $292.2 \pm 0.2$ \\[2pt]
        $\rm X_c$ ($^\circ$) & $217.44177 \pm 0.00001$\\[2pt]
        $\rm Y_c$ ($^\circ$) &  $-0.15259 \pm 0.00001$\\[2pt]
        $\rm V_{sys}$ (\kms) &  $8740 \pm 1$ \\[2pt]
        $\rm inc$ ($^\circ$) &  $85.9 \pm 0.2$ \\[2pt]
        totflux (Jy) & $35.4 \pm 0.7$ \\[2pt]
        $\rm vel_{disp}$ (\kms) & $25 \pm 2$ \\[2pt]
        $\rm Peak_{Flux, gauss}$ & 1 \\[2pt]
        $\rm Mean_{gauss}$ (\arcsec) & $3.8\substack{+0.2\\-0.3}$ \\[2pt]
        $\rm Sigma_{gauss}$ (\arcsec) & $2.9 \pm 0.2$ \\[2pt]
        $\rm Peak_{Flux, gauss}$ & $7\substack{+2\\-1}$ \\[2pt]
        $\rm Mean_{gauss}$ (\arcsec) & $0.9\substack{+0.1\\-0.2}$ \\[2pt]
        $\rm Sigma_{gauss}$ (\arcsec) &  $0.2 \pm 0.1$ \\[2pt]
        $\rm V_{max}$ (\kms) & $189\substack{+3\\-2}$ \\[2pt]
        $\rm R_{turn}$ (\arcsec) & $0.22\substack{+0.08\\-0.07}$ \\[2pt]
        $\rm V_{t}$ (\kms) & $112 \pm 4$ \\[2pt]
        $\rm V_{r}$ (\kms) & $63\substack{+6\\-5}$ \\[2pt]
        $\rm R_{bar}$ (\arcsec) & $4.39\substack{+0.10\\-0.09}$ \\[2pt]
        $\rm Phi_{bar}$ ($^\circ$) & $71 \pm 1$ \\[2pt]       
    
	    \hline
	\end{tabular}
	\label{tab: 593680 rad params}
\end{table*}

\clearpage
\newpage

\section{Velocity Asymmetry Analysis}
\label{appendix: velocity}

Additional analysis conducted to test the velocity asymmetry of our major-axis PVDs. We look for the presence of velocity asymmetry by transforming our major-axis PVDs so that where gas is present (according to our masking process) the intensity is $=1$ and otherwise $=0$. The PVDs are then rotated by 180$^\circ$ and subtracted from the un-rotated, binary PVDs. Each of the panels below illustrates the results of this process for each of our objects.

\begin{figure*}
\centering
\begin{minipage}{.5\textwidth}
  \centering
  \includegraphics[width=.65\linewidth]{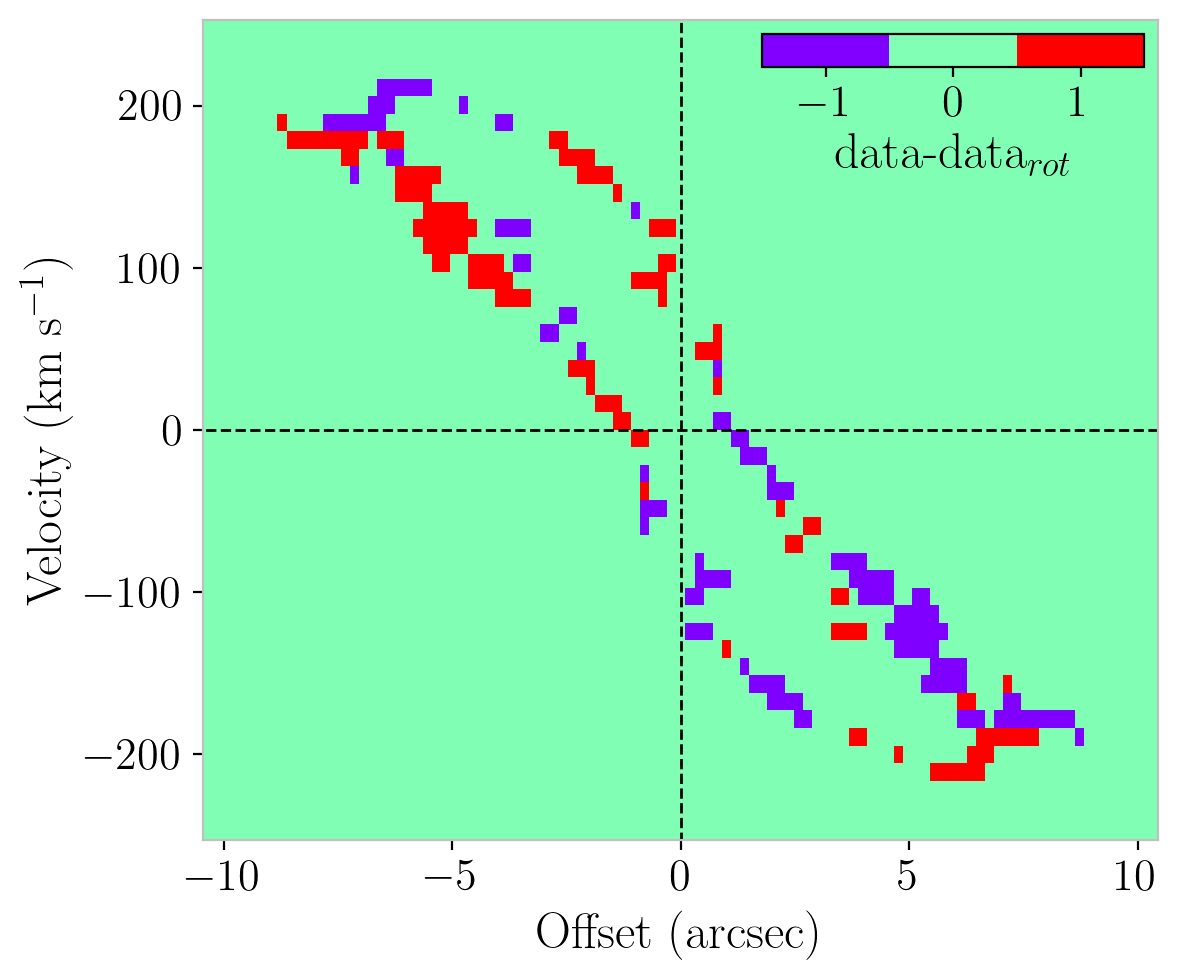}
  \caption{GAMA106389}
\end{minipage}%
\begin{minipage}{.5\textwidth}
  \centering
  \includegraphics[width=.65\linewidth]{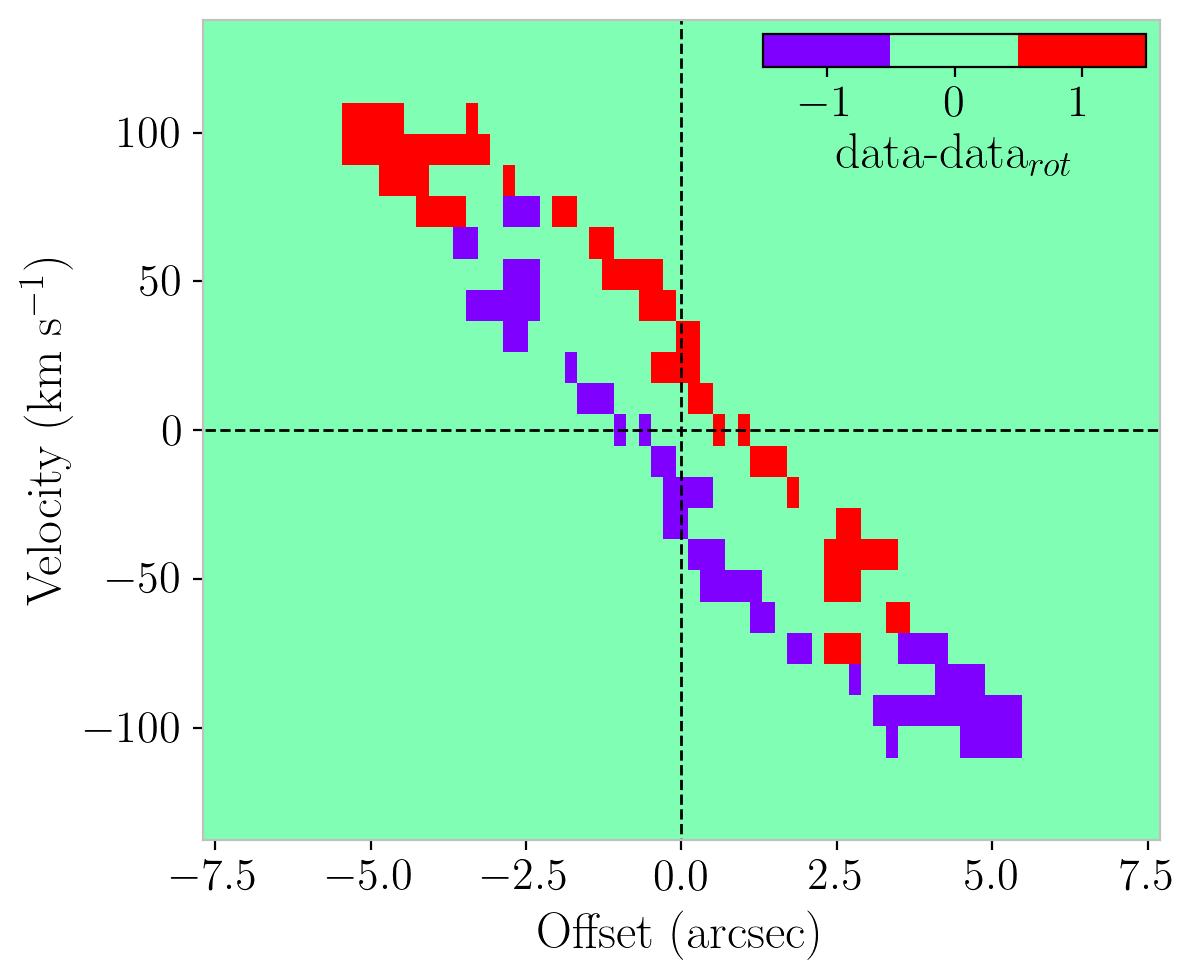}
  \caption{GAMA228432}
\end{minipage}
\newline
\centering
\begin{minipage}{.5\textwidth}
  \centering
  \includegraphics[width=.65\linewidth]{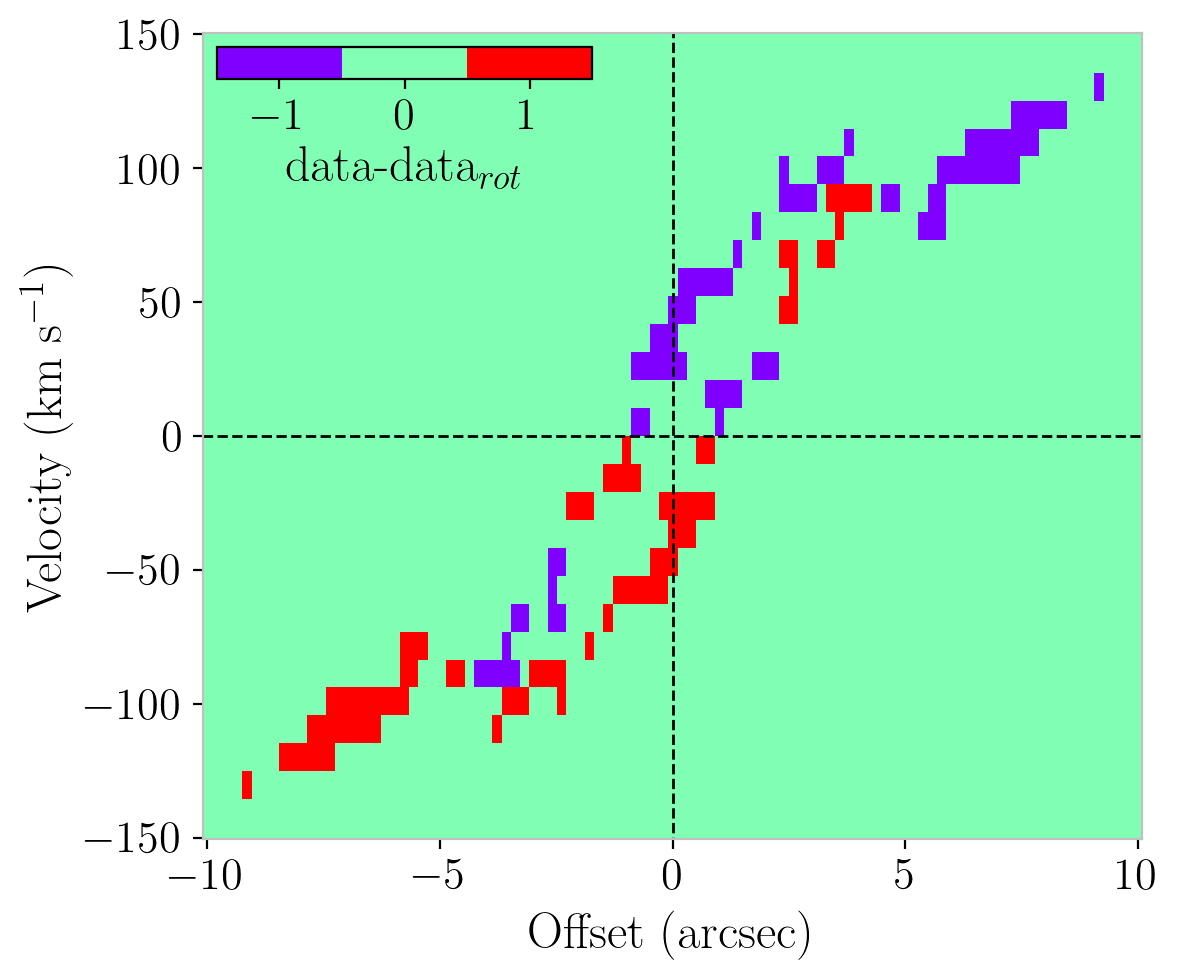}
  \caption{GAMA238125}
\end{minipage}%
\begin{minipage}{.5\textwidth}
  \centering
  \includegraphics[width=.65\linewidth]{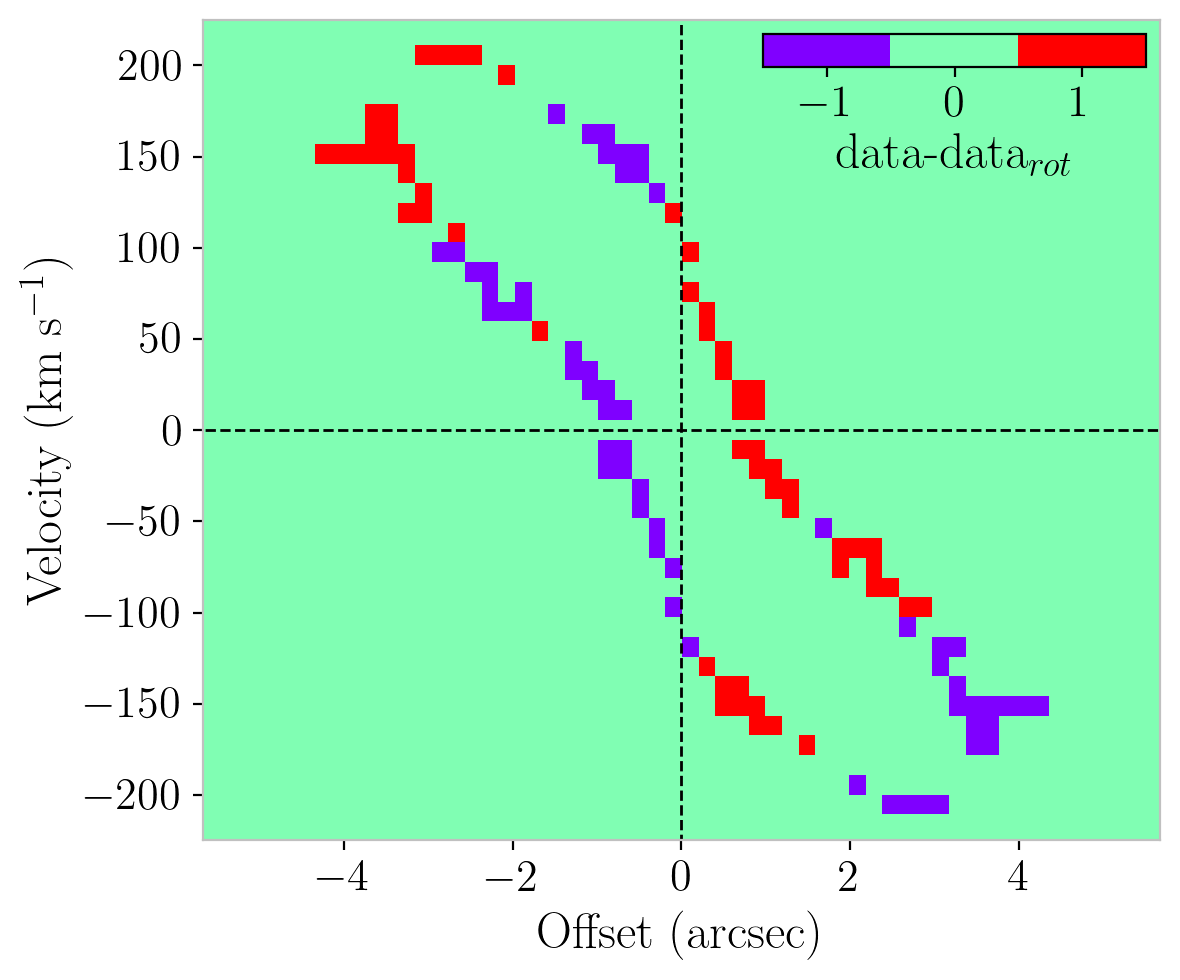}
  \caption{GAMA417678}
\end{minipage}
\newline
\centering
\begin{minipage}{.5\textwidth}
  \centering
  \includegraphics[width=.65\linewidth]{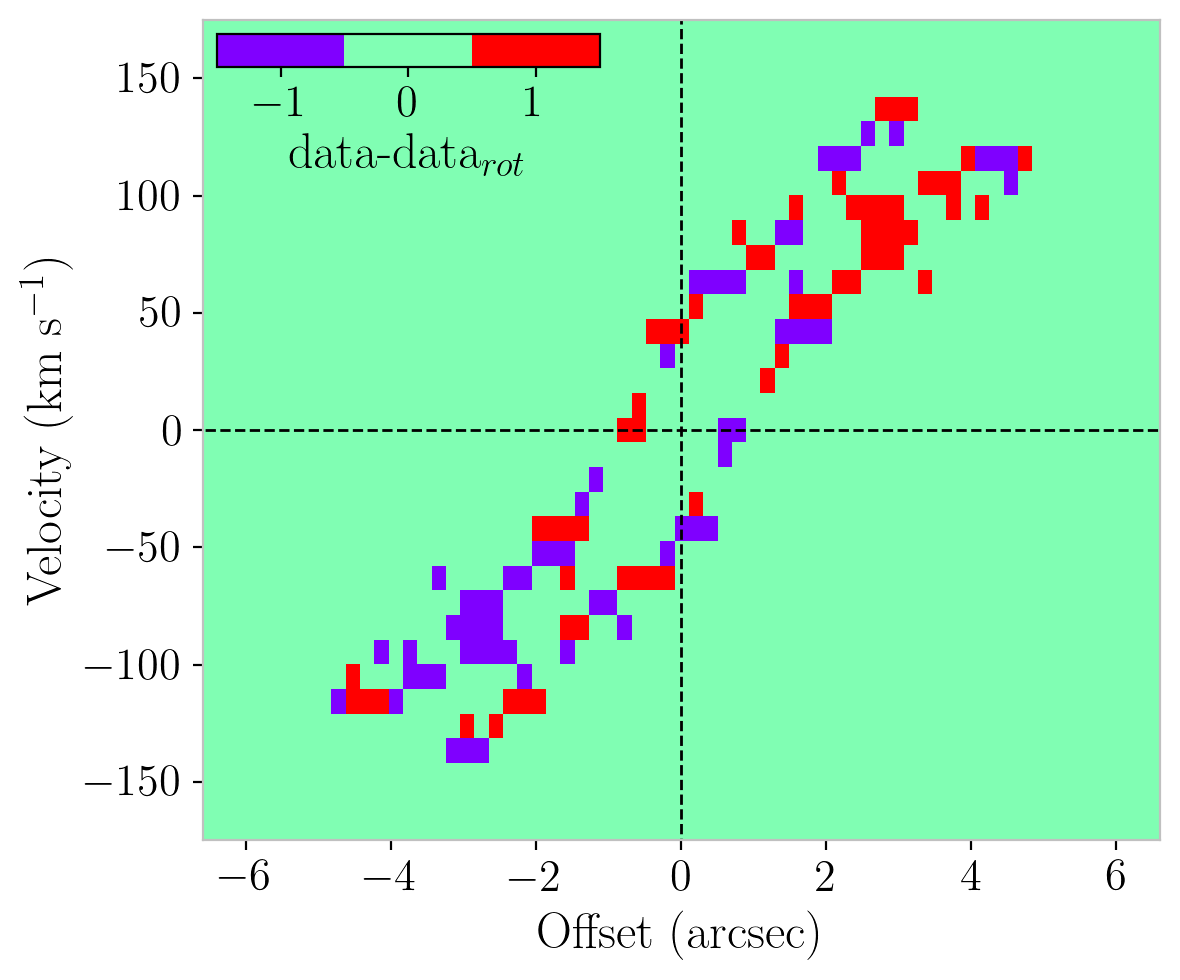}
  \caption{GAMA574200}
\end{minipage}%
\begin{minipage}{.5\textwidth}
  \centering
  \includegraphics[width=.65\linewidth]{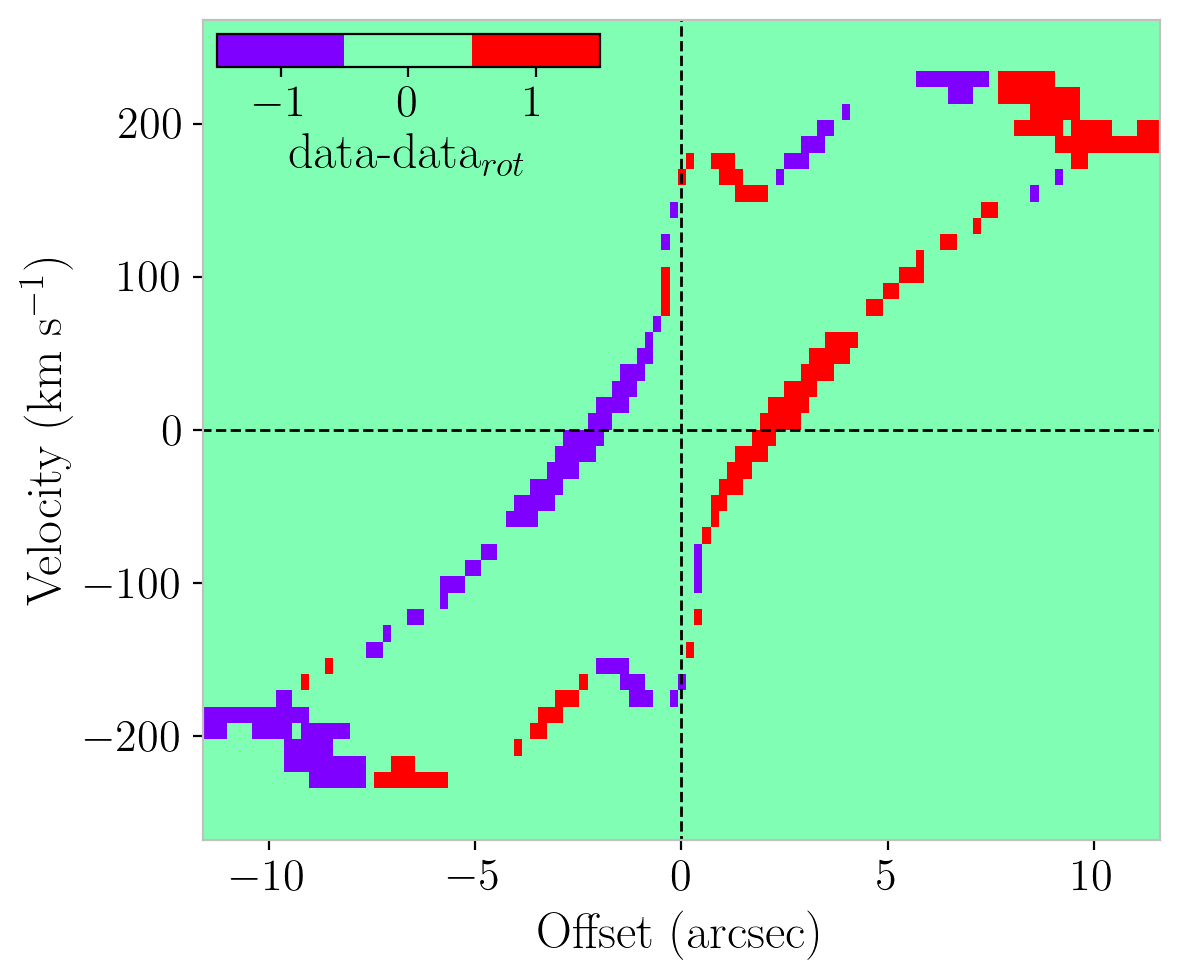}
  \caption{GAMA593680}
\end{minipage}%
\newline
\centering
\begin{minipage}{.5\textwidth}
  \centering
  \includegraphics[width=.65\linewidth]{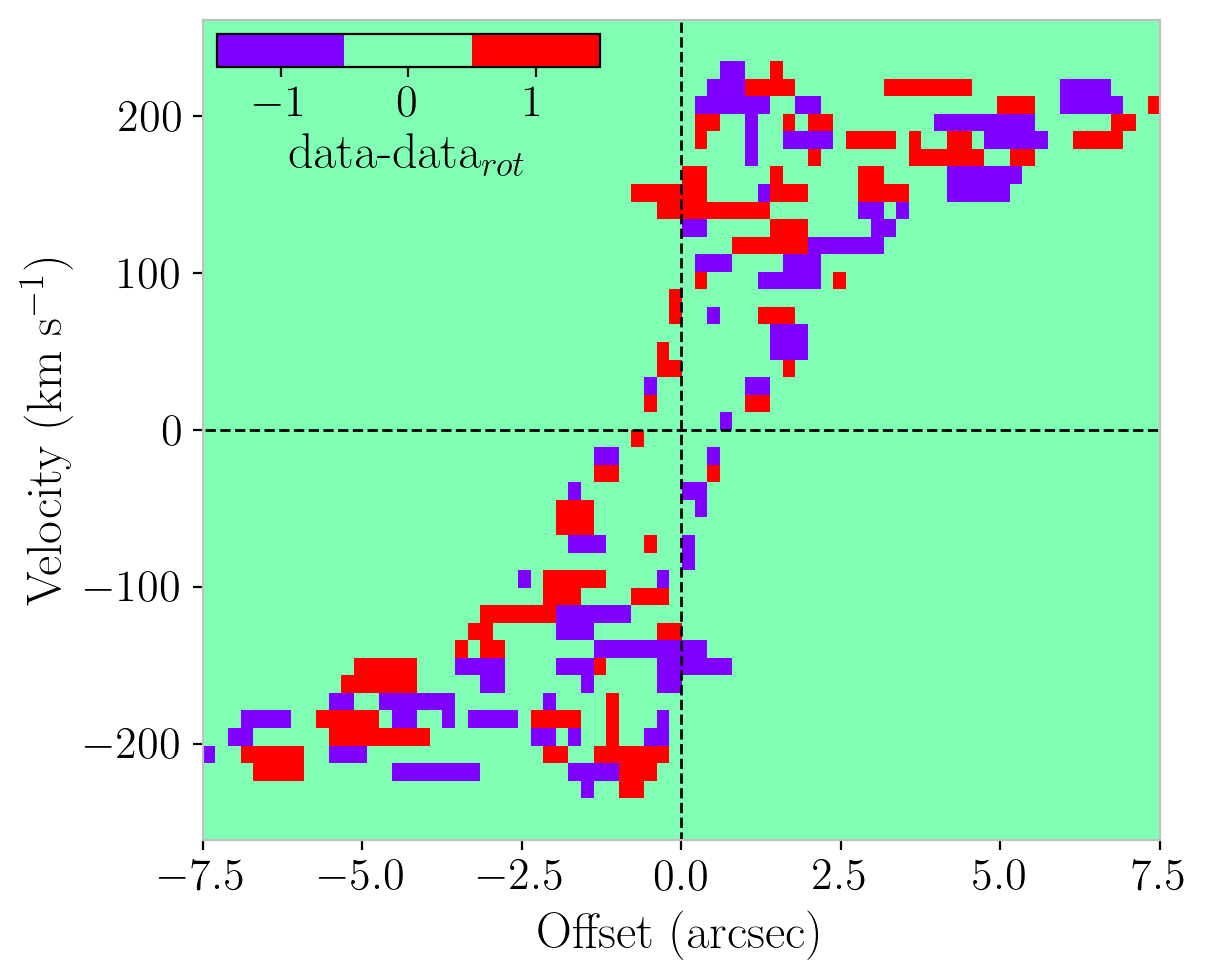}
  \caption{GAMA618906}
\end{minipage}%
\end{figure*}


\bsp	
\label{lastpage}
\end{document}